\documentclass[aps,pra,reprint, amsmath, amssymb,superscriptaddress]{revtex4-1}
 
\usepackage{bm}
\usepackage[retainorgcmds]{IEEEtrantools}
\usepackage{graphicx}
\usepackage{mathrsfs}
\usepackage{amsmath}
\usepackage{amsfonts}
\usepackage{amssymb}
\usepackage{color}
\usepackage{times,txfonts}
\usepackage{nicefrac}
\usepackage{ragged2e}
\usepackage{tikz}
\usepackage{dsfont}
\usepackage{physics}

\usepackage[colorlinks=true,linkcolor=blue,urlcolor=blue,citecolor=blue,pdfusetitle]{hyperref}

\usepackage{blindtext}

\newcommand{\rhono}{\rho_{\rm no}}
\newcommand{\Pno}{P_{\rm no}}
\newcommand{\trans}{^{\text{T}}}

\newcommand{\cmct}{{\bm c}^\dagger \cdot {\bm M}_t \cdot \bm{c}}
\newcommand{\cmctno}{{\bm c}^\dagger \cdot {\bm M}_t^{\rm no} \cdot \bm{c}}

\newcommand{\ymyt}{{{\bm Y}}^\dagger \cdot \bm{\mathcal{M}}_t \cdot {\bm{Y}}}
\newcommand{\rmrt}{{{\bm R}}^\dagger \cdot (i\bm\Omega)\bm{\mathcal{M}}_t \cdot {\bm{R}}}
\newcommand{\ymytno}{{{\bm Y}}^\dagger \cdot \bm{\mathcal{M}}_t^{\rm no} \cdot {\bm{Y}}}
\newcommand{\rmrtno}{{{\bm R}}^\dagger \cdot (i\bm\Omega)\bm{\mathcal{M}}_t^{\rm no} \cdot {\bm{R}}}

\newcommand{\inv}{^{-1}}
\newcommand{\bigzero}{\mbox{\normalfont\Large 0}}

\definecolor{cadmiumgreen}{HTML}{097969}

\newcommand{\appsection}[2]{\setcounter{equation}{0}\setcounter{subsection}{0}
\section*{Appendix #1. #2}
\renewcommand{\theequation}{#1.\arabic{equation}}
              \renewcommand{\thesection}{#1} }

\catcode`\@=11
\def\numberbysection{\@addtoreset{equation}{section}
        \def\theequation{\thesection.\arabic{equation}}}
\numberbysection

\begin{document}

\title{Conditional no-jump dynamics of non-interacting quantum chains}

\author{Michele Coppola}
\email{michele.coppola@univ-lorraine.fr}
\affiliation{LPCT, Université de Lorraine, CNRS, F-54000 Nancy, France}
\author{Dragi Karevski}
\email{dragi.karevski@univ-lorraine.fr}
\affiliation{LPCT, Université de Lorraine, CNRS, F-54000 Nancy, France}
\author{Gabriel T. Landi}
\email{gabriel.landi@rochester.edu}
\affiliation{Department of Physics and Astronomy, University of Rochester, Rochester, New York 14627, USA}

\begin{abstract}
We analyze the open dynamics of quantum systems conditioned on no jumps being detected. 
We first obtain general results relating the no-jump probability and the waiting-time distributions to the conditional evolution of specific system observables. 
These results are applied to single-qubit models, whose conditional dynamics is quite involved and shows a rich set of physical behaviors.
Furthermore, we obtain general expressions for the no-jump dynamics of non-interacting fermionic-bosonic chains undergoing Gaussian-preserving dynamics. We show that the conditional dynamics is determined by a non-linear Riccati-type differential equation for the correlation matrix. 
Finally, we apply our results to chains of hopping particles under inhomogeneous jump rates and boundary driven systems in presence of pairing terms.


\end{abstract}

\maketitle{}

\section{Introduction}
Non-equilibrium open quantum systems continue to raise many physical challenges. In this respect, we may quote the transport phenomena in driven quantum chains, where a non-unitary dynamics is generated by the coupling with external reservoirs (e.g. Lindblad baths) ~\cite{landi2022nonequilibrium}. The quantum trajectory techniques represent a simple tool to access the non-unitary evolution by stochastic averages over single trajectories evolving in time as pure states \cite{daley2014quantum,wiseman1996quantum,dalibard1992wave,gardiner1992wave,carollo2019unraveling}. The evolution of any trajectory is generated by a non-Hermitian effective Hamiltonian, perturbed by the so-called quantum jumps appearing stochastically in time with characteristic rates. The complete statistics of jumps occurring in time, as well as their nature, is captured by the theory of Full Counting Statistics \cite{levitov1993charge,esposito2007fluctuation,esposito2009nonequilibrium,landi2023current}. 
Connected to these concepts, the no-jump probabilities and the waiting-time distributions (WTDs)~\cite{brandes2008waiting,landi2021waiting,landi2023current} represent a powerful theoretical probe to investigate the interplay between non-Hermitian dynamics and jump events. 

In Ref.~\cite{landi2021waiting} one of us approached the problem of computing the WTDs for boundary driven quadratic fermion chains, subject to single-particle gain-loss processes. In that case, quantum jumps may be pictorially seen as the action of some detectors monitoring the exchange of particles between the system and the external environment.
In this paper we expand on those preliminary results in various ways. 
First, we generalize the boundary-driven setup by keeping the coupling with the bath(s) along all the sites of the chain, where the coupling amplitude is modulated by inhomogeneous jump rates. Second, we introduce non-ideal monitoring efficiencies, by assigning each jump channel $j$ a real coefficient $\Lambda_j\in[0,1]$, with $\Lambda_j=1$ for perfect efficiency and $\Lambda_j=0$ when the channel is not monitored at all. Therefore, our setting reduces to the case studied into Ref.~\cite{landi2021waiting} for the specific choice $\Lambda_j=1$. In other terms, we extend the view to the missed detection of some jump events due the intrinsic imperfection of the monitoring apparatus. This study is clearly useful for experimentalists, where the efficiency of the monitoring apparatus is not ideal and may change with the work conditions. The main purpose of this paper is to compute the conditional no-jump dynamics of the density operator, which is the time evolution conditioned on no jumps being detected. In this view, the waiting time distribution $W_k(t)$ is the probability distribution to detect the first jump in channel $k$ after time $t$. Third, our efforts will also be addressed to the no-jump probability $P_{\rm no}(t)$, which is the probability to detect the first jump at time $t$. Fourth, we also derive results for particle non-conserving Hamiltonian operators and, fifth, we study both bosonic and fermionic cases. In conclusion, this work aims to be a complete study of the no-jump dynamics of non-interacting quantum chains. 


In Sec.~\ref{sec:Conditional-no-jump-dynamics} we define the Lindblad evolution and the no-jump dynamics, characterized by damping rates and monitoring efficiencies. We introduce the no-jump probability and the WTDs, generalizing the definitions into Ref.~\cite{landi2021waiting} for non-ideal efficiencies. In Sec.~\ref{Qubit example}, we study two simple applications involving a single qubit, either incoherently or coherently driven. 
In Sec.~\ref{nopairingterms} we approach the many-body problem of a chain of hopping particles with single particle gain-loss processes. Thanks to the Gaussian ansatz, the no-jump dynamics turns to be governed by a Riccati-type differential equation for the correlation matrix. As an application (Sec.~\ref{example1}), we  study a chain of hopping fermions with specific absorption-emission profiles. In Sec.~\ref{sec:pairing} and Sec.~\ref{sec:pairing_b}, we also derive the no-jump dynamics for particle non-conserving Hamiltonians (pairing terms). As an example, we consider the no-jump dynamics of a $2$-site boundary driven system (Sec.~\ref{example2}).
Our main findings are summarized in Sec.~\ref{discuss-concl}, where we draw some future perspectives. Several appendices contain technical details and background.

\section{\label{sec:Conditional-no-jump-dynamics}General framework }

\begin{figure}
    \centering
    \includegraphics[width=0.5\textwidth]{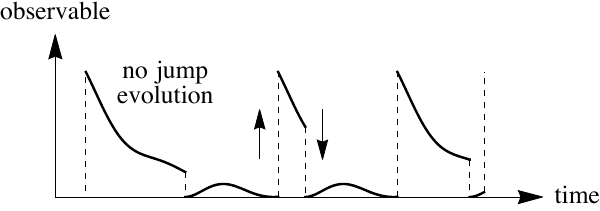}
    \caption{Example of a quantum jump trajectory, as viewed from the perspective of a pictorial system observable.
    Abrupt jumps occur at random times and, in between, the observable evolves smoothly,  governed by the so-called no-jump conditional evolution. 
    }
    \label{fig:jump_trajectory_drawing}
\end{figure}

Let $\rho$ be the density operator of a quantum system whose dynamics is generated by the Liouvillian $\mathcal{L}(\rho)$ 
\begin{equation}\label{M}
    \frac{d\rho}{dt} = \mathcal{L}(\rho)= -i[H,\rho] + \mathcal{D}(\rho),
\end{equation}
where $H$ denotes the Hamiltonian and $\mathcal{D}(\rho)$ is the dissipator
\begin{equation}\label{dissipator1}
\mathcal{D}(\rho)= \sum_{k} D_\rho[ L_k]
= \sum_k  \left(L_k \rho  L_k^\dagger - \frac{1}{2} \{ L_k^\dagger  L_k, \rho\}\right),
\end{equation} 
characterized by a generic set of Lindblad jump operators $L_k$.

In the quantum jump unravelling~\cite{wiseman1996quantum} the terms $L_k\rho L_k^\dagger$ represent quantum jumps, so that the dynamics can be viewed as a series of discrete jumps, followed by a smooth no-jump evolution. (Fig.~\ref{fig:jump_trajectory_drawing}). Each Lindblad operator corresponds to a different channel where the jump might occur, which is signaled by a click in a classical detector.
In this work, we assume that one can independently choose to measure each $L_k$ with different efficiencies $\Lambda_k$, with $\Lambda_{k} = 1$ being perfect monitoring and $\Lambda_{k} = 0$ meaning that channel is not monitored at all. 
We now split Eq.~\eqref{M} as 
\begin{equation}\label{split}
    \mathcal{L} = \mathcal{L}_0 + \mathcal{L}_1, 
\end{equation}
where 
\begin{align}\label{L1}
    \mathcal{L}_1(\rho) &= \sum_{k} \mathcal{J}_{k}\rho
    = \sum_k \Lambda_k L_k \rho L_k^\dagger,
\end{align}
with each jump channel specified by the super-operator $\mathcal{J}_{k}$. 
The conditional no-jump operator is
\begin{equation}\label{L0}
    \mathcal{L}_0(\rho)= -i[H,\rho] + \sum_k \left[(1-\Lambda_k) L_k \rho L_k^\dagger - \frac{1}{2}\{L_k^\dagger L_k,\rho\}\right]. 
\end{equation}
If all jumps are perfectly monitored, $\Lambda_k =1$ $\forall k$, then $\mathcal{L}_0$ can also be written as 
\begin{equation}\label{L0_all_monitored}
    \mathcal{L}_0(\rho) = -i \Big(H_{\rm e} \rho - \rho H_{\rm e}^\dag\Big), 
\end{equation}
where 
\begin{equation}\label{He}
    H_e = H -\frac{i}{2} \sum_k  L_k^\dagger L_k,
\end{equation}
is a kind of non-Hermitian Hamiltonian. 

The solution $\rho(t) = e^{\mathcal{L}t} \rho_0$, called the unconditional dynamics (since it is ignorant about whether any jumps occurred or not) can be expanded in a Dyson series as 
\begin{align}\label{dyson}
    \rho(t) &= e^{\mathcal{L}_0 t} \rho_0 + \int\limits_{0}^t dt_1 e^{\mathcal{L}_0 (t-t_1)} \mathcal{L}_1 e^{\mathcal{L}_0 t_1} \rho_0 \nonumber \\
    &+ \int\limits_{0}^{t} dt_2 \int\limits_{0}^{t_2} dt_1 e^{\mathcal{L}_0 (t-t_2)} \mathcal{L}_1 e^{\mathcal{L}_0 (t_2-t_1)} \mathcal{L}_1 e^{\mathcal{L}_0 t_1}\rho_0 + \ldots,
\end{align} 
where $\rho(t=0)=\rho_0$. The first term $e^{\mathcal{L}_0 t} \rho_0$ is the  evolution conditioned on having no jumps up to time $t$,  the second is conditioned on having exactly one jump at  time $t_1$, etc~\cite{plenio1998quantum,landi2022nonequilibrium}. Let us focus on the first term, i.e. the no-jump conditional dynamics. Since $\mathcal{L}_0$ is not a proper Liouvillian, the evolved state $e^{\mathcal{L}_0 t} \rho_0$ is not normalized. We define the unnormalized density  matrix 
\begin{equation}\label{rho_no}
    \tilde{\rho}_{\rm no}(t) = e^{\mathcal{L}_0 t} \rho_0,
    \qquad 
    \frac{d\tilde{\rho}_{\rm no}}{dt} = \mathcal{L}_0 (\tilde{\rho}_{\rm no}).
\end{equation}
If all channels are monitored ($\Lambda_k=1$), this reduces to 
\begin{equation}\label{solu-full-monit}
    \tilde{\rho}_{\rm no}(t) = e^{-i H_e t} \rho_0 e^{i H_e^\dagger t},
\end{equation}
which represents a non-Hermitian dynamics. 
The conditional evolution is indeed a very natural way of implementing non-Hermitian physics~\cite{Purkayastha2022}.
The trace of $\tilde{\rho}_{\rm no}(t)$ is precisely the probability that no jump occurs between $[0,t]$:
\begin{equation}\label{P_no}
    \Pno(t) = \tr~ \tilde{\rho}_{\rm no}(t).
\end{equation}
The corresponding normalized conditional state is 
\begin{equation}
    {\rho}_{\rm no}(t)= \frac{ \tilde{\rho}_{\rm no}(t)}{\Pno(t)}.
\end{equation}
The probability $\Pno(t)$ satisfies $\Pno(0) = 1$. In the limit $t\to\infty$, it may either tend to zero or to a finite value. 
When $\Pno(\infty)=0$ it means a jump must eventually occur.
Conversely, $\Pno(\infty)\neq0$ means it might never do. This happens when the system has a dark subspace, which is immune to the dissipator in \eqref{dissipator1}. 

The unconditional evolution of any system observable $\mathcal{O}$ can be obtained from Eq.~\eqref{M} and reads
\begin{equation}\label{O_evo_unconditional}
    \frac{d\langle \mathcal{O}\rangle}{dt} = \langle \mathcal{L}^\dagger (\mathcal{O})\rangle, 
\end{equation}
where 
\begin{align}\label{adjoint}
\mathcal{L}^\dagger(\mathcal{O})=& i[H,\mathcal{O}] + \sum_k  \left[L_k^\dagger \mathcal{O} L_k - \frac{1}{2}\{L_k^\dagger L_k,\mathcal{O}\}\right],
\end{align}
is the adjoint dissipator and $\langle \ldots \rangle$ denotes the expectation value over the unconditional state $\rho(t)$. We can also write down a similar evolution equation for the conditional expectation value
\begin{equation}
    \langle \mathcal{O} \rangle_{\rm no} := \tr(\mathcal{O} {\rho}_{\rm no}(t)) =\frac{\tr(\mathcal{O} \tilde{\rho}_{\rm no}(t))}{\Pno(t)},
\end{equation} 
i.e., properly normalized. 
Using Eq.~\eqref{L0} and taking into account the time-dependence of $\Pno(t)$, one finds 
\begin{align}\label{O_evo_conditional}
    \frac{d\langle \mathcal{O}\rangle_{\rm no}}{dt} =& \,\langle \mathcal{L}^\dagger(\mathcal{O})\rangle_{\rm no}\nonumber\\
    & - \sum_k \Lambda_k \left[ \langle L_k^\dagger \mathcal{O} L_k \rangle_{\rm no} - \langle \mathcal{O}\rangle_{\rm no} \langle L_k^\dagger L_k \rangle_{\rm no}\right].
\end{align}
The first term is the same as in the unconditional evolution~\eqref{O_evo_unconditional}, while the others are new contributions stemming from monitoring the jumps.

Next we differentiate Eq.~\eqref{P_no}, substituting $\mathcal{L}_0 = \mathcal{L}- \mathcal{L}_1$. Since $\mathcal{L}$ is traceless, using Eq.~\eqref{L1} we find
\begin{equation}\label{P_no_rhono}
    \frac{d{P}_{\rm no}}{dt} = - \beta(t)\,\Pno,
\end{equation}
where 
\begin{equation}\label{P_no_rhono2}
    \beta(t):=\tr\left(\mathcal{L}_1 (\rho_{\rm no})\right)=\sum_k\Lambda_k  \langle L_k^\dagger L_k \rangle_{\rm no} (t).
\end{equation}
The solution of Eq.~\eqref{P_no_rhono} is 
\begin{equation}\label{P_no_observables}
    P_{\rm no}(t) = \exp\left\{ - \int\limits_0^t dt'\beta(t')\right\}.
\end{equation}
This result relates the no-jump probability with the expectation values of the jump operators. 
It therefore shows that it is not necessary to compute the full state $\rhono$ to obtain $\Pno$, but instead we only need $\langle L_k^\dagger L_k\rangle_{\rm no}$, which can be obtained from Eq.~\eqref{O_evo_conditional}. Eq.~\eqref{P_no_observables} also establishes a criteria for the existence or not of a dark subspace. 
If $\beta^\infty=\lim_{t\to\infty}\beta(t)$ then, in the long-time limit, the no-jump probability behaves as $P_{\rm no}(t)\sim \exp{-\beta^\infty t}$. Thus, $\lim_{t\to\infty}\Pno(t) = 0$ if at least one term $\Lambda_k\langle L_k^\dagger L_k\rangle_{\rm no}^\infty \neq 0$. 
The existence of dark subspaces is thus not only a property of the master equation~\eqref{M}, but also of what is monitored (i.e., the $\Lambda_k$). 
As a sanity check, if nothing is monitored ($\Lambda_k = 0 ~\forall k$) we simply find $\Pno(t)=1$.

Computing the no-jump probability $\Pno(t)$ also plays a key role in the study of the WTDs. The waiting-time distribution $W_{k}(t)$ is defined as
\begin{align}
    W_{k}(t)&:=\tr \Big\{\mathcal{J}_{k} e^{\mathcal{L}_0 t}({\rho}_0)\Big\}, \nonumber\\[0.2cm]
    &=\Pno(t)\tr \Big\{\mathcal{J}_{k} \:{\rho}_{\rm no}(t)\Big\}.
    \\[0.2cm]
    &= \Lambda_k P_{\rm no}(t) \langle L_k^\dagger L_k\rangle_{\rm no}(t).
    \label{WTD_general_observable}
\end{align}
which is the probability distribution to detect the first jump in the channel $k$ at time $t$. The WTDs are normalized as 
\begin{equation}
    \sum_k\int_0^\infty W_{k}(t)\: dt=1.
\end{equation}
From these results we therefore reach the conclusion that both $P_{\rm no}(t)$ and $ W_{k}(t)$ can be computed by the conditional evolution of the Lindblad jump operators $\langle L_k^\dagger L_k\rangle_{\rm no}(t)$.

\section{\label{Qubit example}Qubit example}
Consider a single qubit with Rabi drive $H = \frac{\Omega}{2}\sigma_x$ and $\Omega>0$. Assume a single emission channel described by the Lindblad operator $L = \sqrt{\Gamma} \sigma_-$, with jump frequency $\Gamma$ and monitoring efficiency $\Lambda^-$. The resulting unconditional dynamics is 
\begin{equation}\label{qubit_M}
    \mathcal{L}(\rho) = - i\frac{\Omega}{2} [\sigma_x, \rho] + \Gamma \left[ \sigma_- \rho \sigma_+ - \frac{1}{2} \{\sigma_+\sigma_-,\rho\}\right],
\end{equation}
where the $\sigma$'s are the Pauli matrices. We study the conditional dynamics by applying
Eq.~\eqref{O_evo_conditional}.

\begin{figure}
    \centering
    \includegraphics[width=\columnwidth]{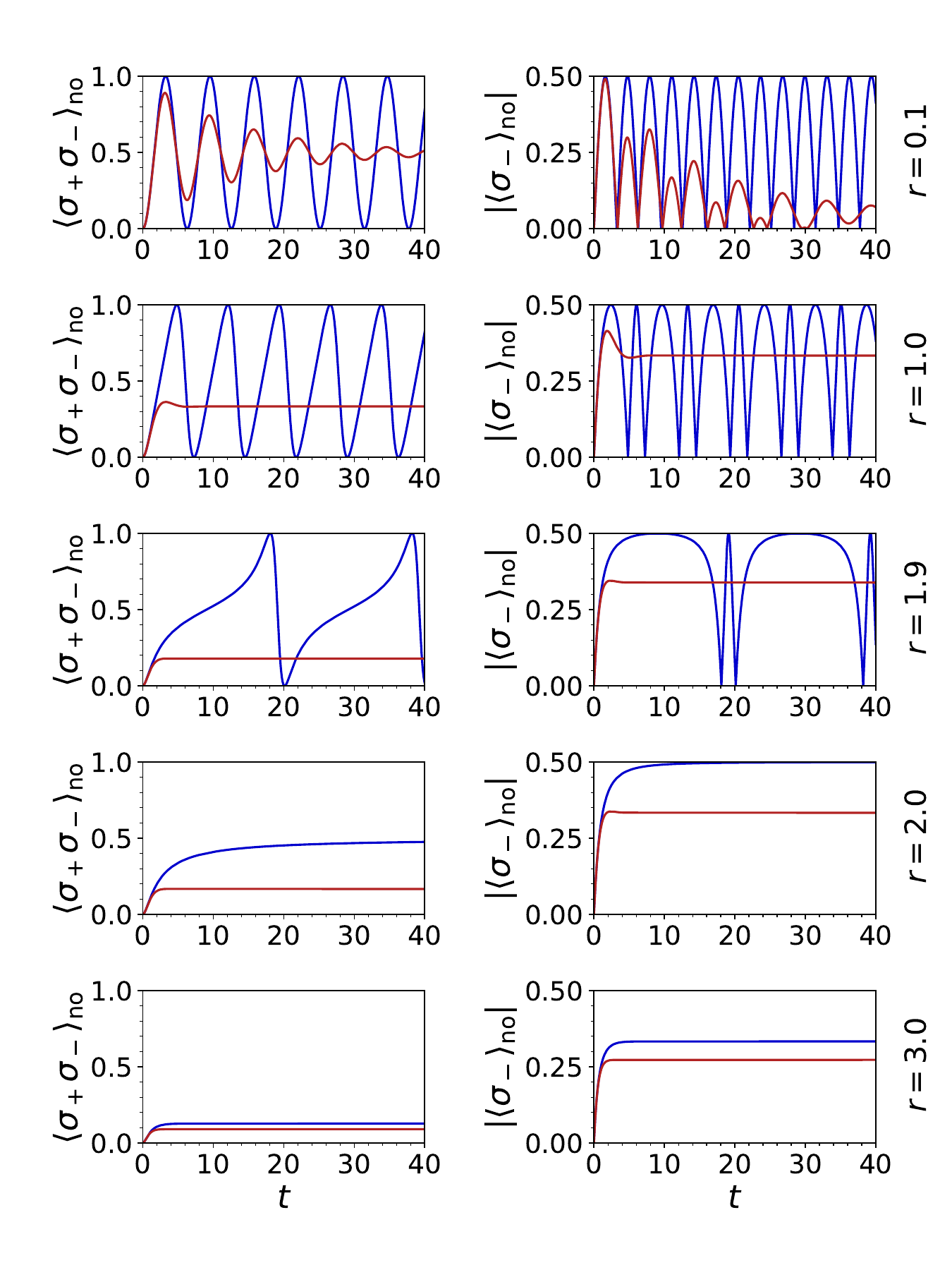}
    \caption{
    Conditional ($\Lambda^-=1$; blue) and unconditional ($\Lambda^-=0$; red) dynamics of a single qubit governed by the master equation~\eqref{qubit_M}. 
    Left: population $q = \langle \sigma_+ \sigma_-\rangle_{\rm no}$.
    Right: coherence $|c| = |\langle \sigma_-\rangle_{\rm no}|$.
    Each row is for $\Omega=1$ and a different value of $r=\Gamma/\Omega = 0.1, 1.0, 1.9, 2.0, 3.0$ (from top to bottom).
    }
    \label{fig:qubit_example1}
\end{figure}

\begin{figure*}[!t]
    \centering
    \includegraphics[scale=0.5]{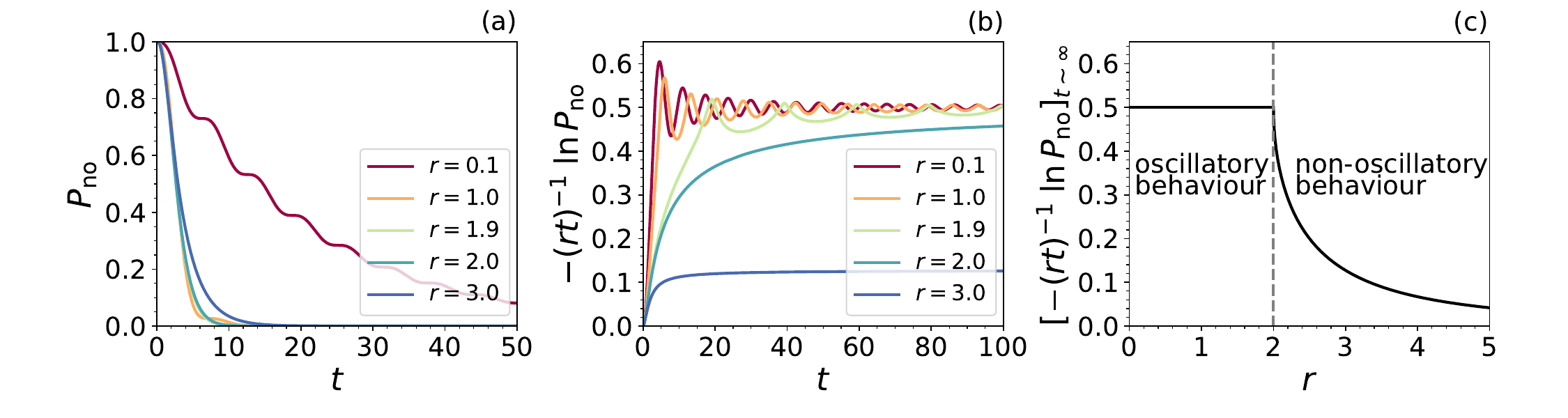}
    \caption{(a): Evolution of the no-jump probability $P_{\rm no}$ for $\Lambda^-=1.0$, resulting from Fig.~\ref{fig:qubit_example1} (we set $\Omega=1.0$ and $r=0.1,1.0,1.9,2.0,3.0$). 
    (b): Evolution of $-\frac{1}{rt} \ln P_{\rm no}$ as a function of time $t$, which approaches a finite value for $t\to \infty$. 
    (c) The quantity $\alpha(r)/r$ [Eq.~\eqref{qubit_example_alpha_r}], representing the long-time behavior of $P_{\rm no}$.
    }
    \label{fig:qubit_example2}
\end{figure*}

The natural operators to consider are the population $\sigma_+\sigma_-$ and the coherence $\sigma_-$. 
We thus define the conditional expectation values
$q = \langle \sigma_+\sigma_-\rangle_{\rm no}$ and
$c = \langle \sigma_- \rangle_{\rm no}$.
Eq.~\eqref{O_evo_conditional} then yields a system of two coupled non-linear equations 
\begin{align}
\label{qubit_equation_1}
    \frac{dq}{dt} &= \frac{i\Omega}{2} (c-c^*) - \Gamma q + \Gamma \Lambda^- q^2, 
    \\[0.2cm]
\label{qubit_equation_2}        
    \frac{dc}{dt} &= \frac{i\Omega}{2} (2q-1) - \frac{\Gamma}{2} c + \Gamma \Lambda^- c q.
\end{align}
Once we have their solution, we immediately obtain the no-jump probability~\eqref{P_no_observables} with
\begin{equation}\label{qubit_Pno}
    \beta(t) = \Lambda^- \Gamma q(t).
\end{equation}
We prepare the system in the state $\ket{\downarrow}_z$ (corresponding to $\sigma_z\ket{\downarrow}_z=-\ket{\downarrow}_z$). The parameters $(\Omega,\Gamma)$  affect the no-jump dynamics in a non-trivial way, but Eqs. \eqref{qubit_equation_1}, \eqref{qubit_equation_2} clearly show that the steady-state is uniquely determined by the rate $r=\Gamma/\Omega>0$, if it exists.

Fig.~\ref{fig:qubit_example1} shows the evolution of $q(t)$ (left) and $|c(t)|$ (right) as a function of time for $\Omega=1$ and different values of the rate $r$. The unconditional dynamics ($\Lambda^- = 0$; no monitoring) shows the typical behavior of a damped oscillator and always relaxes to a steady-state,  
\begin{equation}
   q^{\infty}_{\Lambda^-=0}=\frac{1}{2+r^2},\qquad  \abs{c^{\infty}_{\Lambda^-=0}}=rq^{\infty}_{\Lambda^-=0}.
\end{equation}
The general conditional case $\Lambda^-\in (0,1)$ is more difficult but still shares the same features with the case $\Lambda^-=0$. The oscillations are damped and 
$q^\infty_{\Lambda^-}$ is the real positive solution of the cubic equation
\begin{equation}\label{q_inf_L}
2r^2(\Lambda^-)^2 (q^\infty_{\Lambda^-})^3-3\Lambda^- r^2 (q^\infty_{\Lambda^-})^2+(r^2+2)q^\infty_{\Lambda^-}-1=0,
\end{equation}
with
\begin{equation}\label{c_inf_L}
\abs{c^{\infty}_{\Lambda^-}}=rq^\infty_{\Lambda^-}(1-\Lambda^- q^\infty_{\Lambda^-}).
\end{equation}
Conversely, the full monitoring case ($\Lambda^- = 1$) provides new physical insights. We can identify a critical rate $r_c=2$ which distinguishes two different regimes. If $r<r_c$, $q(t)$ and $|c(t)|$ are periodic functions and show an oscillatory phase; if $r\geq r_c$, the oscillations are suppressed with an emerging asymptotic state,
\begin{align}
    q^{\infty}_{\Lambda^-=1}&=\frac{1}{2}-\frac{1}{2}\sqrt{1-\frac{4}{r^2}},\\
    \abs{c^{\infty}_{\Lambda^-=1}}&=rq^{\infty}_{\Lambda^-=1}(1-q^{\infty}_{\Lambda^-=1}).
\end{align}
The insight is that this transition is associated to the spectrum of the non-Hermitian Hamiltonian~\eqref{He}
\begin{equation}
    H_e = \frac{\Omega}{2} \sigma_x - i\frac{\Gamma}{2} \sigma_+ \sigma_- = 
    \frac{1}{2}\begin{pmatrix}
    -i \Gamma & \Omega \\
    \Omega & 0 
    \end{pmatrix}.
\end{equation}
The no-jump probability is plotted in Fig.~\ref{fig:qubit_example2}~(a) for several values of $r$ and $\Omega=1$, showcasing the differences between the oscillatory phase $r<r_c=2$ and the damped phase, $r>r_c$. 
In the long-time limit the no-jump distribution behaves as 
\begin{equation}\label{qubit_example_alpha_r}
   P_{\rm no}\sim e^{-\alpha(r)t},\qquad  \alpha(r)=\begin{cases}
    r/2 & r < r_c, \\[0.2cm]
     rq^{\infty}_{\Lambda=1} & r \geq r_c,
    \end{cases}
\end{equation}
as shown in Fig.~\ref{fig:qubit_example2}~(b), where we plot $-(rt)^{-1}\ln P_{\rm no}$ as a function of time $t$.
Finally, in Fig.~\ref{fig:qubit_example2}~(c) we plot $\alpha(r)$, which showcases the abrupt transition between the oscillatory and non-oscillatory phases. 

\subsection{Incoherent dynamics with emission and absorption}

Next we consider a two-level system with trivial Hamiltonian $H = \frac{\Omega}{2} \sigma_z$, interacting with a fermionic bath. The coupling is
described by two jump operators 
$L_- = \sqrt{\Gamma^-} \sigma_-$ and 
$L_+ = \sqrt{\Gamma^+} \sigma_+$. $L_+$ and $L_-$ determine, respectively, the emission and the absorption of an excitation.  
We shall adopt the following parametrization
\begin{equation}
    \Gamma^+ = \Gamma f, \qquad 
    \Gamma^- = \Gamma(1-f),
\end{equation}
where $\Gamma>0$ is the coupling strength to the reservoir and $f = (e^{\Omega/T}+1)^{-1} \in [0,1]$ is the Fermi-Dirac distribution of the reservoir (with $T$ being the temperature and $k_B = 1$).
We again apply Eq.~\eqref{O_evo_conditional} to $\mathcal{O} = \sigma_+ \sigma_-$. Let $\Lambda^-$ and $\Lambda^+$ be the monitoring efficiencies for emission and absorption, respectively.
The qubit population conditioned on no jumps being detected, $q = \langle \sigma_+ \sigma_- \rangle_{\rm no}$, evolves according to the non-linear equation 
\begin{equation}\label{qubit_incoherent_population_equation}
    \frac{1}{\Gamma}\frac{dq}{dt} = f - q + (1-f) \Lambda^- q^2 - f \Lambda^+ (1-q)^2 .
\end{equation}
The first two terms are the same as in the unconditional dynamics, while the last two are the new contributions from monitoring. 
Plugging the solution in Eq.~\eqref{P_no_observables} also yields the no-jump probability in the form \eqref{P_no_observables} with 
\begin{equation}
    \beta(t) = \Gamma(1-f)\Lambda^- q(t) + \Gamma f \Lambda^+ (1-q(t)).
\end{equation}

\begin{figure}
    \centering
    \includegraphics[width=\columnwidth]{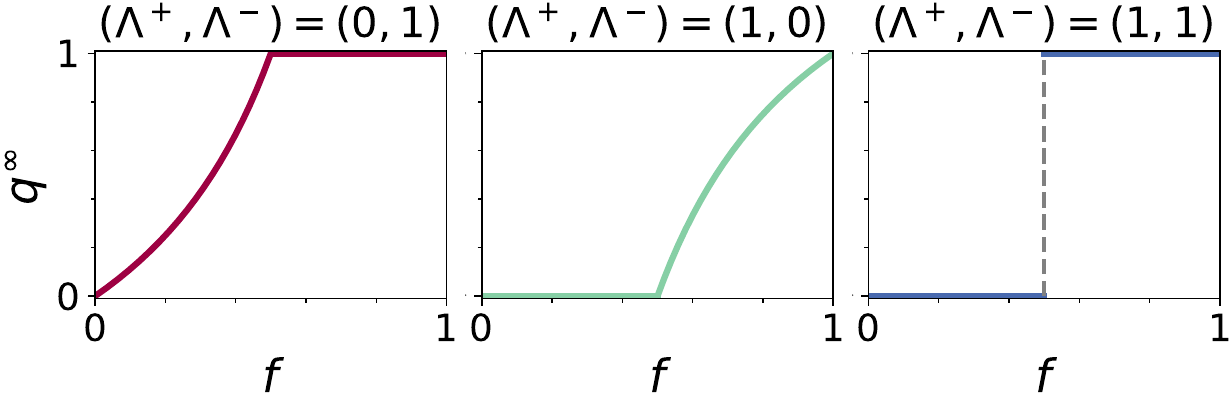}
    \caption{Steady-state population $q^\infty$ of the conditional dynamics~\eqref{qubit_incoherent_population_equation} as a function of the Fermi-Dirac occupation $f$, for different choices of $(\Lambda^+,\Lambda^-)$.  
    }
    \label{fig:qubit_example_incoherent}
\end{figure}

Solving $\frac{dq}{dt}=0$ we find the stationary solution of Eq.~\eqref{qubit_incoherent_population_equation}, which is
\begin{align}
    q^\infty =&\frac{\sqrt{\left(2 f \Lambda ^+-1\right){}^2-4 f\left(1- \Lambda ^+\right)
    \left(\Lambda^-
   -f (\Lambda ^- + \Lambda ^+)\right)}}{2 \left(f \Lambda ^-+f \Lambda ^+-\Lambda ^-\right)}\nonumber\\
   &+\frac{2 f \Lambda ^+-1}{2 \left(f \Lambda ^-+f \Lambda ^+-\Lambda ^-\right)}.
\end{align}
The result is independent of $\Gamma$. The cases $\Lambda^+,\Lambda^- = 0,1$ behave as follows (see also Fig.~\ref{fig:qubit_example_incoherent}). If $\Lambda^+ = \Lambda^- = 0$ we get the unconditional steady-state $q^\infty = f$. If we only monitor emissions, but not absorptions ($\Lambda^- = 1$ and $\Lambda^+ = 0$) we get 
\begin{equation}\label{monitor_emissions}
    q^\infty = 
    \begin{cases}
    \frac{f}{1-f} & f < 1/2, \\[0.2cm]
    1 & f > 1/2
    \end{cases},
\end{equation}
which is the red curve in Fig.~\ref{fig:qubit_example_incoherent}. 
The rationale is that, since the evolution is conditioned on no emission events having occurred, it is more likely that an excitation is present, leading to a population  $q^\infty \geqslant f$. In the opposite scenario ($\Lambda^+ =1$ and $\Lambda^-=0$) we get 
\begin{equation}\label{monitor_absorptions}
    q^\infty = 
    \begin{cases}
    0 & f < 1/2, \\[0.2cm]
    \frac{2f-1}{f} & f > 1/2,
    \end{cases}
\end{equation}
which is the green curve in Fig.~\ref{fig:qubit_example_incoherent}. Here the observer knows with certainty that no excitations are absorbed, and this leads to a population  $q^\infty \leqslant f$, i.e. smaller than in the unconditional case. 
Finally, if $\Lambda^- = \Lambda^+ = 1$, we get a step function (blue curve), which reflects the non-trivial interplay of conditioning on both channels. 


\section{\label{nopairingterms}No-jump dynamics of non-interacting particles}
The previous examples show that even for a single qubit the conditional no-jump dynamics is quite involved. 
In order to provide further analytical insights, we now specialize to the case of many-body one dimensional systems with Gaussian-preserving dynamics. 

More specifically, we consider a chain with $L$ sites, each described by an annihilation operator $c_i$,  either bosonic or fermionic,
\begin{align}
    \label{ferm_rules}\{c_i,c_j^\dag\} &= \delta_{ij}, \quad \{c_i,c_j\} =0,\qquad\text{fermions}\\
    \label{bos_rules}\comm{c_i}{c^\dag_j} &= \delta_{ij}, \quad \comm{c_i}{c_j}=0. \qquad\quad\text{bosons}
\end{align}
The system is subject to the Lindblad master equation \eqref{M} with dissipator of the form 
\begin{equation}\label{dissipator_X}
    \mathcal{D}(\rho) = \sum_{i=1}^L \left(\Gamma_i^- D_\rho[X_i] + \Gamma_i^+ D_\rho [X_i^\dagger]\right).
\end{equation}
The dissipator~\eqref{dissipator_X} consists of $2L$ jump  operators 
$\left\{\sqrt{\Gamma^-_i}X_i\,,\,\sqrt{\Gamma^+_i}X_i^\dag\right\}\,\,\forall i\in\{1,\dots,L\}$ and rates $\Gamma_i^\pm \geqslant 0$. By hypothesis, the Lindblad operators are linear in the creation/annihilation operators; that is, $\bm{X} = \bm\alpha^\dagger \bm{c}$, for $\bm{c}^\dag=(c_1^\dag,\dots,c_L^\dag)$ and some unitary matrix $\bm\alpha$.
The form of the dissipator~\eqref{dissipator_X} is thus 
\begin{IEEEeqnarray}{rCl}
\label{L-2}
     \mathcal{D}({\rho}) &=& \sum\limits_{ij=1}^L (\bm{\gamma}^-)_{ij}\left[ {c}_j {\rho} {c}_i^\dagger - \frac{1}{2} \{{c}_i^\dagger {c}_j, {\rho}\}\right] 
   \nonumber \\
    & &+ \sum\limits_{ij=1}^L (\bm{\gamma}^+)_{ij}\left[ {c}_i^\dagger {\rho} {c}_j - \frac{1}{2} \{{c}_j {c}_i^\dagger, {\rho}\}\right]. 
\end{IEEEeqnarray}
where 
$\bm{\gamma}^-=\bm{\alpha}\bm{\Gamma}^-\bm{\alpha}^\dag$ and 
$\bm{\gamma}^+=\bm{\alpha}\bm{\Gamma}^+\bm{\alpha}^\dag$.
Similarly, if $\Lambda^+_i,\,\Lambda^-_i\in[0,1]$ are the monitoring efficiencies for $X_i^\dag,\,X_i$ then the superoperators $\mathcal{L}_0$ and $\mathcal{L}_1$ read
\begin{IEEEeqnarray}{rCl}
\label{L0-2}
     \mathcal{L}_0({\rho}) &=& -i [{H},{\rho}] + \sum\limits_{ij=1}^L \left[ [\bm{\gamma}^-(\mathds{1}-\bm{\lambda}^-)]_{ij} {c}_j {\rho} {c}_i^\dagger - \frac{(\bm{\gamma}^-)_{ij}}{2} \{{c}_i^\dagger {c}_j, {\rho}\}\right] 
   \nonumber \\
    & &+ \sum\limits_{ij=1}^L \left[ [\bm{\gamma}^+(\mathds{1}-\bm{\lambda}^+)]_{ij} {c}_i^\dagger {\rho} {c}_j - \frac{(\bm{\gamma}^+)_{ij}}{2} \{{c}_j {c}_i^\dagger, {\rho}\}\right], 
\end{IEEEeqnarray}
\begin{equation}
\label{L0-3}
\mathcal{L}_1({\rho}) = \sum\limits_{ij=1}^L \left(\bm{\gamma}^-\bm{\lambda}^-\right)_{ij} {c}_j {\rho} {c}_i^\dagger + \sum\limits_{ij=1}^L \left(\bm{\gamma}^+\bm{\lambda}^+\right)_{ij} {c}_i^\dagger {\rho} {c}_j,
\end{equation}
where
$\bm{\lambda}^+=\bm{\alpha}\bm{\Lambda}^+\bm{\alpha}^\dag$, $\bm{\lambda}^-=\bm{\alpha}\bm{\Lambda}^-\bm{\alpha}^\dag$ and $H$ is the Hamiltonian of the system. By hypothesis, we assume that $H$ is quadratic in the creation and annihilation operators.

The aim of the next sections is to study the no-jump dynamics~\eqref{L0-2} of non-interacting quantum chains, that means the evolved state ${\tilde\rho}_{\rm no}(t)=e^{\mathcal{L}_0t}\rho_0$ and the normalized state ${\rho}_{\rm no}(t)={\tilde\rho}_{\rm no}(t)/\tr\left({\tilde\rho}_{\rm no}(t)\right)$. The two-point functions $\langle {c}_i^\dagger {c}_j \rangle_{\rm no}(t)=\tr\left({c}_i^\dagger {c}_j{\rho}_{\rm no}(t)\right)$ give access to the no-jump probability and the WTDs, according to Eqs.~(\ref{P_no_rhono},\ref{P_no_rhono2},\ref{WTD_general_observable},\ref{L0-3}).

We start studying the simplest case where $H=\sum_{ij}h_{ij}c^\dag_i c_j$ and then we shall generalize to include pairing terms $c_i c_j$. As we shall see, the pair creation makes the no-jump dynamics more expensive in terms of computational resources.

\section{No-jump dynamics without pairing terms - fermionic and bosonic chains}
In this section, we shall study the no-jump evolution~\eqref{L0-2} with the quadratic Hamiltonian
\begin{equation}\label{H}
     H = \bm{c}^\dag \bm h \bm{c},\qquad \bm h = \bm h^\dag.
\end{equation}
For the moment, we do not include pair creation terms in Eq.~\eqref{H}, which will be discussed below, in Sec.~\ref{sec:pairing} and Sec.~\ref{sec:pairing_b}.

\subsection{Full monitoring} 
If each channel is perfectly monitored, $\bm{\Lambda}^+ =\bm{\Lambda}^-= \mathds{1}$ and the no-jump Liouvillian~\eqref{L0-2} reduces to 
\begin{equation}\label{Lind0_part}
    \mathcal{L}_0({\rho})=-i [{H},{\rho}] -\frac{1}{2}\sum_{ij=1}^L(\bm{\gamma}^- \mp \bm{\gamma}^+)_{ij}\{{c}_i^\dagger {c}_j, {\rho}\}-\tr(\bm{\gamma}^+){\rho}.
\end{equation}
Here the $-$ sign is for fermions and $+$ sign for bosons. Since we can study both fermionic and bosonic systems with minimal modifications, hereinafter the sign on top will refer to fermions and the one on bottom to bosons. This convention allows us to write formulas in compact form. 

Eq.~\eqref{Lind0_part} is tantamount to Eq.~\eqref{L0_all_monitored} with the non-Hermitian Hamiltonian 
\begin{equation}
    {H}_{\rm e}=\sum_{ij=1}^L(\bm h_{\rm e})_{ij} {c}_i^\dag {c}_j-i\tr(\bm{\gamma}^+)/2,
\end{equation}
where
\begin{equation}
    \bm h_{\rm e} = \bm h-\frac{i}{2}(\bm{\gamma}^- \mp \bm{\gamma}^+),
\end{equation}
with $-$ sign for fermions and $+$ sign for bosons.

\subsection{\label{fermionic_derivation}Dynamics under partial monitoring}
Here we shall focus on the most general case, where all channels are not perfectly monitored $\bm{\lambda}^+,\bm{\lambda}^- \neq \mathds{1}$ and the no-jump Liouvillian takes the form \eqref{L0-2}. Differently from the full monitoring case, in order to solve the dynamics, we need  additional hypotheses on the initial states. 

The evolution generated by the Lindbladian~\eqref{M} with the Hamiltonian~\eqref{H} and the dissipator~\eqref{L-2} is Gaussian preserving. Assuming the initial state $\rho_0$ is Gaussian, the density operator at time $t$ is
\begin{equation}
    {\rho}(t) := e^{\mathcal{L}t}({\rho}_0)= \frac{e^{-\cmct}}{Z_t},
\end{equation}
where 
\begin{align}
    Z_t &= \tr(e^{-\cmct})= \left[{\rm det}\left(\mathds{1} \pm e^{-\bm{M}_t}\right)\right]^{\pm 1},\\
       \bm{M}_t&=\ln \bigg(\frac{\mathds{1}\mp \bm{C}}{\bm{C}}\bigg),
\end{align}
and 
\begin{equation}
    \bm{C}(t)_{ij} = \tr\Big\{ {c}_j^\dagger {c}_i~ {\rho}(t)\Big\}=\langle {c_j}^\dagger {c}_i\rangle(t),
\end{equation}
is the correlation matrix. The matrix $\bm{C}(t)$ must satisfy specific Lyapunov equation for unconditional dynamics \cite{turkeshi2021diffusion,coppola2023wigner} 
\begin{equation}\label{Lyap_full_L}
    \frac{d\bm{C}}{dt} = - (\bm{W C} + \bm{C W}^\dagger) + \bm{\gamma}^+,
\end{equation}
with 
\begin{equation}\label{doppiaW}
    \bm{W}=i \bm{h} + (\bm{\gamma}^- \pm \bm{\gamma}^+)/2.
\end{equation}

However, the goal is to solve the more general non trace-preserving dynamics \eqref{L0-2}. Here we shall demonstrate that, if the system is propperly prepared in a Gaussian state at time $t=0$, the dynamics~\eqref{L0-2} is solved by the Gaussian ansatz
\begin{equation}\label{rho_no_ansatz}
    {\tilde{\rho}}_{\rm no}(t) := e^{\mathcal{L}_0 t} ({\rho}_0) = \frac{e^{-\cmctno}}{\zeta_t},
\end{equation}
for some time-dependent matrix $\bm{M}_t^{\rm no}$ and $c$-number $\zeta_t$, which must satisfy specific differential equations, to be determined. The properly normalized state is then 
\begin{align}\label{gaussian-ansatz}
   {\rho}_{\rm no}(t) :&= \frac{{\tilde\rho}_{\rm no}(t)}{P_{\rm no}(t)} = \frac{e^{-\cmctno}}{Z_t^{\rm no}},\\
   \label{zt_gaussian-ansatz} Z_t^{\rm no} &= \tr(e^{-\cmctno})=\left[{\rm det}\left(\mathds{1} \pm e^{-\bm{M}_t^{\rm no}}\right)\right]^{\pm 1}.
\end{align}
The two-point functions 
\begin{align}
\bm{C}_{\rm no}(t)_{ij} &=\tr\Big\{ {c}_j^\dagger {c}_i~ {\rho}_{\rm no}(t)\Big\}=\langle c_j^\dagger {c}_i\rangle_{{\rm no}}(t),\\
&= [(e^{\bm{M}_t^{\rm no}}\pm \mathds{1})^{-1}]_{ij}. 
\end{align}
fully characterize the Gaussian state ${\rho}_{\rm no}(t)$.
In Appendix A, we show that the ansatz~\eqref{rho_no_ansatz} solves the dynamical map~\eqref{L0-2}. In particular, we derive two independent differential equations for $\zeta_t$ and the conditional covariance matrix $\bm{C}_{\rm no}(t)$,
\begin{align}\label{riccati}
    \frac{d\bm{C}_{\rm no}}{dt} =& - (\bm W \bm C_{\rm no} + \bm C_{\rm no} \bm W^\dagger) + \bm{\gamma}^+ \pm \bm C_{\rm no} (\bm{\gamma}^-\bm{\lambda}^-) \bm C_{\rm no} \nonumber\\
    & - (\mathds{1}\mp \bm C_{\rm no}) (\bm{\gamma}^+ \bm{\lambda}^+) (\mathds{1}\mp \bm C_{\rm no}),\\[0.2cm]
    \label{zeta_riccati} \frac{d}{dt}\ln{\zeta}_t =& \tr\left\{ \bm{\gamma}^+-\bm{\gamma}^-(\mathds{1}-\bm{\lambda}^-) \frac{\bm{C}_{\rm no}(t)}{\mathds{1}\mp \bm{C}_{\rm no}(t)} \right\},
\end{align}
where $\bm W$ is the matrix into Eq.~\eqref{doppiaW}.
Eq.~\eqref{riccati} is a Riccati-type differential equation and represents one of the most important results of this work. The solution of Eq.~\eqref{riccati} provides the conditional evolution under partial monitoring.  
If there is no monitoring, $\bm \Lambda^+ = \bm \Lambda^-= 0$, $\bm{C}_{\rm no}=\bm{C}$ and we recover the Lyapunov equation \eqref{Lyap_full_L} for the unconditional dynamics. Conversely, monitoring introduces two extra terms. 
These are non-linear and individually positive semi-definite.
In the language of quantum optics, they are called 'information matrices', as they establish how our knowledge about the state of the system changes given the information that no jump occurred~\cite{Genoni2016,Belenchia2022}.
If $\bm \Lambda^+ = \bm \Lambda^- = \mathds{1}$, the result can also be written more compactly as 
\begin{equation}
    \frac{d\bm C_{\rm no}}{dt} = - (\bm K \bm C_{\rm no} + \bm C_{\rm no} \bm K^\dagger) - \bm C_{\rm no} (\bm{\gamma}^+ \mp \bm{\gamma}^-) \bm C_{\rm no},
\end{equation}
where $\bm K=\bm W\mp\bm{\gamma}^+= i \bm h + (\bm{\gamma}^-\mp\bm{\gamma}^+)/2$.

The evolution of the no-jump probability $P_{\rm no}(t)$ is given by Eq.~\eqref{P_no_observables}, where $\beta(t)$ is specified by Eqs.~(\ref{P_no_rhono2},\ref{L0-3}),
\begin{equation}\label{Pno_final}
    \beta(t)= \tr \Big[ \bm{\gamma}^+ \bm{\lambda}^+ (\mathds{1}\mp \bm C_{\rm no}(t)) + \bm{\gamma}^- \bm{\lambda}^- \bm C_{\rm no}(t)\Big].
\end{equation}
At long time
\begin{equation}
    \beta(t)\stackrel{t\sim \infty}{\simeq} t~\tr\Big\{\bm{\gamma}^+ \bm{\lambda}^+ (\mathds{1}\mp \bm C_{\rm no}^{\infty}) + \bm{\gamma}^- \bm{\lambda}^- \bm C_{\rm no}^{\infty}\Big\}, 
\end{equation}
where $\bm C_{\rm no}^{\infty}$ is the stationary solution of Eq.~\eqref{riccati}. Whence, it follows that $P_{\rm no}$ goes exponentially to $0$ for large times.

The no-jump probability $P_{\rm no}(t)$ therefore depends on the entire history of the conditional covariance matrix $\bm C_{\rm no}$, weighed by the emission monitoring rates $\bm{\gamma}^- \bm{\lambda}^-$, plus the entire history of the matrix $(\mathds{1}\mp \bm C_{\rm no})$ weighed by the absorption monitoring rates $ \bm{\gamma}^+ \bm{\lambda}^+$. For fermion chains with $\bm{\gamma}^+ \bm{\lambda}^+=\bm{\gamma}^- \bm{\lambda}^-$, it is possible to talk about a {\em universal behavior} since the no-jump probability $ P_{\rm no}(t)$ does not depend on the initial state $\rho_0$ or the Hamiltonian~\eqref{H}. The insight is that the no-jump probability is not sensitive to the kind of detected quantum jump, whether fermions are absorbed or lost. As an example, we could consider standard one-body gain-loss processes, i.e. $\bm{\gamma}^+ \bm{\lambda}^+$ and $\bm{\gamma}^- \bm{\lambda}^-$ are diagonal. In such a scenario, the probability of absorbing a fermion on site $j$ at time $t$ is given by $dt\, (\bm{\gamma}^+ \bm{\lambda}^+)_{j} \,(1-\langle c^\dag_j c_j\rangle_{\rm no})$, where $dt\, (\bm{\gamma}^+ \bm{\lambda}^+)_{j}$ is the probability of selecting the absorption channel $j$ and $1-\langle c^\dag_j c_j\rangle_{\rm no}$ is the probability that the site $j$ is empty; the probability of ejecting a fermion from site $j$ at time $t$ is given by $dt\, (\bm{\gamma}^- \bm{\lambda}^-)_j \,\langle c^\dag_j c_j\rangle_{\rm no}$, where $dt\, (\bm{\gamma}^- \bm{\lambda}^-)_j$ is the probability of selecting the dissipative channel $j$ and $\langle c^\dag_j c_j\rangle_{\rm no}$ is the probability that the site $j$ is filled. Hence, the probability that a quantum jump occurs on site $j$ is the sum of the two probabilities. Therefore, when the gain and loss rates coincide, i.e. $\bm{\gamma}^+ \bm{\lambda}^+=\bm{\gamma}^- \bm{\lambda}^-$, the probability is independent on the local density.  

\section{\label{example1}Example: tight-binding model}

As an application, we consider a chain of hopping fermions with open boundary conditions (OBC) and Hamiltonian 
\begin{equation}\label{Tight-b}
    H = -\sum\limits_{i=1}^{L-1} \left(c_i^\dag c_{i+1}+c_{i+1}^\dag c_{i}\right),
\end{equation}
being a typical example of ballistic transport. 
To illustrate the interesting physics that may emerge, we consider the following non-standard configuration for the reservoirs.
The first site is assumed to be coupled to an injection (gain) bath with jump operator $c_1^\dagger$ and rate $\Gamma_1^+ = \Gamma$.
All other sites are connected to emission (loss) baths with jump operators $c_i$ and rates $\Gamma^-_j=\Gamma$ $\forall j\in\{2,\dots,L\}$.
By hypothesis, we assume perfect monitoring on site $1$ (i.e. $\Lambda_1^+ = 1$) and a non-ideal monitoring efficiency $\Lambda^-_j=\Lambda^-$ $\forall j\in\{2,\dots,L\}$. 
In this view, the no-jump conditional dynamics is given by
\begin{align}\label{L0_example}
    \mathcal{L}_0({\rho}) =& -i [{H},{\rho}] -\frac{\Gamma}{2}\{ c_1 c_1^\dagger,\rho\}\nonumber\\
    &+\Gamma\sum_{j=2}^{L}\left((1-\Lambda^-){c}_j{\rho}{c}_j^\dag - \frac{1}{2}\{c_j^\dagger c_j,{\rho}\}\right).
\end{align}
Suppose the system is prepared in the vacuum state $\ket{0}$ and undergoes a jump in the channel $(1,+)$ at time $t=0$. $\rho(0)={c}_1^\dag\ket{0}\bra{0}c_1$ is still a Gaussian state and represents a paradigmatic example for transport protocols, where a wavepacket is inserted in an empty chain and one studies the propagation.
According to the quasi-particle picture, an excitation  on site $1$ spreads ballistically to the RHS of the chain and is destroyed with rate $\Gamma$ on all the other sites, eventually reaching the second boundary after time $L/2$~\cite{coppola2023wigner,castro2016emergent,ruggiero2020quantum,doyon2020lecture,bastianello2019generalized,capizzi2022domain,scopa2022exact,fagotti2017higher,scopa2021exact,dubail2017conformal,collura2018analytic,collura2020domain,alba2021generalized,bouchoule2022generalized,bulchandani2017solvable,bulchandani2018bethe,doyon2018soliton,schemmer2019generalized,malvania2021generalized,collura2012entangling,wendenbaum2013hydrodynamic,cao2019entanglement,jin2021interplay,bouchoule2020effect,dast2014quantum,alba2022noninteracting,alba2022hydrodynamics,carollo2022dissipative}. 
In the meantime, any jump event which occurs on site $1$ is automatically detected. 

Since $\Lambda^- \neq 1$, not all emission clicks are detected; for this reason, we expect $P_{\rm no}$ to decrease faster for larger efficiencies $\Lambda^-$. The no-jump probability always approaches exponentially $0$ for large times. Since we are interested in the thermodynamic limit, we focus on the evolution in the time window $[0,L/2]$. In this way, we avoid all the secondary processes due to the finite size effects, with the wavepackets moving back and forth multiple times within the chain. 
\begin{figure}
    \centering
    \includegraphics[width=\columnwidth]{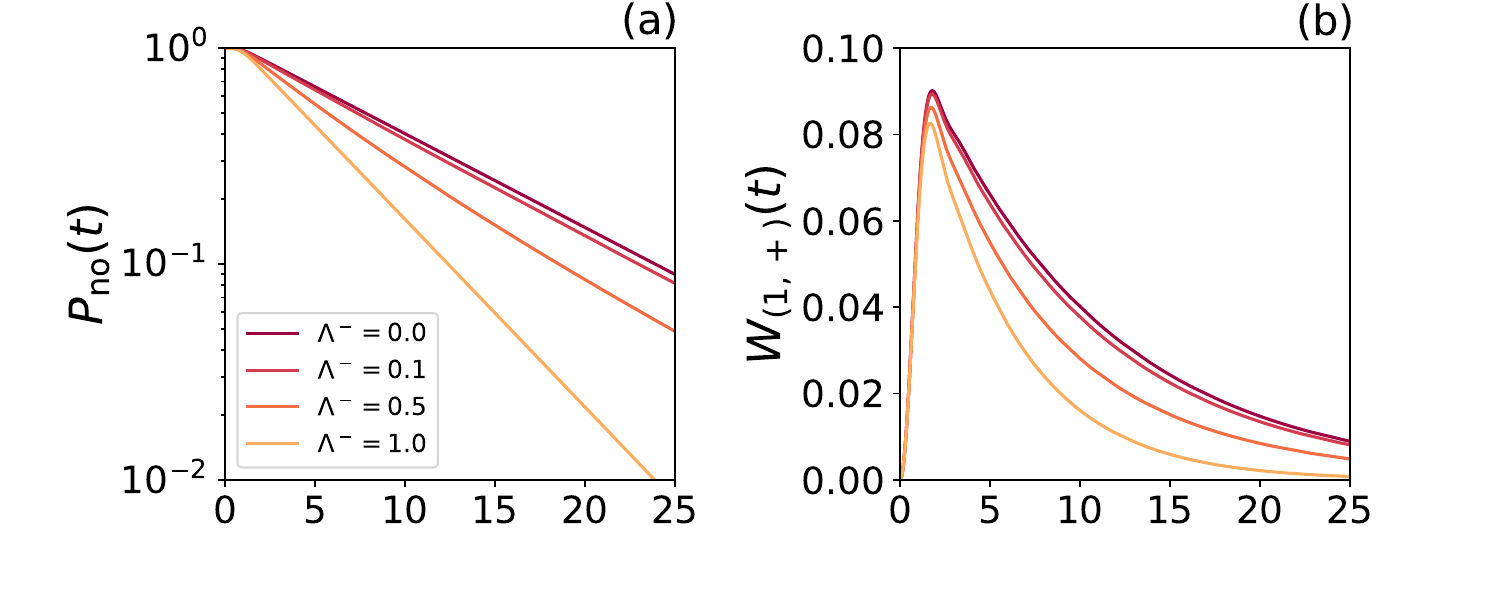}
    \caption{
    Conditional dynamics for a quantum chain of hopping fermions, $\Gamma=0.1$ and several values of the monitoring efficiency $\Lambda^-=\{0,0.1,0.5,1.0\}$. (a) no-jump probability $P_{\rm no}(t)$ as a function of time in semi-log scale; (b) waiting-time distribution $W_{(1,+)}(t)$ between two consecutive jumps in the channel $(1,+)$. }
    \label{fig:WTD1}
\end{figure}

In Fig.~\ref{fig:WTD1}~(a), we plot the no-jump probability $P_{\rm no}(t)$ for $\Gamma=0.1$, system size $L=50$ and efficiency $\Lambda^-=\{0,0.1,0.5,1.0\}$. 
At time $t=0$, $P_{\rm no}(t)=1$ as expected and, at early times, $P_{\rm no}(t)$ is roughly independent of $\Lambda^-$. 
This phenomenon finds a simple explanation in generalized hydrodynamics. For the specific tight-binding Hamiltonian~\eqref{Tight-b}, any particle located on site $1$ needs at least $\Delta t=0.5$ to reach the consecutive site and eventually be ejected. According to this rough picture, for $t<0.5$ the particle cannot be ejected, independently on the monitoring efficiency. At long time, Fig.~\ref{fig:WTD1}~(a) clearly shows the exponential decay of $P_{\rm no}$, which is dominated by the parameter $\Lambda^-$ and the steady-state $\bm C^\infty_{\rm no}$.

In Fig.~\ref{fig:WTD1}~(b) we plot the waiting-time distribution $W_{(1,+)}(t)$ between two consecutive jumps. In fact, $W_{(1,+)}(t)$ is the conditional probability distribution to observe the second injection on site $1$ at time $t$, known that one fermion has been absorbed on site $1$ at time $t=0$.
From Eq.~\eqref{WTD_general_observable}, the waiting-time distribution $W_{(1,+)}(t)$ reads
\begin{align}
    W_{(1,+)}(t)&=\Gamma\Pno(t)\left(1- \bm{C}_{\rm no}(t)_{11}\right),
\end{align}
The behavior of $W_{(1,+)}(t)$ clearly reflects the fermionic nature~\eqref{ferm_rules}. In fact, due to the finite capacity, $W_{(1,+)}(t)$ is null at $t=0$ and then increases. 
As for the no-jump probability, $W_{(1,+)}(t)$ is also independent on $\Lambda^-$ at early times. 
\begin{figure}
    \centering
    \includegraphics[width=\columnwidth]{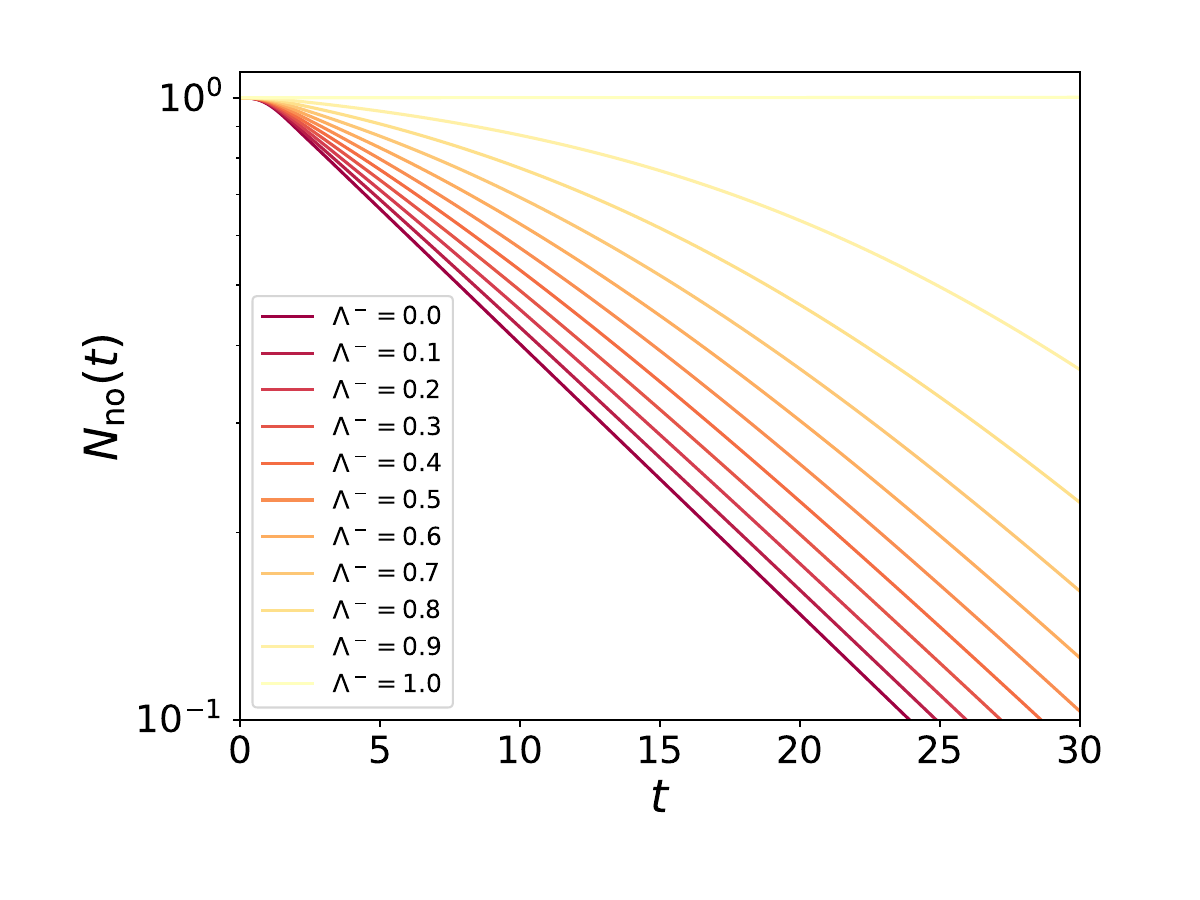}
    \caption{
    Conditional no-jump evolution of the number of particles $N_{\rm no}(t)=\sum_{j=1}^L \langle c_j^\dag c_j \rangle_{\rm no}(t)$ as a function of time $t$ in semi-log scale. Here the time evolution of $N_{\rm no}(t)$ for several monitoring efficiencies $\Lambda^-=0.1n\,\,\,\,\forall n\in\{0,\dots,10\}$. }
    \label{fig:numero_tot}
\end{figure}

In Fig.~\ref{fig:numero_tot}, we show the evolution of $N_{\rm no}(t)=\sum_{j=1}^L \langle c_j^\dag c_j \rangle_{\rm no}(t)$ as a function of time $t$. For $\Lambda^-=1$, $N_{\rm no}=1$ is a constant of motion because any occurred jump event is automatically detected thanks to the perfect efficiency of the monitoring apparatus. In the opposite case $\Lambda^-=0$, once the particle leaves the site $1$, the statistics of the emission events follows a Poisson distribution, with $N_{\rm no}(t)\simeq e^{-\Gamma t}$ being the probability of finding the particle in the chain at time $t$. In general, $N_{\rm no}$ increases with the efficiency $\Lambda^-$, since for large values of $\Lambda^-$ any particle loss is more likely detected. It is interesting to observe that, in the long time limit, $N_{\rm no}(t)\propto e^{-\Gamma t}$ for any efficiency $\Lambda^-$, suggesting that a the monitoring efficiency may affect the dynamics of $N_{\rm no}$ only at finite times. 

\begin{figure}
    \centering
    \includegraphics[width=\columnwidth]{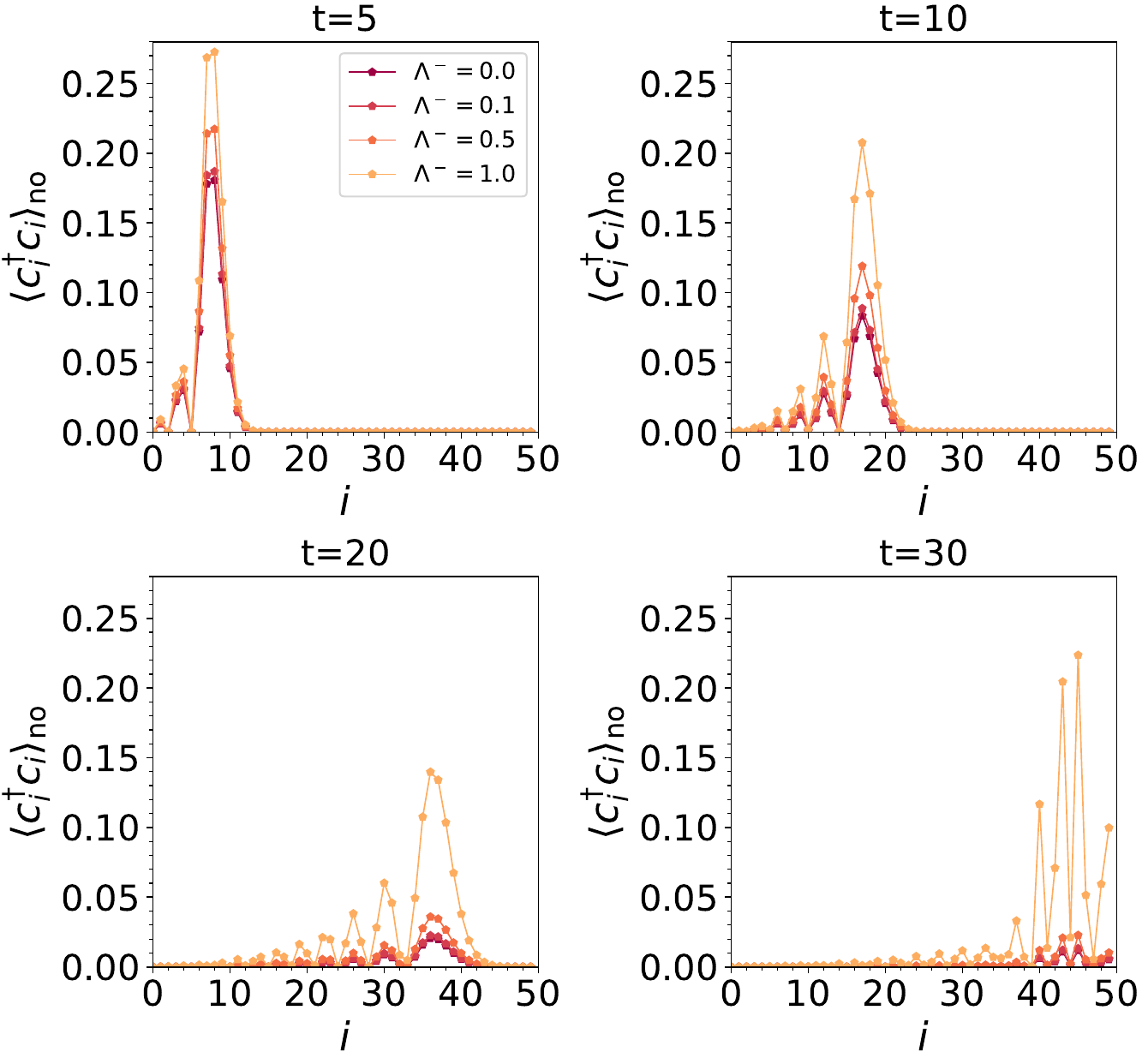}
    \caption{ Time evolution of the local occupation 
    for $\Gamma=0.1$ and monitoring efficiency $\Lambda^-=\{0,0.1,0.5,1.0\}$. As expected, the larger the efficiency the bigger the fermionic occupation, since any loss event cannot be detected in the time window $[0,t]$. Starting from $t\sim 25$, we observe finite size effects and the front wave is reflected according to the open boundary conditions. }
    \label{fig:magn}
\end{figure}

Fig.~\ref{fig:magn} shows the conditional time evolution of the local occupation $\langle c^\dag_i c_i\rangle_{\rm no}$. 
At time $t=0$, we have $\langle c^\dag_1 c_1\rangle_{\rm no}=1$ since we inject a fermion on site $1$. After quenching the state, the system evolves with the no-jump operator~\eqref{L0_example}. As expected, the higher the efficiency $\Lambda^-$, the bigger the local density since the missed detection of loss processes is less likely.

\section{\label{sec:pairing}{No-jump dynamics with pairing terms - fermionic chains}}

We now generalize the dynamics to include pairing terms ${c}_i {c}_j$ in the Hamiltonian operator. For clarity, we prefer to treat the fermionic and bosonic cases in two different sections.

In this section, we shall focus on fermionic chains~\eqref{ferm_rules}. We introduce the $2L$ Majorana operators 
\begin{equation}\label{majorana}
    {y}_{n}^1 = {c}_n^\dagger + c_n, 
    \qquad 
    {y}_{n}^2 = i({c}_n^\dagger - {c}_n).
    \qquad n = 1,\ldots, L
\end{equation}
These operators are Hermitian (${y}_k^{1\dagger} = {y}^1_k$, ${y}_k^{2\dagger} = {y}^2_k$) and satisfy the Clifford algebra
\begin{equation}
    \label{clifford}
    \{ {y}^1_k, {y}^1_\ell\} = 2 \delta_{k\ell},\qquad \{ {y}^2_k, {y}^2_\ell\} = 2 \delta_{k\ell},\qquad \{ {y}^1_k, {y}^2_\ell\} = 0.
\end{equation}
From this it follows that $({y}_k^1)^2=({y}_k^2)^2 = {\mathds{1}}$.
Below, we work with the vector 
${\bm{Y}}^\dag = ({y}_1^1,\ldots,{y}_L^1, {y}_1^2, \ldots, {y}_L^2)$, 
which then satisfies 
\begin{equation}
    \{Y_i,Y_j\} = 2\delta_{ij},
\end{equation}
and ${\bm{Y}}^\dagger \cdot {\bm{Y}} = 2L$. 
The most general quadratic Hamiltonian is
\begin{equation}\label{ham_pair}
    H = \frac{1}{4} {\bm{Y}}^\dagger \cdot \bm{\mathcal{T}} \cdot {\bm{Y}},
\end{equation}
where $\bm{\mathcal{T}}$ is a $2L\times 2L$ Hermitian matrix. In Appendix B we show that, without loss of generality, we can assume $\bm{\mathcal{T}}$ to be anti-symmetric. On the other hand, in Appendix C we prove that the dissipator~\eqref{L-2} may be put in the form 
\begin{equation}\label{pairing_dissipator}
\mathcal{D}(\rho) = \sum\limits_{i,j=1}^{2L} \Upsilon_{ij} \left[ Y_i  \rho  Y_j - \frac{1}{2} \{ Y_j  Y_i,  \rho\}\right],
\end{equation}
where 
\begin{equation}\label{pairing_upsilon_structure}
    \bm\Upsilon = \frac{1}{4} \begin{pmatrix}
    \bm{\gamma}^+ + (\bm{\gamma}^-)\trans & i\Big[\bm{\gamma}^+ - (\bm{\gamma}^-)\trans\Big] \\[0.2cm]
    -i\Big[\bm{\gamma}^+ - (\bm{\gamma}^-)\trans\Big] & \bm{\gamma}^+ + (\bm{\gamma}^-)\trans
    \end{pmatrix}.
\end{equation}
The same holds for the no-jump dynamics~\eqref{L0-2},
\begin{equation}\label{L0_mod}
    \mathcal{L}_0({\rho}) = -i [{H},{\rho}]
    + \sum\limits_{i,j=1}^{2L} \left[[\bm \Upsilon-\bm \Phi]_{ij} Y_i  \rho  Y_j - \frac{ (\bm \Upsilon)_{ij}}{2} \{ Y_j  Y_i,  \rho\}\right],
\end{equation}
where 
\begin{equation}\label{phimatrix}
    \hspace{-0.2cm}\bm\Phi  = \frac{1}{4} \begin{pmatrix}
    \bm{\gamma}^+\bm{\lambda}^+ + (\bm{\gamma}^-\bm{\lambda}^-)\trans & i\Big[\bm{\gamma}^+\bm{\lambda}^+ - (\bm{\gamma}^-\bm{\lambda}^-)\trans\Big] \\[0.2cm]
    -i\Big[\bm{\gamma}^+\bm{\lambda}^+ - (\bm{\gamma}^-\bm{\lambda}^-)\trans\Big] & \bm{\gamma}^+\bm{\lambda}^+ + (\bm{\gamma}^-\bm{\lambda}^-)\trans
    \end{pmatrix}.
\end{equation}
Notice that, in general, $\bm \Upsilon^\dagger = \bm \Upsilon$, $\bm \Phi^\dagger = \bm \Phi$, but $\bm \Upsilon\trans \neq \bm \Upsilon$, $\bm \Phi\trans \neq \bm \Phi$. 

\subsection{Full monitoring}
When all channels are perfectly monitored, $\bm{\Lambda}^+ =\bm{\Lambda}^-= \mathds{1}$ and $\bm\Upsilon=\bm\Phi$. The no-jump operator~\eqref{L0_mod} reduces to 
\begin{equation}\label{L0_mod2}
    \mathcal{L}_0({\rho}) = -i [{H},{\rho}] -\frac{1}{2} \sum\limits_{i,j=1}^{2L} (\bm\Upsilon)_{ij} \{ Y_j  Y_i,  \rho\}.
\end{equation}
After some algebra, $\mathcal{L}_0$ takes the form \eqref{L0_all_monitored} with effective Hamiltonian
\begin{equation}
    {H}_{e}={\bm{Y}}^\dagger \cdot \frac{\bm{\mathcal{T}} +i(\bm{\Upsilon}-\bm\Upsilon^{\trans})}{4} \cdot {\bm{Y}} -i\frac{\tr(\bm\Upsilon)}{2}.
\end{equation}

\subsection{\label{partial_ferm}Dynamics under partial monitoring}
We now derive the conditional dynamics under partial monitoring, following closely the calculations (and the notation) of Sec.~\ref{fermionic_derivation}. 
We assume that the initial state is Gaussian.
The unconditional dynamics~\eqref{M} with the Hamiltonian \eqref{ham_pair}, the dissipator \eqref{pairing_dissipator} and the matrix \eqref{pairing_upsilon_structure} is Gaussian preserving,
\begin{equation}
    {\rho}(t) := e^{\mathcal{L}t}({\rho}_0)= \frac{e^{\frac{1}{4} \ymyt}}{Z_t},
\end{equation}
where $\bm{\mathcal{M}}_t$ is Hermitian and anti-symmetric,
\begin{align}
    Z_t &= \tr\Big\{ e^{\frac{1}{4} \ymyt} \Big\} = \sqrt{\det( \mathds{1} + e^{\bm{\mathcal{M}}_t})},\\
    \bm{\mathcal{M}}_t&=\ln\bigg(\frac{\mathds{1}-\bm\Theta}{\mathds{1}+\bm\Theta}\bigg),
\end{align}
and
\begin{equation}
    \bm\Theta(t)_{ij} = \frac{1}{2}\tr\Big\{[ Y_i, Y_j]\rho(t)\Big\}= \frac{1}{2} \langle [ Y_i, Y_j]\rangle(t),
\end{equation}
is the correlation matrix, 
which is Hermitian ($\bm\Theta^\dagger = \bm\Theta$) and anti-symmetric ($\bm\Theta\trans = - \bm\Theta$) as well. The correlation matrix $\bm\Theta$ satisfies the Lyapunov differential equation~\cite{Purkayastha2022}
\begin{equation}\label{lyap_pair}
    \frac{d\bm\Theta}{dt} = - \big(\bm{\mathcal{W}} \bm\Theta + \bm\Theta \bm{\mathcal{W}}^\dagger\big) + \bm{\mathcal{F}},
\end{equation}
with
\begin{align}
    \label{doppiaW_maj}\bm{\mathcal{W}} &= i \bm{\mathcal{T}} + \bm\Upsilon + \bm\Upsilon\trans,\\
    \label{F_maj}\bm{\mathcal{F}} &= 2 (\bm\Upsilon\trans - \bm\Upsilon).
\end{align}

Here we show that the conditional no-jump dynamics~\eqref{L0_mod} is solved by the Gaussian ansatz 
\begin{equation}\label{rho_tilde_copp}
    {\tilde{\rho}}_{\rm no}(t) := e^{\mathcal{L}_0 t} ({\rho}_0) = \frac{e^{\frac{1}{4} \ymytno}}{\zeta_t},
\end{equation}
for some time-dependent matrix $\bm{\mathcal{M}}_t^{\rm no}$ and $c$-number $\zeta_t$, which must satisfy specific differential equations, to be determined. Again, the properly normalized state is 
\begin{align}\label{gaussian-ansatz2}
   {\rho}_{\rm no}(t) &:= \frac{{\tilde\rho}_{\rm no}(t)}{P_{\rm no}(t)} = \frac{e^{\frac{1}{4} \ymytno}}{Z_t^{\rm no}},\\
    Z_t^{\rm no} &= \tr\Big\{e^{\frac{1}{4} \ymytno}\Big\}=\sqrt{\det( \mathds{1} + e^{\bm{\mathcal{M}}^{\rm no}_t})}.
\end{align}
The correlation matrix 
\begin{align}\label{theta_no_fermions}
\bm{\Theta}_{\rm no}(t)_{ij} &= -[\tanh(\bm{\mathcal{M}}_t^{\rm no}/2)]_{ij}
= \frac{1}{2} \langle [ Y_i, Y_j]\rangle_{\rm no}(t),
\end{align}
fully characterizes the state ${\rho}_{\rm no}(t)$. The matrix $\bm{\Theta}_{\rm no}$ can also be put in the block form
\begin{align}
\bm{\Theta}_{\rm no}=\begin{pmatrix}\label{block_form}
    \bm{\Theta}_{1,\rm no} & -i\bm{\Theta}_{3,\rm no} \\[0.2cm]
    i\bm{\Theta}_{3,\rm no}^\dag & \bm{\Theta}_{2,\rm no}
    \end{pmatrix},
\end{align}
where 
\begin{align}
    \bm{\Theta}_{1,\rm no} &= \bm{C}_{\rm no}\trans-\bm{C}_{\rm no}-\bm{S}_{\rm no}-\bm{S}_{\rm no}^\dag,\\
    \bm{\Theta}_{2,\rm no} &=\bm{C}_{\rm no}\trans-\bm{C}_{\rm no}+\bm{S}_{\rm no}+\bm{S}_{\rm no}^\dag,\\
    \bm{\Theta}_{3,\rm no} &=\bm{C}_{\rm no}\trans+\bm{C}_{\rm no}+\bm{S}_{\rm no}-\bm{S}_{\rm no}^\dag-\mathds{1},
\end{align}
and
\begin{align}
        \bm{C}_{\rm no}(t)_{ij}&=\tr\Big\{ {c}_j^\dagger {c}_i~ {\rho}_{\rm no}(t)\Big\}=\langle c_j^\dagger {c}_i\rangle_{{\rm no}}(t),\\
    \bm{S}_{\rm no}(t)_{ij}&=\tr\Big\{ {c}_j^\dagger {c}_i^\dagger~ {\rho}_{\rm no}(t)\Big\}=\langle c_j^\dagger {c}_i^\dagger\rangle_{{\rm no}}(t),
\end{align}
are $L\times L$ matrices. Due to the fermionic rules~\eqref{ferm_rules}, $\bm{C}_{\rm no}=\bm{C}_{\rm no}^\dag$ and $\bm{S}_{\rm no}=-\bm{S}_{\rm no}\trans$. As sanity check, we see that $\bm{\Theta}_{\rm no}=\bm{\Theta}_{\rm no}^\dag$ and $\bm{\Theta}_{\rm no}=-\bm{\Theta}_{\rm no}\trans$. 

In Appendix D, we show that Eq.~\eqref{rho_tilde_copp} solves the conditional no-jump dynamics~\eqref{L0_mod}. We provide two independent differential equations for $\zeta_t$ and the covariance matrix $\bm{\Theta}_{\rm no}(t)$,
\begin{align}\label{pairing_riccati}
    \frac{d\bm\Theta_{\rm no}}{dt} =& - \big(\bm{\mathcal{W}} \bm\Theta_{\rm no} + \bm\Theta_{\rm no} \bm{\mathcal{W}}^\dagger\big) + \bm{\mathcal{F}} - (\mathds{1}-\bm\Theta_{\rm no}) \bm\Phi\trans (\mathds{1}-\bm\Theta_{\rm no})\nonumber\\
    & + (\mathds{1}+\bm\Theta_{\rm no})\bm\Phi (\mathds{1}+\bm\Theta_{\rm no}),\\[0.2cm]
    \label{pairing_zeta_riccati}\frac{d}{dt}\ln{\zeta}_t = & \tr\bigg(\bm\Upsilon-[\bm\Upsilon-\bm\Phi] \frac{\mathds{1}+\bm\Theta_{\rm no}}{\mathds{1}-\bm\Theta_{\rm no}}\bigg),
\end{align}
where $\bm{\mathcal{W}}$ and $\bm{\mathcal{F}}$ are the matrices into Eqs.~(\ref{doppiaW_maj},\ref{F_maj}), while $\bm\Upsilon$ and $\bm\Phi$ are specified by Eqs.~(\ref{pairing_upsilon_structure},\ref{phimatrix}).
Eq.~\eqref{pairing_riccati} is a Riccati-type equation and represents another important result of this work, which generalizes Eq.~\eqref{riccati} for fermionic chains with particle non-conserving Hamiltonians. If there is no monitoring, $\bm \Lambda^+ = \bm \Lambda^-= 0$, $\bm{\Theta}_{\rm no}=\bm{\Theta}$ and we recover the Lyapunov equation \eqref{lyap_pair} for the unconditional dynamics. When $\bm \Lambda^+ = \bm \Lambda^- = \mathds{1}$, Eq. \eqref{pairing_riccati} can also be written more compactly as
\begin{equation}
   \frac{d\bm\Theta_{\rm no}}{dt} = -i[\bm{\mathcal{T}},\bm\Theta_{\rm no}] + \frac{1}{2}\bm{\mathcal{F}} - \frac{1}{2}\bm\Theta_{\rm no} \bm{\mathcal{F}}\bm\Theta_{\rm no}.
\end{equation}
Again, the evolution of the no-jump probability $P_{\rm no}(t)$ is given by Eq.~\eqref{P_no_observables}, where $\beta(t)$ is specified by Eq.~\eqref{Pno_final} choosing the $-$ sign. In Appendix D, we show that $\beta(t)$ can also be rewritten in the compact form 
\begin{equation}
 \beta(t) = \tr\Big[ \bm\Phi(\mathds{1}+\bm\Theta_{\rm no}(t))\Big].
\end{equation}
In the long-time limit,
\begin{equation}
    P_{\rm no}(t)\sim \exp\left(-\tr\left[ \bm\Phi(\mathds{1}+\bm\Theta_{\rm no}^{\infty})\right]t\right), \qquad  t\sim \infty,
\end{equation}
where $\bm\Theta_{\rm no}^{\infty}$ is the stationary solution of the Riccati equation \eqref{pairing_riccati}. 

\section{\label{sec:pairing_b}{No-jump dynamics with pairing terms - bosonic chains}}
In this section, we focus on the conditional no-jump dynamics of non-interacting bosons with particle non-conserving Hamiltonian operators, in analogy with Sec.~\ref{sec:pairing}. 

Assuming the $\bm c$'s to be bosonic operators~\eqref{bos_rules}, we define the $2L$ quadratures, 
\begin{equation}\label{quadratures}
    {x}_{n} = {c}_n^\dagger + c_n, 
    \qquad 
    {p}_{n} = i({c}_n^\dagger - {c}_n).
    \qquad n = 1,\ldots, L
\end{equation}
which are Hermitian ($x_k=x_k^\dag$, $p_k=p_k^\dag$) and satisfy the algebra
\begin{equation}
    \label{algebra_bos}
    \comm{x_k}{p_\ell} = 2i \delta_{k\ell}, \qquad \comm{x_k}{x_\ell} =0,\qquad \comm{p_k}{p_\ell} =0.
\end{equation}
Below, we work with the vector 
${\bm{R}}^\dag = (x_1,\ldots,x_L, p_1, \ldots, p_L)$, 
which then satisfies 
\begin{equation}
    \comm{R_i}{R_j} = 2i\Omega_{ij},
\end{equation}
where $\bm\Omega$ is the symplectic form
\begin{equation}\label{symplectic}
    \bm\Omega = (i\sigma_y) \otimes \mathds{1}_{L} = 
    \begin{pmatrix}
    \bigzero
  & \vline & \mathds{1}_{L} \\
\hline
-\mathds{1}_{L} & \vline &
\bigzero
\end{pmatrix},
\end{equation}
$\mathds{1}_{L}$ is the $L\times L$ identity matrix and $(i\bm\Omega)^2 = \mathds{1}$. 
The most general quadratic Hamiltonian is
\begin{equation}\label{ham_pair_bos}
    H = \frac{1}{4} {\bm{R}}^\dagger \cdot \bm{\mathcal{T}} \cdot {\bm{R}},
\end{equation}
where $\bm{\mathcal{T}}$ is a $2L\times 2L$ Hermitian matrix. In Appendix E we show that, without loss of generality, we can assume $\bm{\mathcal{T}}$ to be symmetric. Similarly to Appendix C, we can prove that the dissipator~\eqref{L-2} may be put in the form 
\begin{equation}\label{pairing_dissipator_bos}
\mathcal{D}(\rho) = \sum\limits_{i,j=1}^{2L} \Upsilon_{ij} \Big[ R_i  \rho  R_j - \frac{1}{2} \{ R_j  R_i,  \rho\}\Big].
\end{equation}
The same holds for the partial monitoring dynamics~\eqref{L0-2},
\begin{equation}\label{L0_mod_bos}
    \mathcal{L}_0({\rho}) = -i [{H},{\rho}] + \sum\limits_{i,j=1}^{2L} \Big[[\bm \Upsilon-\bm \Phi]_{ij} R_i  \rho  R_j - \frac{ (\bm \Upsilon)_{ij}}{2} \{ R_j  R_i,  \rho\}\Big],
\end{equation}
where $\bm \Upsilon$ and $\bm \Phi$ are given by Eqs.~(\ref{pairing_upsilon_structure},\ref{phimatrix}), since these matrices are not sensitive to bosonic or fermionic algebra. 

\subsection{Full monitoring}
When all channels are perfectly monitored, $\bm{\Lambda}^+ =\bm{\Lambda}^-= \mathds{1}$ and $\bm\Upsilon=\bm\Phi$. The no-jump Liouvillian reduces to 
\begin{equation}
    \mathcal{L}_0({\rho}) = -i [{H},{\rho}] -\frac{1}{2} \sum\limits_{i,j=1}^{2L} (\bm\Upsilon)_{ij} \{ R_j  R_i,  \rho\},
\end{equation}
After some algebra, $\mathcal{L}_0$ takes the form \eqref{L0_all_monitored} with effective Hamiltonian
\begin{equation}
    {H}_{e}={\bm{R}}^\dagger \cdot \frac{\bm{\mathcal{T}} -i(\bm{\Upsilon}+\bm\Upsilon^{\trans})}{4} \cdot {\bm{R}} -\frac{i}{4}\tr(\bm\gamma^+-\bm\gamma^-).
\end{equation}

\subsection{Dynamics under partial monitoring}
We now derive the conditional dynamics under partial monitoring, following closely the calculations (and the notation) of Sec.~\ref{partial_ferm}. 
We assume that the initial state is Gaussian.
The unconditional dynamics~\eqref{M} with the quadratures~\eqref{quadratures}, the Hamiltonian~\eqref{ham_pair_bos}, the dissipator~\eqref{pairing_dissipator_bos} and the matrix~\eqref{pairing_upsilon_structure} is Gaussian preserving, which means
\begin{equation}
    {\rho}(t) := e^{\mathcal{L}t}({\rho}_0)= \frac{e^{\frac{1}{4} \rmrt}}{Z_t},
\end{equation}
where $(i\bm\Omega)\bm{\mathcal{M}}_t$ is Hermitian and symmetric,
\begin{align}
    Z_t &= \tr\Big\{ e^{\frac{1}{4} \rmrt} \Big\} = \frac{1}{\sqrt{\det( \mathds{1} - e^{\bm{\mathcal{M}}_t})}},\\
    e^{\bm{\mathcal{M}}_t}&=\frac{\bm\Theta(i\bm\Omega)-\mathds{1}}{\bm\Theta(i\bm\Omega)+\mathds{1}},
\end{align}
and 
\begin{align}
\bm\Theta(t)_{ij} &= \frac{1}{2}\tr\Big\{\{ R_i, R_j\}\rho(t)\Big\}= \frac{1}{2} \langle \{ R_i, R_j\}\rangle(t),
\end{align} 
is the correlation matrix. The matrix $\bm\Theta$ is Hermitian ($\bm\Theta^\dagger = \bm\Theta$) and symmetric ($\bm\Theta\trans = \bm\Theta$) as well. It can be proved that the correlation matrix $\bm\Theta$ satisfies the Lyapunov differential equation
\begin{equation}\label{lyap_pair_bos}
    \frac{d\bm\Theta}{dt} = - \big(\bm{\mathcal{W}} \bm\Theta + \bm\Theta \bm{\mathcal{W}}^\dagger\big) + \bm{\mathcal{F}},
\end{equation}
with 
\begin{align}
    \label{doppiaW_bos}\bm{\mathcal{W}} &= -\bm\Omega \bm{\mathcal{T}} - (i\bm\Omega)(\bm\Upsilon - \bm\Upsilon\trans),\\
    \label{F_bos}\bm{\mathcal{F}} &= 2 (i\bm\Omega)(\bm\Upsilon\trans + \bm\Upsilon)(i\bm\Omega).
\end{align}

In Appendix F, we show that the conditional no-jump dynamics is solved by the Gaussian ansatz 
\begin{equation}\label{rho_tilde_copp_bos}
    {\tilde{\rho}}_{\rm no}(t) := e^{\mathcal{L}_0 t} ({\rho}_0) = \frac{e^{\frac{1}{4} \rmrtno}}{\zeta_t},
\end{equation}
for some matrix $\bm{\mathcal{M}}_t^{\rm no}$ and some $c$-number $\zeta_t$, which must satisfy specific differential equations, to be determined. Again, the properly normalized state is 
\begin{align}\label{gaussian-ansatz2_bos}
   {\rho}_{\rm no}(t) &:= \frac{{\tilde\rho}_{\rm no}(t)}{P_{\rm no}(t)} = \frac{e^{\frac{1}{4} \rmrtno}}{Z_t^{\rm no}},\\
    Z_t^{\rm no} &= \tr(e^{\frac{1}{4} \rmrtno})= \frac{1}{\sqrt{\det( \mathds{1} - e^{\bm{\mathcal{M}}^{\rm no}_t})}}.
\end{align}
The correlation matrix
\begin{equation}\label{theta_no_bosons}
\hspace{-0.35cm}\bm{\Theta}_{\rm no}(t)_{ij} = -[\coth(\bm{\mathcal{M}}_t^{\rm no}/2)\,(i\bm\Omega)]_{ij}= \frac{1}{2} \langle \{ R_i, R_j\}\rangle_{\rm no}(t), 
\end{equation}
fully characterizes the normalized state ${\rho}_{\rm no}(t)$. The matrix $\bm{\Theta}_{\rm no}$ can also be put in the same block form~\eqref{block_form} with 
\begin{align}
    \bm{\Theta}_{1,\rm no} &= \bm{C}_{\rm no}\trans+\bm{C}_{\rm no}+\bm{S}_{\rm no}+\bm{S}_{\rm no}^\dag+\mathds{1},\\
    \bm{\Theta}_{2,\rm no} &=\bm{C}_{\rm no}\trans-\bm{C}_{\rm no}-\bm{S}_{\rm no}+\bm{S}_{\rm no}^\dag,\\
    \bm{\Theta}_{3,\rm no} &=\bm{C}_{\rm no}\trans+\bm{C}_{\rm no}-\bm{S}_{\rm no}-\bm{S}_{\rm no}^\dag+\mathds{1},
\end{align}
where 
\begin{align}
        \bm{C}_{\rm no}(t)_{ij}&=\tr\Big\{ {c}_j^\dagger {c}_i~ {\rho}_{\rm no}(t)\Big\}=\langle c_j^\dagger {c}_i\rangle_{{\rm no}}(t),\\
    \bm{S}_{\rm no}(t)_{ij}&=\tr\Big\{ {c}_j^\dagger {c}_i^\dagger~ {\rho}_{\rm no}(t)\Big\}=\langle c_j^\dagger {c}_i^\dagger\rangle_{{\rm no}}(t),
\end{align}
are $L\times L$ matrices. Due to the bosonic commutation rules~\eqref{bos_rules}, $\bm{C}_{\rm no}=\bm{C}_{\rm no}^\dag$ and $\bm{S}_{\rm no}=\bm{S}_{\rm no}\trans$. As sanity check, we see that $\bm{\Theta}_{\rm no}=\bm{\Theta}_{\rm no}^\dag$ and $\bm{\Theta}_{\rm no}=\bm{\Theta}_{\rm no}\trans$. 

In Appendix F, we provide two independent differential equations for $\zeta_t$ and the conditional covariance matrix $\bm{\Theta}_{\rm no}(t)$,
\begin{align}\label{pairing_riccati_bos}
    \hspace{-0.5cm}\frac{d\bm\Theta_{\rm no}}{dt} =& - \big(\bm{\mathcal{W}} \bm\Theta_{\rm no} + \bm\Theta_{\rm no} \bm{\mathcal{W}}^\dagger\big) + \bm{\mathcal{F}}\nonumber\\
    & - (i\bm\Omega-\bm\Theta_{\rm no})\, \bm\Phi\trans \,(i\bm\Omega-\bm\Theta_{\rm no})\nonumber\\
    & - (i\bm\Omega+\bm\Theta_{\rm no})\,\bm\Phi \,(i\bm\Omega+\bm\Theta_{\rm no}),\\[0.2cm]
    \label{pairing_zeta_riccati_bos}\hspace{-0.5cm}\frac{d}{dt}\ln{\zeta}_t = & \tr\bigg(\bm\Upsilon(i\bm\Omega)+[\bm\Upsilon-\bm\Phi] \frac{\bm\Theta_{\rm no}(i\bm\Omega)+\mathds{1}}{\bm\Theta_{\rm no}(i\bm\Omega)-\mathds{1}}(i\bm\Omega)\bigg),
\end{align}
where $\bm{\mathcal{W}}$ and $\bm{\mathcal{F}}$ are the matrices into Eqs.~(\ref{doppiaW_bos},\ref{F_bos}), while $\bm\Upsilon$ and $\bm\Phi$ are specified by Eqs.~(\ref{pairing_upsilon_structure},\ref{phimatrix}).
Eq.~\eqref{pairing_riccati_bos} is again a Riccati-type equation and represents another important result of this work. Eq.~\eqref{pairing_riccati_bos} generalizes Eq.~\eqref{riccati} for bosonic chains with particle non-conserving Hamiltonians. If there is no monitoring, $\bm \Lambda^+ = \bm \Lambda^-= 0$, $\bm{\Theta}_{\rm no}=\bm{\Theta}$ and we recover the Lyapunov equation \eqref{lyap_pair_bos} for the unconditional dynamics. 
\begin{figure*}[!t]
    \centering
    \includegraphics[scale=0.32]{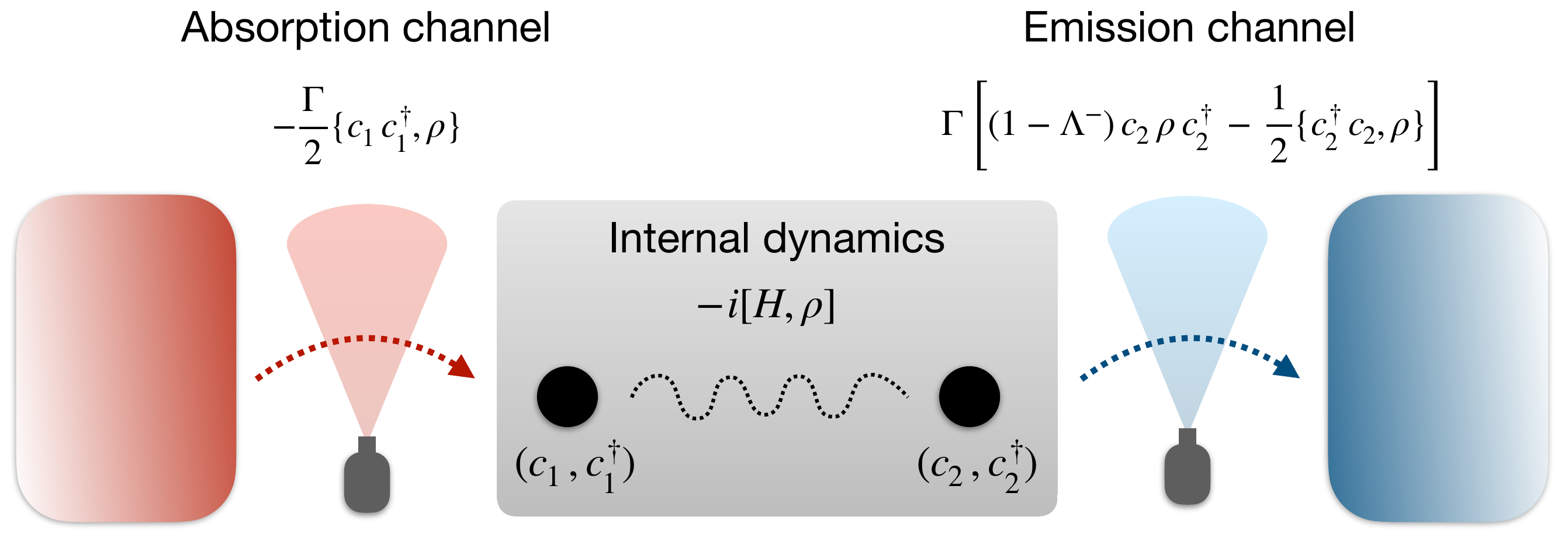}
    \caption{Sketch of the 2-body problem: we study the no-jump dynamics~\eqref{no-jump_last_ex} for a boundary driven system, which undergoes a non-ideal measurement process on the emission channel, with monitoring efficiency $\Lambda^-$. The internal dynamics is generated by the Hamiltonian $H$, which takes the form~\eqref{fermion_ham_ex} for fermions and the form~\eqref{boson_ham_ex} for bosons.}
    \label{fig:hot_cold}
\end{figure*}
When $\bm \Lambda^+ = \bm \Lambda^- = \mathds{1}$, Eq. \eqref{pairing_riccati_bos} can also be written more compactly as
\begin{equation}
   \frac{d\bm\Theta_{\rm no}}{dt} = -\bm\Theta_{\rm no}\bm{\mathcal{T}}\bm\Omega- (\bm{\mathcal{T}}\bm\Omega)\trans \bm\Theta_{\rm no}+ \frac{1}{2}\bm{\mathcal{F}} - \frac{1}{2}\bm\Theta_{\rm no} (i\bm\Omega)\bm{\mathcal{F}}(i\bm\Omega) \bm\Theta_{\rm no}.
\end{equation}

The evolution of the no-jump probability $P_{\rm no}(t)$ is given by Eq.~\eqref{P_no_observables}, where $\beta(t)$ is specified by Eq.~\eqref{Pno_final} choosing the $+$ sign. In Appendix F, we show that $\beta(t)$ can also be rewritten in the compact form 
\begin{equation}\label{no-jump-prob-pairing}
 \beta(t) = \tr\Big[ \bm\Phi(i\bm\Omega+\bm\Theta_{\rm no}(t))\Big].
 \end{equation}
In the long-time limit, 
\begin{equation}
    P_{\rm no}(t)\sim \exp\left(-\tr\left[ \bm\Phi(i\bm\Omega+\bm\Theta_{\rm no}^{\infty})\right]t\right), \qquad  t\sim \infty,
\end{equation}
where $\bm\Theta_{\rm no}^{\infty}$ is the stationary solution of the Riccati equation \eqref{pairing_riccati_bos}. 

Sec.~\ref{sec:pairing_b} concludes the study of the conditional no-jump dynamics of non-interacting quantum chains. In the next section, we shall consider a specific setting where we all the analytical results provided in this work can be successfully applied. 

\begin{widetext}
\section{\label{example2}Example: boundary driven system}
As an application of the no-jump dynamics with pairing terms (see Sec.~\ref{sec:pairing} and Sec.~\ref{sec:pairing_b}), we study a simple boundary driven system of bosons (B) and fermions (F). More specifically, we couple a 2-site system to two reservoirs, with an absorption channel on site $1$ and an emission channel on site $2$. We assume jump rates $\Gamma^+_1=\Gamma^-_2=\Gamma$ and two local detectors for the particle exchange. In particular, we assume the detector on the hot bath to work with perfect efficiency $\Lambda^+=1$ and the other one on the cold reservoir with efficiency $\Lambda^-\in(0,1)$. The goal is to study the no-jump dynamics generated by the dynamical map
\begin{equation}\label{no-jump_last_ex}
    \mathcal{L}_0({\rho}) = -i [{H},{\rho}]
    - \frac{\Gamma}{2}\{c_1 c_1^\dag,\rho\}+ \Gamma \left[(1-\Lambda^-)\,c_2\rho c_2^\dag - \frac{1}{2}\{c_2^\dag c_2,\rho\}\right],
\end{equation}
where $H$ is the prototypical Hamiltonian 
\begin{align}
 \label{fermion_ham_ex}   H=&\,c_1^\dag c_2 + c_2^\dag c_1 + g(c_1 c_2 - c_1^\dag c_2^\dag),\hspace{1.3cm} \text{(F)}\\
    \nonumber \\
  \label{boson_ham_ex}    H=&\, c_1^\dag c_2 + c_2^\dag c_1 + g(c_1 c_2 + c_1^\dag c_2^\dag),\hspace{1.3cm} \text{(B)}
\end{align} 
and $g$ is the amplitude of the pairing terms. The full experimental setup is schematically represented in Fig.~\ref{fig:hot_cold}. 
As usual, we aim to compute the no-jump probability 
\begin{equation}\label{no-jump_ex}
    P_{\rm no}(t) = \begin{cases}
        \exp\left(-\int_0^t ds\,\tr\bigg[ \bm\Phi\bigg(\mathds{1}+\bm\Theta_{\rm no}(s)\bigg)\,\bigg]\,\right),\hspace{1.5cm}\text{(F)}\\ 
        \\
        \exp\left(-\int_0^t ds\,\tr\bigg[ \bm\Phi\bigg(i\bm\Omega+\bm\Theta_{\rm no}(s)\bigg)\,\bigg]\,\right),\hspace{1.3cm}\text{(B)}
    \end{cases}
\end{equation}
and the WTDs
\begin{align}\label{W2_ex}
    W_{(2,-)}(t) &= \Gamma \Lambda^- P_{\rm no}(t)\, \langle c_2^\dag c_2 \rangle_{\rm no}(t)\nonumber\\
    &= \Gamma \Lambda^-P_{\rm no}(t)\times\begin{cases}
        \,\frac{1}{2}\left(1+i\bm\Theta_{\rm no}(t)_{24}\right),\hspace{1.8cm}\text{(F)}\\ \\
        \,\frac{1}{4}\left(\bm\Theta_{\rm no}(t)_{22}+\bm\Theta_{\rm no}(t)_{44}-2\right),\hspace{0.35cm}\text{(B)}
    \end{cases}
\end{align}
\begin{align}\label{W1_ex}
    W_{(1,+)}(t) &= \Gamma P_{\rm no}(t)\, \langle c_1 c_1^\dag \rangle_{\rm no}(t)\nonumber\\
    &= \Gamma P_{\rm no}(t)\times\begin{cases}
        \,\frac{1}{2}\left(1-i\bm\Theta_{\rm no}(t)_{13}\right),\hspace{1.8cm}\text{(F)}\\ \\
        \,\frac{1}{4}\left(\bm\Theta_{\rm no}(t)_{11}+\bm\Theta_{\rm no}(t)_{33}+2\right),\hspace{0.35cm}\text{(B)}
    \end{cases}
\end{align}
which are formulated in terms of the $4\times 4$ correlation matrix $\bm\Theta$, whose definition and dynamics are given by Eqs.~(\ref{theta_no_fermions},\ref{pairing_riccati}) for fermions and Eqs.~(\ref{theta_no_bosons},\ref{pairing_riccati_bos}) for bosons. We remember that $P_{\rm no}(t)$ is the probability of not detecting any jump in $[0,t]$, while $W_{(2,-)}(t)$ and $W_{(1,+)}(t)$ are the probability distributions to observe the first jump in the channel $(2,-)$ and $(1,+)$, respectively. Due to the normalization condition, the WTDs satisfy 
\begin{equation}
    \int_0^\infty dt \,\left(W_{(1,+)}(t) + W_{(2,-)}(t)\right) = 1.
\end{equation}
\end{widetext}

We assume the system is initially prepared in the vacuum state $\ket{00}$, that is a Gaussian state and hence we can apply the theoretical results of Sec.~\ref{sec:pairing} and Sec.~\ref{sec:pairing_b}. 

\begin{figure}
    \centering
    \includegraphics[width=\columnwidth]{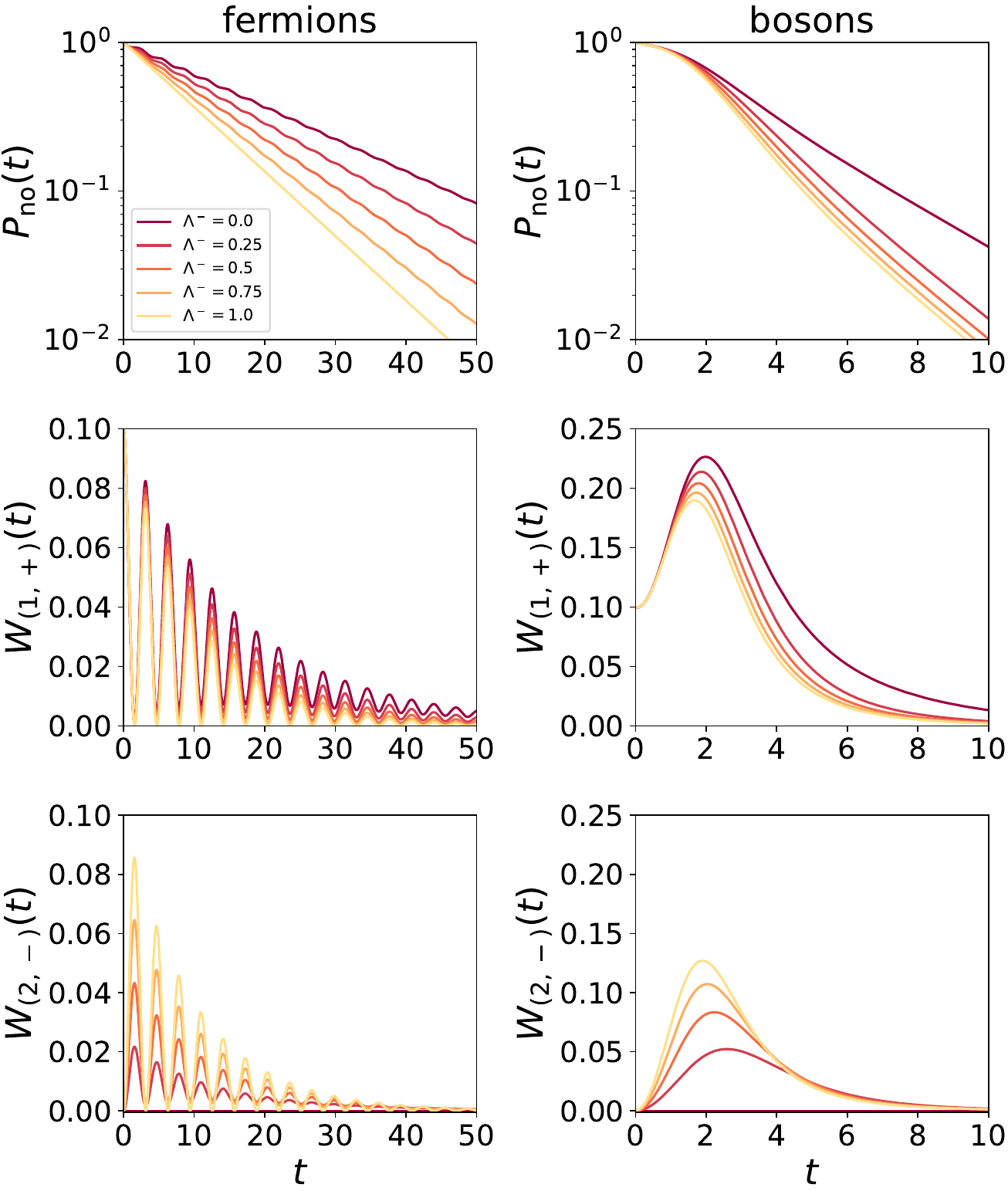}
    \caption{Conditional dynamics~\eqref{no-jump_last_ex} for a quantum chain of non-interacting particles (fermionic and bosonic case), $\Gamma=0.1$, $g=1$ and monitoring efficiency $\Lambda^-=\{0,0.25,0.5,0.75,1\}$. The no-jump probability and the WTDs always show an exponential decay. }
    \label{fig:driven}
\end{figure}
\begin{figure}
    \centering
    \includegraphics[width=\columnwidth]{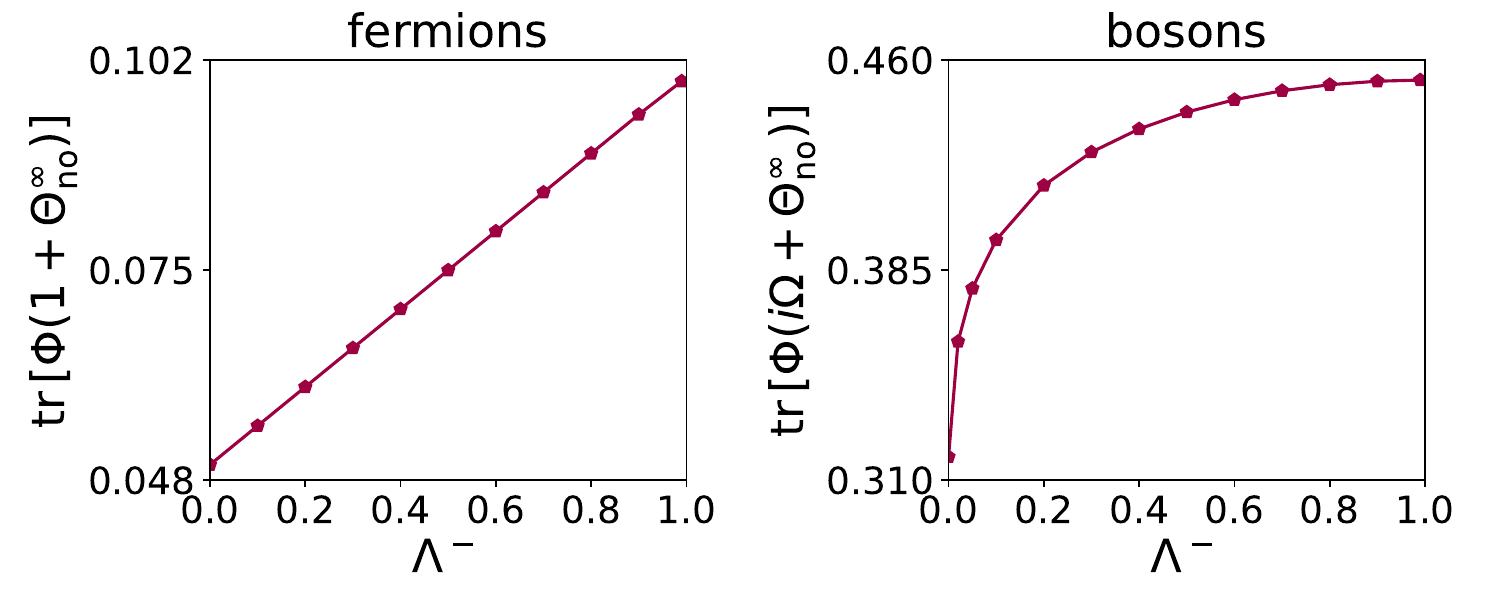}
    \caption{ We show the coefficient governing the exponential decay of the no-jump probability~\eqref{no-jump_ex} and the WTDs~(\ref{W2_ex},\ref{W1_ex}), as a function of the monitoring efficiency $\Lambda^-$ (bosonic and fermionic case).}
    \label{fig:exp_coeff}
\end{figure}

In Fig.~\ref{fig:driven}, we show the no-jump probability~\eqref{no-jump_ex} and the WTDs~(\ref{W2_ex},\ref{W1_ex}) for $\Gamma=0.1$, $g=1$ and different values of the monitoring efficiency $\Lambda^-$. The two columns refer to the fermionic and bosonic dynamics, respectively. The no-jump probability is a decreasing function of $\Lambda^-$; the higher the monitoring efficiency, more likely particle loss events are detected. 
If the bosonic WTDs show a fast decay, revealing that the system quickly reaches a stationary regime where the pair creation-annihilation processes are balanced by the non-detected emissions, the fermionic WTDs exhibit a damped oscillatory behaviour with the same periodicity of the local occupation, with maxima and minima reflecting the back and forth motion of the particle along the system. The exponential decay of $P_{\rm no}(t)$, as well as the decay of $W_{(1,+)}(t)$ and $W_{(2,-)}(t)$, may be extrapolated from the stationary state, characterized by $\bm\Theta_{\rm no}^{\infty}$, which always exists except for the fermionic case under perfect monitoring ($\Lambda^-=1$). In addition, for the sake of completeness, Fig.~\ref{fig:exp_coeff} shows $\tr\big[ \bm\Phi\big(\mathds{1}+\bm\Theta_{\rm no}^\infty\big)\,\big]$ and $\tr\big[ \bm\Phi\big(i\bm\Omega+\bm\Theta_{\rm no}^\infty\big)\,\big]$ as a function of the parameter $\Lambda^-$, which are the coefficients governing the exponential decay of the fermionic and bosonic no-jump probabilities; since the results in Fig.~\ref{fig:exp_coeff} are the numerical solutions of highly nonlinear equations, the functions $\tr\big[ \bm\Phi\big(\mathds{1}+\bm\Theta_{\rm no}^\infty\big)\,\big]$ and $\tr\big[ \bm\Phi\big(i\bm\Omega+\bm\Theta_{\rm no}^\infty\big)\,\big]$ are difficult to handle. However, Fig.~\ref{fig:exp_coeff} apparently shows a linear growth for $\tr\big[ \bm\Phi\big(\mathds{1}+\bm\Theta_{\rm no}^\infty\big)\,\big]$ as a function of $\Lambda^-$. In any case, the decay coefficient is increasing with $\Lambda^-$, as we expect if approaching the ideal monitoring regime. 

\section{\label{discuss-concl}Discussion and conclusion}
This work aims to be a complete guide to the conditional no-jump dynamics of free particles, providing analytical results for the no-jump probability and the WTDs.

After presenting generic results, we computed the no-jump dynamics of single qubits, either incoherently or coherently driven, whose time evolution turns to be very involved. 

As a main result, we successfully generalized the findings into Ref.~\cite{landi2021waiting}, deriving analytical results for the conditional no-jump dynamics of non-interacting chains under inhomogeneous one-body gain-loss processes. We also studied both bosonic and fermionic chains with particle non-conserving Hamiltonian operators. The hypothesis of Gaussianity is crucial to solve the no-jump dynamics, which turns to be governed by a Riccati-type differential equation for the correlation matrix. The solution of the Riccati equation gives access to the no-jump probability and the WTDs. As an example, we studied a tight-binding model with specific gain-loss profiles and finally we moved to boundary driven systems with pairing terms in the Hamiltonian operator. Generalizing the findings into Ref.~\cite{landi2021waiting} to non-interacting bosons also paves the way to quantum optics, with applications in optomechanical systems~\cite{aspelmeyer2014cavity,bowen2015quantum}.

\section*{Acknowledgments}
The authors acknowledge the financial support the French ANR funding UNIOPEN (Grant No. ANR-22-CE30-0004-01). M.C. thanks the LUE for funding a  travel grant (Aide \`a la mobilit\'e internationale DrEAM). 

\begin{widetext}
\newpage 
\appsection{A}{No-jump dynamics without pairing terms: explicit solution}

In this section, we assume the system is prepared in a Gaussian state and we solve the no-jump dynamics generated by \eqref{L0-2}.
To compute $d{\tilde{\rho}}_{\rm no}/dt$ we make use of the BCH formula
\begin{equation}\label{BCH}
    \frac{d e^{ K(t)}}{dt} = \left\{ \frac{ dK}{dt} + \frac{1}{2!} \left[ K,\frac{ dK}{dt}\right] + \frac{1}{3!} \left[ K,\left[ K,\frac{ dK}{dt}\right]\right] + \ldots\right\} e^{ K},
\end{equation}
for any Hilbert operator $K$. Combining the BCH formula, the Gaussian ansatz \eqref{rho_no_ansatz} and the result
\begin{equation}\label{fermionic_commutation}
    [{{\bm c}}^\dagger \cdot \bm{X} \cdot {\bm{c}},~
    {{\bm c}}^\dagger \cdot \bm{Y} \cdot {\bm{c}}] = {{\bm c}}^\dagger \cdot [\bm{X},\bm{Y}] \cdot {\bm{c}}.
\end{equation}
we get 
\begin{equation}
    \frac{d}{dt} e^{-\cmctno} = {{\bm c}}^\dagger \cdot \left(\frac{d e^{-\bm{M}^{\rm no}_t}}{dt} e^{\bm{M}^{\rm no}_t} \right) {\bm{c}}~ e^{-\cmctno}.
\end{equation}
Hence 
\begin{equation}\label{drho_no}
    \frac{d {\tilde{\rho}}_{\rm no}}{dt}   = \left[ {\bm{c}}^\dagger \cdot \left(\frac{d e^{-\bm{M}^{\rm no}_t}}{dt} e^{\bm{M}^{\rm no}_t} \right) {\bm{c}} - \frac{d}{dt}\ln{\zeta}_t  \right] {\tilde\rho}_{\rm no}.
\end{equation}
Next, we compute $\mathcal{L}_0({\tilde{\rho}}_{\rm no})$. We use the BCH formulas 
\begin{align}
    e^{-\cmctno}{\bm{c}} &= (e^{\bm{M}^{\rm no}_t} {\bm{c}})~e^{-\cmctno}, \\
    e^{-\cmctno} {\bm{c}}^\dagger &=  ({\bm{c}}^\dagger e^{-\bm{M}^{\rm no}_t})~e^{-\cmctno}.
\end{align}
to push the factors of $e^{-\cmctno}$ to the right. 
This  leads to the following list of factors:
\begin{IEEEeqnarray}{rCl}
\label{result_1}
    [{H}, {\tilde{\rho}}_{\rm no}] &=& {\bm{c}}^\dagger \cdot \big( \bm{h} - e^{-\bm{M}^{\rm no}_t} \bm{h} e^{\bm{M}^{\rm no}_t}\big) \cdot {\bm{c}} ~ {\tilde{\rho}}_{\rm no},
\\[0.2cm]
\label{result_2}
    \sum\limits_{ij} [\bm{\gamma}^-(\mathds{1}-\bm{\lambda}^-)]_{ij}  {c}_j {\tilde{\rho}}_{\rm no} {c}_i^\dagger &=& \mp {\bm{c}}^\dagger \cdot \Big(e^{-\bm{M}^{\rm no}_t} \bm{\gamma}^-(\mathds{1}-\bm{\lambda}^-) \Big) \cdot {\bm{c}}~ {\tilde{\rho}}_{\rm no} +\tr\big( \bm{\gamma}^-(\mathds{1}-\bm{\lambda}^-) e^{-\bm{M}^{\rm no}_t}\big)~{\tilde{\rho}}_{\rm no},    
\\[0.2cm]
\label{result_3}
     \sum\limits_{ij} (\bm{\gamma})_{ij}^- \{{c}_i^\dagger {c}_j , {\tilde{\rho}}_{\rm no}\} &=&  {\bm{c}}^\dagger \cdot \Big[ \bm{\gamma}^- + e^{-\bm{M}^{\rm no}_t} \bm{\gamma}^- e^{\bm{M}^{\rm no}_t} \Big] \cdot {\bm{c}} ~{\tilde{\rho}}_{\rm no}, 
\\[0.2cm]
\label{result_4}
    \sum\limits_{ij} [\bm{\gamma}^+(\mathds{1}-\bm{\lambda}^+)]_{ij} {c}_i^\dagger {\tilde{\rho}}_{\rm no} {c}_j &=& {\bm{c}}^\dagger \cdot \Big( \bm{\gamma}^+(\mathds{1}-\bm{\lambda}^+) e^{\bm{M}^{\rm no}_t} \Big) \cdot {\bm{c}} ~{\tilde{\rho}}_{\rm no}, 
\\[0.2cm]
\label{result_5}
     \sum\limits_{ij} (\bm{\gamma})_{ij}^+ \{ {c}_j {c}_i^\dagger, {\tilde{\rho}}_{\rm no} \} &=& \mp {\bm{c}}^\dagger \cdot \Big( \bm{\gamma}^+ + e^{-\bm{M}^{\rm no}_t} \bm{\gamma}^+ e^{\bm{M}^{\rm no}_t} \Big) \cdot {\bm{c}}~{\tilde{\rho}}_{\rm no}+2 \tr\big(\bm{\gamma}^+\big)~{\tilde{\rho}}_{\rm no}
\end{IEEEeqnarray}
Comparing  with Eq.~\eqref{drho_no}, we see that both take the form $\kappa~{\tilde{\rho}}_{\rm no} + {\bm{c}}^\dagger \cdot \bm{\mathcal{K}} \cdot {\bm{c}}~{\tilde{\rho}}_{\rm no} $, for c-numbers $\kappa$ and Hermitian matrices $\bm{\mathcal{K}}$. 
The Gaussian ansatz is indeed a solution, provided that $\bm M_t$ and $\zeta_t$ satisfy the differential equations
\begin{equation}\label{zeta1}
     \frac{d}{dt}\ln{\zeta}_t = \tr\Big\{ \bm{\gamma}^+ -\bm{\gamma}^-(\mathds{1}-\bm{\lambda}^-) e^{-\bm{M}^{\rm no}_t} \Big\},
\end{equation}
\begin{equation}\label{M1}
    \frac{de^{-\bm{M}^{\rm no}_t}}{dt} e^{\bm{M}^{\rm no}_t} =
    -i (\bm h - e^{-\bm{M}^{\rm no}_t} \bm h e^{\bm{M}^{\rm no}_t}) 
    \mp e^{-\bm{M}^{\rm no}_t} \bm{\gamma}^-(\mathds{1}-\bm{\lambda}^-)- \frac{1}{2} (\bm{\gamma}^- + e^{-\bm{M}^{\rm no}_t} \bm{\gamma}^- e^{\bm{M}^{\rm no}_t}) + \bm{\gamma}^+(\mathds{1}-\bm{\lambda}^+) e^{\bm{M}^{\rm no}_t}\pm \frac{1}{2} (\bm{\gamma}^+ + e^{-\bm{M}^{\rm no}_t} \bm{\gamma}^+ e^{\bm{M}^{\rm no}_t}).
\end{equation}
The matrix $\bm{M}^{\rm no}_t$ in Eq.~\eqref{rho_no_ansatz} is not convenient to work with. Instead, it is better to work with the conditional covariance matrix $(\bm{C}_{\rm no})_{ij} = \tr\Big\{ {c}_j^\dagger {c}_i~ {\rho}_{\rm no}(t)\Big\}=\langle {c_j}^\dagger {c}_i\rangle_{{\rm no}}$. Given the Gaussian ansatz \eqref{gaussian-ansatz}, there is an important relation between the matrix $\bm M_t$ and its correlation matrix $\bm{C}_{\rm no}$~\cite{peschel2003calculation,peschel2009reduced,peschel2012entanglement}, 
\begin{equation}\label{CM_relation}
    \bm{C}_{\rm no}(t) = \left(\frac{1}{e^{\bm{M}^{\rm no}_t}\pm \mathds{1}}\right),
    \qquad 
    e^{\bm{M}^{\rm no}_t} = \frac{\mathds{1}\mp \bm{C}_{\rm no}(t)}{\bm{C}_{\rm no}(t)}.
\end{equation}
Eq. \eqref{CM_relation} implies 
\begin{equation}\label{dC_dt}
    \frac{d\bm{C}_{\rm no}}{dt} = (\mathds{1}\mp \bm{C}_{\rm no}) \frac{d e^{-\bm{M}^{\rm no}_t}}{dt} (\mathds{1}\mp \bm{C}_{\rm no}).
\end{equation}
In the following, we will rewrite Eq. \eqref{M1} in terms of Eq.~\eqref{dC_dt}, getting a differential equation for $\bm C_{\rm no}$. After simplifying the results, Eq. \eqref{M1} reads
\begin{equation}
    \frac{d\bm C_{\rm no}}{dt} = - (\bm W \bm C_{\rm no} + \bm C_{\rm no} \bm W^\dagger) + \bm{\gamma}^+ \pm \bm C_{\rm no} (\bm{\gamma}^-\bm{\lambda}^-) \bm C_{\rm no} - (\mathds{1}\mp \bm C_{\rm no}) (\bm{\gamma}^+ \bm{\lambda}^+) (\mathds{1}\mp \bm C_{\rm no}),
\end{equation}
where $\bm W$ is the matrix into Eq.~\eqref{doppiaW}. 
\\

\textbf{No-jump probability}: Assuming that $\bm C_{\rm no}(t)$ is known, the solution of Eq.~\eqref{zeta_riccati} can then be written as 
\begin{equation}\label{zeta_sol1}
    \zeta_t =
    Z_0 \exp\left\{ 
    - \int\limits_0^t dt' \tr\bigg[ \bm{\gamma}^- (\mathds{1}-\bm{\lambda}^-) \frac{\bm C_{\rm no}(t')}{\mathds{1} \mp \bm C_{\rm no}(t')} - \bm{\gamma}^+\bigg]
    \right\}.
\end{equation}
This, together with 
\begin{equation}\label{Zt_formula}
    Z_t^{\rm no} = \tr(e^{-\cmctno}) = \left[{\rm det}\left(\mathds{1} \pm e^{-\bm{M}^{\rm no}_t}\right)\right]^{\pm 1} =\left[{\rm det}\left(\mathds{1}\mp \bm C_{\rm no}(t)\right)\right]^{\mp 1},
\end{equation}
determines the no-jump probability
\begin{equation}
    P_{\rm no}(t)=\tr~ \tilde{\rho}_{\rm no}(t)=\frac{\zeta_t}{Z_t^{\rm no}}.
\end{equation}
However, Eq.~\eqref{zeta_sol1} is complicated  to handle, for two reasons: it depends on an integral over all previous times and it requires knowledge of the inverse matrix $(\mathds{1}\mp\bm C_{\rm no})$. As sanity check, we can use the results into Eqs.~(\ref{zeta_sol1},\ref{Zt_formula}) to show that $\frac{d}{dt}\ln(P_{\rm no}(t))=\frac{d}{dt}\ln(\zeta_t/Z_t^{\rm no})=\beta(t)$, where $\beta(t)$ is the same as Eq.~\eqref{Pno_final}. As a first step, we use Eq.~\eqref{riccati} to rewrite the right-hand side of Eq.~\eqref{zeta1}. After a series of manipulations, we get
\begin{equation}\label{bho1}
    \tr\Big\{ \bm{\gamma}^- (\mathds{1}-\bm{\lambda}^-) e^{-\bm{M}^{\rm no}_t} - \bm{\gamma}^+\Big\}
    = 
    - \tr\left[(\mathds{1}\mp \bm C_{\rm no})\inv \frac{d{\bm C}_{\rm no}}{dt}\right] - \tr\Big\{ \bm{\gamma}^+ \bm{\lambda}^+ (\mathds{1}\mp \bm C_{\rm no}) + \bm{\gamma}^- \bm{\lambda}^- \bm C_{\rm no}\Big\}.
\end{equation}
Thanks to the Jacobi's formula, for any matrix $\bm A$, the identity $\frac{d}{dt} \ln \det (\bm A) = \tr\left(\bm A\inv \frac{d{\bm A}}{dt}\right)$ holds. Applying the Jacobi's formula to Eq.~\eqref{Zt_formula},
\begin{equation}\label{bho2}
    \frac{d}{dt} \ln Z_t^{\rm no}=\tr\left[(1\mp \bm C_{\rm no})\inv \frac{d{\bm C}_{\rm no}}{dt}\right].
\end{equation}
Combining Eqs.~(\ref{bho1},\ref{bho2}),
\begin{equation}
    \frac{d}{dt} \ln \zeta_t/Z_t^{\rm no} = \frac{d}{dt}\ln P_{\rm no}=\tr \Big\{ \bm{\gamma}^+ \bm{\lambda}^+ (\mathds{1}\mp \bm C_{\rm no}) + \bm{\gamma}^- \bm{\lambda}^- \bm C_{\rm no}\Big\},
\end{equation}
which coincides with the differential equation~\eqref{P_no_rhono}, where $\beta(t)=\tr \Big\{ \bm{\gamma}^+ \bm{\lambda}^+ (\mathds{1}\mp \bm C_{\rm no}) + \bm{\gamma}^- \bm{\lambda}^- \bm C_{\rm no}\Big\}$ is the same as Eq.~\eqref{Pno_final}.  

\appsection{B}{Majorana fermions and quadratic forms}
Quadratic forms ${\bm{Y}}^\dagger \cdot \bm{\mathcal{T}} \cdot {\bm{Y}}$ involving Majorana fermions $\bm{Y}$ can always be defined using anti-symmetric ($\bm{\mathcal{T}}\trans = -\bm{\mathcal{T}}$) and hermitian matrices ($\bm{\mathcal{T}}^\dagger = \bm{\mathcal{T}}$). Indeed, if $\bm{\mathcal{T}}$ is not anti-symmetric, we can always write 
\begin{equation}\label{pairing_anti_symmetric_relation}
    {\bm{Y}}^\dagger \cdot \bm{\mathcal{T}} \cdot {\bm{Y}} = \tr(\bm{\mathcal{T}}) + \frac{1}{2} {\bm{Y}}^\dagger\cdot (\bm{\mathcal{T}}-\bm{\mathcal{T}}\trans) \cdot {\bm{Y}}.
\end{equation}
In this appendix, we focus on the structure of the $2L\times2L$ matrix $\bm{\mathcal{T}}$. The most general Gaussian-preserving Hamiltonian is of the form 
\begin{equation}\label{pairing_H_c}
    {H} = \sum\limits_{n,m} \Big\{ h_{n m} {c}_n^\dagger {c}_m + \frac{1}{2} \Big( G_{nm} {c}_n^\dagger {c}_m^\dagger + G_{mn}^* {c}_n {c}_m\Big) \Big\},
\end{equation}
where $\bm h = \bm h^\dagger$ and $\bm G\trans = - \bm G$ (but  may still be complex).
In terms of Majorana operators, this can be written as 
\begin{equation}\label{pairing_H_majorana}
    H = \frac{1}{4} {\bm{Y}}^\dagger \cdot \bm{\mathcal{K}} \cdot {\bm{Y}},
\end{equation}
where
\begin{equation}
    \bm{\mathcal{K}} = \begin{pmatrix}
    \bm h + (\bm G+\bm G^\dagger)/2 & i \bm h - i (\bm G-\bm G^\dagger)/2 \\[0.2cm]
    -i \bm h - i (\bm G-\bm G^\dagger)/2 & \bm h -(\bm G+\bm G^\dagger)/2
    \end{pmatrix}= \mathds{1}\otimes \bm h - \bm \sigma_y \otimes \bm h +  \bm \sigma_z \otimes \left(\frac{\bm G+\bm G^\dagger}{2}\right) - i \bm \sigma_x \otimes \left(\frac{\bm G-\bm G^\dagger}{2}\right).
\end{equation}
Since $\bm h\trans = \bm h^*$ and $\bm G^\dagger = -\bm G^*$,
\begin{equation}
    \frac{1}{2}(\bm{\mathcal{K}}-\bm{\mathcal{K}}\trans) = \frac{1}{2}\begin{pmatrix}
    (\bm h-\bm h^*) + (\bm G-\bm G^*) & i(\bm h+\bm h^*) -i(\bm G+\bm G^*) \\[0.2cm]
    -i(\bm h+\bm h^*) -i(\bm G+\bm G^*) & (\bm h-\bm h^*) -(\bm G-\bm G^*)
    \end{pmatrix},\qquad\qquad\tr(\bm{\mathcal{K}})=2\tr(\bm h).
\end{equation}
If $\bm h$ and $\bm G$ are real, the anti-symmetrized $\bm{\mathcal{K}}$ reduces to 
\begin{equation}
    \frac{1}{2}(\bm{\mathcal{K}}-\bm{\mathcal{K}}\trans) = i \begin{pmatrix} 
    0 & \bm h - \bm G \\[0.2cm]
    - \bm h - \bm G & 0 
    \end{pmatrix}
    = -\bm \sigma_y \otimes \bm h - i \bm \sigma_x \otimes \bm G,
\end{equation}
and finally
\begin{equation}
    H = \frac{1}{4} {\bm{Y}}^\dagger \cdot \bm{\mathcal{T}} \cdot {\bm{Y}}+ \frac{1}{2}\tr(\bm h),\hspace{2cm}  \bm{\mathcal{T}}=i \begin{pmatrix} 
    0 & \bm h - \bm G \\[0.2cm]
    - \bm h - \bm G & 0 
    \end{pmatrix}.
\end{equation}

\appsection{C}{Dissipation matrix}
The goal of this section is to get the structure of the dissipation matrix $\bm\Upsilon$, which must verify the equivalence between the dissipator \eqref{L-2} and \eqref{pairing_dissipator} for fermions, or similarly the dissipator \eqref{L-2} and \eqref{pairing_dissipator_bos} for bosons. We start decomposing $\bm\Upsilon$ in a block form, following the same ordering of the Majorana fermions $\bm{Y}$ or the quadratures $\bm R$:
\begin{equation}
    \bm\Upsilon = \begin{pmatrix}
    \bm A & \bm B \\[0.2cm]
    \bm C & \bm D
    \end{pmatrix}.
\end{equation}
One may verify that the dissipator \eqref{L0_mod} is a bilinear superoperator,
\begin{IEEEeqnarray}{rCl}
\mathcal{D}(\rho) &=& \sum\limits_{ij} \Bigg\{ 
(\bm A-\bm D+i\bm B+i\bm C)_{ij} \left[ {c}_i^\dagger {\rho} {c}^\dagger_j - \frac{1}{2} \{{c}^\dagger_j {c}_i^\dagger, {\rho}\}\right] + (\bm A-\bm D-i\bm B -i\bm C)_{ij} \left[ {c}_i {\rho} {c}_j - \frac{1}{2} \{{c}_j {c}_i, {\rho}\}\right] \nonumber 
\\[0.2cm]
&&\qquad + (\bm A+\bm D+i\bm B-i\bm C)_{ji} \left[ {c}_j {\rho} {c}_i^\dagger - \frac{1}{2} \{{c}_i^\dagger {c}_j, {\rho}\}\right] + (\bm A+\bm D-i\bm B+i\bm C)_{ij} \left[ {c}_i^\dagger {\rho} {c}_j - \frac{1}{2} \{{c}_j {c}_i^\dagger, {\rho}\}\right].
\end{IEEEeqnarray}
In the usual thermal case, where the dissipator is given by Eq. \eqref{L-2}, 
we get 
\begin{align}
    \begin{cases}
    \bm A - \bm D + i(\bm B+\bm C) = 0 \\
    \bm A - \bm D - i(\bm B+\bm C) = 0 \\
    \bm A + \bm D + i(\bm B-\bm C) = (\bm{\gamma}^-)\trans \\
    \bm A + \bm D - i(\bm B-\bm C) = \bm{\gamma}^+
    \end{cases}
\end{align}
which implies 
\begin{equation}
    \bm A = \bm D = \frac{\bm{\gamma}^+ + (\bm{\gamma}^-)\trans}{4},
    \qquad 
    \bm B = - \bm C = i\frac{\bm{\gamma}^+ - (\bm{\gamma}^-)\trans}{4}.
\end{equation}
Note that, if $\bm{\gamma}^+$ and $\bm{\gamma}^-$ are two symmetric matrices then
\begin{equation}
    \bm\Upsilon+\bm\Upsilon\trans = \frac{1}{2} \begin{pmatrix} \bm{\gamma}^- + \bm{\gamma}^+ & 0 \\[0.2cm] 0 & \bm{\gamma}^- + \bm{\gamma}^+
    \end{pmatrix},
    \qquad 
    \bm\Upsilon-\bm\Upsilon\trans = \frac{i}{2} \begin{pmatrix} 0 & \bm{\gamma}^+ - \bm{\gamma}^- \\[0.2cm]
    \bm{\gamma}^- - \bm{\gamma}^+ & 0 \end{pmatrix}.
\end{equation}

\appsection{D}{No-jump dynamics with pairing terms: explicit solution for fermionic chains}
In this section, we shall prepare the system in a Gaussian state and we shall solve the dynamics generated by the no-jump operator~\eqref{L0_mod}.
Firstly, we note that Eq.~\eqref{fermionic_commutation} is naturally translated to majorana operators: given two anti-symmetric matrices $\bm A$ and $\bm B$, 
\begin{equation}
    [{\bm{Y}}^\dagger \cdot \bm A \cdot {\bm{Y}}, {\bm{Y}}^\dagger \cdot \bm B \cdot {\bm{Y}}] = 4 {\bm{Y}}^\dagger \cdot[\bm A,\bm B]\cdot {\bm{Y}}.
\end{equation}
The BCH formula \eqref{BCH} leads to 
\begin{equation}
    \frac{d}{dt}e^{\frac{1}{4} \ymytno}={\bm{Y}}^\dagger \cdot \left[\frac{1}{4}\frac{de^{\bm{\mathcal{M}}^{\rm no}_t}}{dt}e^{-\bm{\mathcal{M}}^{\rm no}_t}\right]\cdot{\bm{Y}}e^{\frac{1}{4} \ymytno}.
\end{equation}
Analogously to~\eqref{drho_no}, we then get 
\begin{equation}\label{major1}
    \frac{d {\tilde{\rho}}_{\rm no}}{dt}   ={\bm{Y}}^\dagger \cdot \left[\frac{1}{4}\frac{de^{\bm{\mathcal{M}}^{\rm no}_t}}{dt}e^{-\bm{\mathcal{M}}^{\rm no}_t}\right]\cdot{\bm{Y}}{\tilde{\rho}}_{\rm no} - \left(\frac{d}{dt}\ln{\zeta}_t\right){\tilde{\rho}}_{\rm no}.
\end{equation}
Any matrix $\bm A$ satisfies the obvious identity 
\begin{equation}\label{identity_deriv}
    \bm A^{-1} \frac{d\bm A}{dt} = - \frac{d\bm A^{-1}}{dt}\bm A,
\end{equation}
which implies the matrix $\frac{d e^{\bm{\mathcal{M}}^{\rm no}_t}}{dt}e^{-\bm{\mathcal{M}}^{\rm no}_t}$ is  anti-symmetric. 
Thus, $d{\tilde\rho}_{\rm no}/dt$ has the form 
$\kappa{\tilde\rho}_{\rm no} + \bm{Y}^\dagger \cdot \bm{\mathcal{K}} \cdot \bm{Y}{\tilde\rho}_{\rm no}$, for a c-number $\kappa$ and an anti-symmetric matrix $\bm{\mathcal{K}}$. 
Hence, when we write any right-hand side below, we will always make sure to cast them in the same form. The next step is to compute the Liouvillian terms, which we do using the BCH formula \eqref{BCH},
\begin{equation}
    e^{\frac{1}{4}\ymytno}{\bm{Y}} = e^{-\bm{\mathcal{M}}^{\rm no}_t}{\bm{Y}}e^{\frac{1}{4}\ymytno}.
\end{equation}
This leads, for example, to 
\begin{equation}
   -i [ H,{\tilde{\rho}}_{\rm no}]={\bm{Y}}^\dagger \cdot\Bigg[ -\frac{i}{4}\bigg(\bm{\mathcal{T}}-e^{\bm{\mathcal{M}}^{\rm no}_t}\bm{\mathcal{T}}e^{-\bm{\mathcal{M}}^{\rm no}_t}\bigg)\Bigg] \cdot {\bm{Y}}{\tilde{\rho}}_{\rm no},
\end{equation}
which is the analog of Eq.~\eqref{result_1}.
As in Eq.~\eqref{major1}, the resulting matrix is anti-symmetric, since both $\bm{\mathcal{M}}^{\rm no}_t$ and $\bm{\mathcal{T}}$ are anti-symmetric.
Next we turn to the dissipators. 
We have 
\begin{equation}
    \sum_{ij}[\bm\Upsilon-\bm\Phi]_{ij} Y_i{\tilde{\rho}}_{\rm no}  Y_j = {\bm{Y}}^\dagger \cdot [\bm\Upsilon-\bm\Phi] e^{-\bm{\mathcal{M}}^{\rm no}_t} \cdot {\bm{Y}}{\tilde{\rho}}_{\rm no},
\end{equation}
\begin{equation}
     \sum\limits_{ij} (\bm \Upsilon)_{ij} \{ Y_j  Y_i, {\tilde{\rho}}_{\rm no}\} = 4 \tr(\bm \Upsilon)~{\tilde{\rho}}_{\rm no} - {\bm{Y}}^\dagger \cdot \left[ \bm\Upsilon + e^{\bm{\mathcal{M}}^{\rm no}_t} \bm\Upsilon e^{-\bm{\mathcal{M}}^{\rm no}_t} \right] \cdot {\bm{Y}}~{\tilde{\rho}}_{\rm no}.
\end{equation}
These terms, however, do not have anti-symmetric matrices in their quadratic forms (even if we combine them to form the full dissipator~\eqref{pairing_dissipator}). 
We must therefore use Eq.~\eqref{pairing_anti_symmetric_relation}, which then leads to 
\begin{equation}\label{pairing_jump_terms}
    \sum_{ij}[\bm\Upsilon-\bm\Phi]_{ij} Y_i{\tilde{\rho}}_{\rm no}  Y_j = {\bm{Y}}^\dagger \cdot \Bigg[\frac{ [\bm \Upsilon-\bm \Phi] e^{-\bm{\mathcal{M}}^{\rm no}_t} - e^{\bm{\mathcal{M}}^{\rm no}_t} [\bm \Upsilon-\bm\Phi]\trans}{2}\Bigg] \cdot {\bm{Y}}{\tilde{\rho}}_{\rm no} + \tr([\bm\Upsilon-\bm\Phi] e^{-\bm{\mathcal{M}}^{\rm no}_t}){\tilde{\rho}}_{\rm no},
\end{equation}
\begin{equation}
     \sum\limits_{ij} (\bm \Upsilon)_{ij} \{ Y_j  Y_i, {\tilde{\rho}}_{\rm no}\} = 2\tr(\bm\Upsilon){\tilde{\rho}}_{\rm no} -{\bm{Y}}^\dagger \cdot \Bigg[ \frac{\bm\Upsilon - \bm\Upsilon\trans}{2}+ e^{\bm{\mathcal{M}}^{\rm no}_t} \frac{ \bm\Upsilon - \bm\Upsilon\trans}{2} e^{-\bm{\mathcal{M}}^{\rm no}_t} \Bigg] \cdot {\bm{Y}}{\tilde{\rho}}_{\rm no} .
\end{equation}
These now are in the proper form. The no-jump operator \eqref{L0_mod} applied to the Gaussian ansatz becomes 
\begin{align}\label{before_last}
     \mathcal{L}_0({\tilde{\rho}}_{\rm no}) =& \bm{Y}^\dagger \cdot \Bigg[ \frac{ [\bm\Upsilon-\bm\Phi] e^{-\bm{\mathcal{M}}^{\rm no}_t} - e^{\bm{\mathcal{M}}^{\rm no}_t} [\bm\Upsilon-\bm\Phi]\trans}{2} + \frac{\bm\Upsilon-\bm\Upsilon\trans}{4} + e^{\bm{\mathcal{M}}^{\rm no}_t} \frac{\bm\Upsilon - \bm\Upsilon\trans}{4} e^{-\bm{\mathcal{M}}^{\rm no}_t} \Bigg] \cdot \bm{Y}~{\tilde{\rho}}_{\rm no}\nonumber\\
    &+{\bm{Y}}^\dagger \cdot\Bigg[ -\frac{i}{4}\bigg(\bm{\mathcal{T}}-e^{\bm{\mathcal{M}}^{\rm no}_t}\bm{\mathcal{T}}e^{-\bm{\mathcal{M}}^{\rm no}_t}\bigg)\Bigg] \cdot {\bm{Y}}{\tilde{\rho}}_{\rm no} + \Big[ \tr([\bm\Upsilon-\bm\Phi] e^{-\bm{\mathcal{M}}^{\rm no}_t}) - \tr(\bm\Upsilon)\Big] {\tilde{\rho}}_{\rm no}.
\end{align}
Combining Eq. \eqref{before_last} with Eq. \eqref{major1},
\begin{equation}\label{zeta-pairing-evolution}
    -\frac{d}{dt}\ln{\zeta}_t = \tr([\bm\Upsilon-\bm\Phi] e^{-\bm{\mathcal{M}}^{\rm no}_t}) - \tr(\bm\Upsilon).
\end{equation}
\begin{equation}\label{pairing_eM_eq_tmp2}
     \frac{d e^{\bm{\mathcal{M}}^{\rm no}_t}}{dt} = -i \Big(\bm{\mathcal{T}}e^{\bm{\mathcal{M}}^{\rm no}_t} -e^{\bm{\mathcal{M}}^{\rm no}_t}\bm{\mathcal{T}}\Big) +2[\bm\Upsilon-\bm\Phi] - 2 e^{\bm{\mathcal{M}}^{\rm no}_t} [\bm\Upsilon-\bm\Phi]\trans e^{\bm{\mathcal{M}}^{\rm no}_t} + (\bm\Upsilon-\bm\Upsilon\trans) e^{\bm{\mathcal{M}}^{\rm no}_t} + e^{\bm{\mathcal{M}}^{\rm no}_t} (\bm\Upsilon - \bm\Upsilon\trans).
\end{equation}
Again, the matrix $\bm{\mathcal{M}}^{\rm no}_t$ is not so convenient to work with. For this reason, we recall the definition of the conditional covariance matrix $(\bm{\Theta}_{\rm no})_{ij} = \frac{1}{2}\tr\Big\{[ Y_i, Y_j]\rho_{\rm no}(t)\Big\}= \frac{1}{2} \langle [ Y_i, Y_j]\rangle_{\rm no}(t)$.  Given the Gaussian ansatz \eqref{gaussian-ansatz2}, there is an important relation between the matrix $\bm{\mathcal{M}}^{\rm no}_t$ and its correlation matrix $\bm{\Theta}_{\rm no}$~\cite{bravyi2004lagrangian,eisler2015partial}, 
\begin{equation}\label{pairing_theta_M_relation}
    \bm\Theta_{\rm no} = -\tanh(\bm{\mathcal{M}}^{\rm no}_t/2),
    \qquad 
    e^{\bm{\mathcal{M}}^{\rm no}_t} = \frac{\mathds{1}-\bm\Theta_{\rm no}}{\mathds{1}+\bm\Theta_{\rm no}}.
\end{equation}
One can also readily show that 
\begin{equation}\label{pairing_theta_M_derivative}
    \frac{d\bm\Theta_{\rm no}}{dt} =- \frac{1}{2} (\mathds{1} + \bm\Theta_{\rm no}) \frac{d e^{\bm{\mathcal{M}}^{\rm no}_{t}}}{dt} (\mathds{1}+\bm\Theta_{\rm no}).
\end{equation}
Using Eq.~\eqref{pairing_theta_M_derivative} to rewrite Eq. \eqref{pairing_eM_eq_tmp2}, we get
\begin{equation}
    \frac{d\bm\Theta_{\rm no}}{dt} = - \big(\bm{\mathcal{W}} \bm\Theta_{\rm no} + \bm\Theta_{\rm no} \bm{\mathcal{W}}^\dagger\big) + \bm{\mathcal{F}} - (\mathds{1}-\bm\Theta_{\rm no}) \bm\Phi\trans (\mathds{1}-\bm\Theta_{\rm no}) + (\mathds{1}+\bm\Theta_{\rm no})\bm\Phi (\mathds{1}+\bm\Theta_{\rm no}),
\end{equation}
where $\bm{\mathcal{W}}$ and $\bm{\mathcal{F}}$ are the matrices into Eqs.~(\ref{doppiaW_maj},\ref{F_maj}).
\\ 

\textbf{No-jump probability}: According to Eq. \eqref{zeta-pairing-evolution}, $\zeta_t$ satisfies 
\begin{equation}\label{normalization_pairing1}
    \frac{d}{dt} \ln \zeta_t = - \tr\Big\{(\bm\Upsilon - \bm\Phi) e^{-{\bm{\mathcal{M}}_t^{\rm no}}} - \bm\Upsilon\Big\},
\end{equation}
Conversely, the partition function satisfies 
\begin{equation}\label{normalization_pairing2}
    \frac{d}{dt} \ln Z_t^{\rm no} = \frac{1}{2} \tr \left\{ (1+e^{\bm{\mathcal{M}}^{\rm no}_t})^{-1} \frac{de^{\bm{\mathcal{M}}^{\rm no}_t}}{dt} \right\}.
\end{equation}
Combining Eqs.~(\ref{pairing_theta_M_relation},~\ref{pairing_eM_eq_tmp2},~\ref{normalization_pairing1},~\ref{normalization_pairing2}), the anti-symmetry of $\bm{\mathcal{M}}^{\rm no}_t$ together with the invariance of the trace under transpose, we arrive at 
\begin{equation}
    \frac{d}{dt} \ln \zeta_t/Z_t^{\rm no} =\tr\Big\{ \bm\Phi (\mathds{1}+\bm\Theta_{\rm no})\Big\}.
\end{equation}
The no-jump probability finally acquires the compact form
\begin{equation}
 P_{\rm no}(t) = \exp\left\{ - \int\limits_0^t dt'~\tr\Big[ \bm\Phi(\mathds{1}+\bm\Theta_{\rm no}(t'))\Big] \right\}.
 \end{equation}

\appsection{E}{Quadrature operators and quadratic forms}
Quadratic forms ${\bm{R}}^\dagger \cdot \bm{\mathcal{T}} \cdot {\bm{R}}$ involving quadrature operators $\bm{R}$ can always be defined using symmetric ($\bm{\mathcal{T}}\trans =\bm{\mathcal{T}}$) and hermitian matrices ($\bm{\mathcal{T}}^\dagger = \bm{\mathcal{T}}$). Indeed, if $\bm{\mathcal{T}}$ is not symmetric, we can always write 
\begin{equation}\label{pairing_anti_symmetric_relation_bos}
    \bm{R}^\dagger \cdot \bm{\mathcal{T}} \cdot \bm{R} = -\tr\left(\bm{\mathcal{T}}(i\bm\Omega)\right) + \frac{1}{2} \bm{R}^\dagger\cdot (\bm{\mathcal{T}}+\bm{\mathcal{T}}\trans) \cdot \bm{R},
\end{equation}
where $\bm\Omega$ is the symplectic matrix~\eqref{symplectic}. In this appendix, we focus on the structure of the $2L\times2L$ matrix $\bm{\mathcal{T}}$. The most general Gaussian-preserving Hamiltonian is of the form 
\begin{equation}
    {H} = \sum\limits_{n,m} \Big\{ h_{n m} {c}_n^\dagger {c}_m + \frac{1}{2} \Big( G_{nm} {c}_n^\dagger {c}_m^\dagger + G_{mn}^* {c}_n {c}_m\Big) \Big\},
\end{equation}
where the $\bm c$'s are bosonic operators, $\bm h = \bm h^\dagger$ and $\bm G\trans = \bm G$. Using the quadrature operators $\bm R$, the Hamiltonian may be written as
\begin{equation}\label{pairing_H_quadratures}
    H = \frac{1}{4} {\bm{R}}^\dagger \cdot \bm{\mathcal{K}} \cdot {\bm{R}},
\end{equation}
where
\begin{equation}
    \bm{\mathcal{K}} = \begin{pmatrix}
    \bm h + (\bm G+\bm G^\dagger)/2 & i \bm h - i (\bm G-\bm G^\dagger)/2 \\[0.2cm]
    -i \bm h - i (\bm G-\bm G^\dagger)/2 & \bm h -(\bm G+\bm G^\dagger)/2
    \end{pmatrix}= \mathds{1}\otimes \bm h - \bm \sigma_y \otimes \bm h +  \bm \sigma_z \otimes \left(\frac{\bm G+\bm G^\dagger}{2}\right) - i \bm \sigma_x \otimes \left(\frac{\bm G-\bm G^\dagger}{2}\right).
\end{equation}
Since $\bm h\trans = \bm h^*$ and $\bm G^\dagger = \bm G^*$,
\begin{equation}
    \frac{1}{2}(\bm{\mathcal{K}}+\bm{\mathcal{K}}\trans) = \frac{1}{2}\begin{pmatrix}
    (\bm h+\bm h^*) + (\bm G+\bm G^*) & i(\bm h-\bm h^*) -i(\bm G-\bm G^*) \\[0.2cm]
    -i(\bm h-\bm h^*) -i(\bm G-\bm G^*) & (\bm h+\bm h^*) -(\bm G+\bm G^*)
    \end{pmatrix},\qquad\qquad-\frac{1}{2}\tr\left((\bm{\mathcal{K}}-\bm{\mathcal{K}}\trans)(i\bm\Omega)\right)=-2\tr(\bm h).
\end{equation}
If $\bm h$ and $\bm G$ are real, the symmetrized $\bm{\mathcal{K}}$ reduces to 
\begin{equation}
    \frac{1}{2}(\bm{\mathcal{K}}+\bm{\mathcal{K}}\trans) = \begin{pmatrix} 
    \bm h + \bm G & 0 \\[0.2cm]
    0 & \bm h - \bm G 
    \end{pmatrix}
    = \mathds{1}_2 \otimes \bm h + \bm \sigma_z \otimes \bm G,
\end{equation}
and finally
\begin{equation}
    H = \frac{1}{4} {\bm{R}}^\dagger \cdot \bm{\mathcal{T}} \cdot {\bm{R}}- \frac{1}{2}\tr(\bm h),\hspace{2cm}  \bm{\mathcal{T}}=\begin{pmatrix} 
    \bm h + \bm G & 0 \\[0.2cm]
    0 & \bm h - \bm G 
    \end{pmatrix}.
\end{equation}

\appsection{F}{No-jump dynamics with pairing terms: explicit solution for bosonic chains}
In this section, we shall prepare the system in a Gaussian state and we shall solve the dynamics generated by the no-jump operator~\eqref{L0_mod_bos}.
Firstly, we note that the commutator between two quadratic forms is
\begin{equation}\label{commutator_quadratic_form_quadratures}
    [{\bm{R}}^\dagger \cdot (i\bm\Omega)\bm A \cdot {\bm{R}}, {\bm{R}}^\dagger \cdot (i\bm\Omega)\bm B \cdot {\bm{R}}] = 4 {\bm{R}}^\dagger \cdot(i\bm\Omega)[\bm A,\bm B]\cdot {\bm{R}},
\end{equation}
where $(i\bm\Omega)\bm A$ and $(i\bm\Omega)\bm B$ are two symmetric matrices.
The BCH formula \eqref{BCH} and Eq.~\eqref{commutator_quadratic_form_quadratures} lead to 
\begin{equation}
    \frac{d}{dt}e^{\frac{1}{4} \rmrtno}={\bm{R}}^\dagger \cdot \left[\frac{i\bm\Omega}{4}\frac{de^{\bm{\mathcal{M}}^{\rm no}_t}}{dt}e^{-\bm{\mathcal{M}}^{\rm no}_t}\right]\cdot{\bm{R}}\,e^{\frac{1}{4} \rmrtno}.
\end{equation}
The time-derivative of the no-jump operator reads
\begin{equation}\label{major1_quadratures}
    \frac{d {\tilde{\rho}}_{\rm no}}{dt}   ={\bm{R}}^\dagger \cdot \left[\frac{i\bm\Omega}{4}\frac{de^{\bm{\mathcal{M}}^{\rm no}_t}}{dt}e^{-\bm{\mathcal{M}}^{\rm no}_t}\right]\cdot{\bm{R}}\,{\tilde{\rho}}_{\rm no} - \left(\frac{d}{dt}\ln{\zeta}_t\right){\tilde{\rho}}_{\rm no}.
\end{equation}
Since the matrix $(i\bm\Omega)\bm{\mathcal{M}}^{\rm no}_t$ is symmetric, it is possible to prove that the matrix $\Big[\frac{i\bm\Omega}{4}\frac{de^{\bm{\mathcal{M}}^{\rm no}_t}}{dt}e^{-\bm{\mathcal{M}}^{\rm no}_t}\Big]$ is symmetric as well. Using the identity $(i\bm \Omega)\trans=-(i\bm \Omega)$, the transpose matrix becomes
\begin{equation}
    \left[(i\bm\Omega)\frac{de^{\bm{\mathcal{M}}^{\rm no}_t}}{dt}e^{-\bm{\mathcal{M}}^{\rm no}_t}\right]\trans=-e^{-(\bm{\mathcal{M}}^{\rm no}_t)\trans}\frac{de^{(\bm{\mathcal{M}}^{\rm no}_t)\trans}}{dt}(i\bm\Omega).
\end{equation}
Thanks to the identity~\eqref{identity_deriv}, $(i\bm\Omega)^2=1$ and the Taylor expansion,
\begin{align}
    \left[(i\bm\Omega)\frac{de^{\bm{\mathcal{M}}^{\rm no}_t}}{dt}e^{-\bm{\mathcal{M}}^{\rm no}_t}\right]\trans&=(i\bm\Omega)\left[(i\bm\Omega)\frac{de^{-(\bm{\mathcal{M}}^{\rm no}_t)\trans}}{dt}(i\bm\Omega)\right]\left[(i\bm\Omega)e^{(\bm{\mathcal{M}}^{\rm no}_t)\trans}(i\bm\Omega)\right]\\
    &=(i\bm\Omega)\frac{de^{-(i\bm\Omega)(\bm{\mathcal{M}}^{\rm no}_t)\trans (i\bm\Omega)}}{dt}\,e^{(i\bm\Omega)(\bm{\mathcal{M}}^{\rm no}_t)\trans (i\bm\Omega)}.
\end{align}
Since the matrix $(i\bm\Omega)\bm{\mathcal{M}}^{\rm no}_t$ is symmetric, 
\begin{equation}\label{identity_Mno_bos}
    \bm{\mathcal{M}}^{\rm no}_t = -(i\Omega)\left(\bm{\mathcal{M}}^{\rm no}_t\right)\trans(i\Omega)
\end{equation}
we conclude that $\Big[\frac{i\bm\Omega}{4}\frac{de^{\bm{\mathcal{M}}^{\rm no}_t}}{dt}e^{-\bm{\mathcal{M}}^{\rm no}_t}\Big]\trans=\Big[\frac{i\bm\Omega}{4}\frac{de^{\bm{\mathcal{M}}^{\rm no}_t}}{dt}e^{-\bm{\mathcal{M}}^{\rm no}_t}\Big]$. Thus, $d{\tilde\rho}_{\rm no}/dt$ takes the form 
$\kappa{\tilde\rho}_{\rm no} + \bm{R}^\dagger \cdot \bm{\mathcal{K}} \cdot \bm{R}\,{\tilde\rho}_{\rm no}$, for a c-number $\kappa$ and a symmetric matrix $\bm{\mathcal{K}}$. 
Hence, when we rewrite the generator~\eqref{L0_mod_bos} by using the Gaussian ansatz, we will always make sure to cast it in the same form. The next step is to compute the Liouvillian terms by using the BCH formula \eqref{BCH},
\begin{equation}
    e^{\frac{1}{4}\rmrtno}{\bm{R}} = e^{-\bm{\mathcal{M}}^{\rm no}_t}{\bm{R}}e^{\frac{1}{4}\rmrtno}.
\end{equation}
This leads, for example, to 
\begin{equation}
   -i [ H,{\tilde{\rho}}_{\rm no}]={\bm{R}}^\dagger \cdot\Bigg[ -\frac{i}{4}\bigg(\bm{\mathcal{T}}-e^{\bm{-(\mathcal{M}}^{\rm no}_t)\trans}\bm{\mathcal{T}}e^{-\bm{\mathcal{M}}^{\rm no}_t}\bigg)\Bigg] \cdot {\bm{R}}\,{\tilde{\rho}}_{\rm no},
\end{equation}
which is the unitary contribution to the dynamics.
As in Eq.~\eqref{major1_quadratures}, the resulting matrix $\bigg(\bm{\mathcal{T}}-e^{\bm{-(\mathcal{M}}^{\rm no}_t)\trans}\bm{\mathcal{T}}e^{-\bm{\mathcal{M}}^{\rm no}_t}\bigg)$ is symmetric. Next we turn to the dissipators,
\begin{equation}
    \sum_{ij}[\bm\Upsilon-\bm\Phi]_{ij} R_i{\tilde{\rho}}_{\rm no}  R_j = {\bm{R}}^\dagger \cdot [\bm\Upsilon-\bm\Phi] e^{-\bm{\mathcal{M}}^{\rm no}_t} \cdot {\bm{R}}\,{\tilde{\rho}}_{\rm no},
\end{equation}
\begin{equation}
     \sum\limits_{ij} (\bm \Upsilon)_{ij} \{ R_j  R_i, {\tilde{\rho}}_{\rm no}\} = 4i \tr(\bm \Upsilon\bm\Omega)~{\tilde{\rho}}_{\rm no} + {\bm{R}}^\dagger \cdot \left[ \bm\Upsilon + e^{-(\bm{\mathcal{M}}^{\rm no}_t)\trans} \bm\Upsilon e^{-\bm{\mathcal{M}}^{\rm no}_t} \right] \cdot {\bm{R}}\,{\tilde{\rho}}_{\rm no}.
\end{equation}
These terms, however, do not have symmetric matrices in their quadratic forms. We must therefore use Eq.~\eqref{pairing_anti_symmetric_relation_bos}, which leads to 
\begin{equation}\label{pairing_jump_terms_bos}
    \sum_{ij}[\bm\Upsilon-\bm\Phi]_{ij} R_i{\tilde{\rho}}_{\rm no}  R_j = {\bm{R}}^\dagger \cdot \Bigg[\frac{ [\bm \Upsilon-\bm \Phi] e^{-\bm{\mathcal{M}}^{\rm no}_t} + e^{-(\bm{\mathcal{M}}^{\rm no}_t)\trans} [\bm \Upsilon-\bm\Phi]\trans}{2}\Bigg] \cdot {\bm{R}}\,{\tilde{\rho}}_{\rm no}- \tr\left([\bm \Upsilon-\bm \Phi] e^{-\bm{\mathcal{M}}^{\rm no}_t}(i\bm\Omega)\right)\,{\tilde{\rho}}_{\rm no},
\end{equation}
\begin{equation}
     \sum\limits_{ij} (\bm \Upsilon)_{ij} \{ R_j  R_i, {\tilde{\rho}}_{\rm no}\} = 2\tr\left(\bm\Upsilon(i\bm\Omega)\right)\,{\tilde{\rho}}_{\rm no} +{\bm{R}}^\dagger \cdot \Bigg[ \frac{\bm\Upsilon + \bm\Upsilon\trans}{2}+ e^{-(\bm{\mathcal{M}}^{\rm no}_t)\trans} \frac{ \bm\Upsilon + \bm\Upsilon\trans}{2} e^{-\bm{\mathcal{M}}^{\rm no}_t} \Bigg] \cdot {\bm{R}}\,{\tilde{\rho}}_{\rm no} .
\end{equation}
The no-jump operator \eqref{L0_mod_bos} applied to the Gaussian ansatz becomes 
\begin{align}\label{before_last_bos}
     \mathcal{L}_0({\tilde{\rho}}_{\rm no}) =& \bm{R}^\dagger \cdot \Bigg[ \frac{ [\bm \Upsilon-\bm \Phi] e^{-\bm{\mathcal{M}}^{\rm no}_t} + e^{-(\bm{\mathcal{M}}^{\rm no}_t)\trans} [\bm \Upsilon-\bm\Phi]\trans}{2} -\frac{\bm\Upsilon + \bm\Upsilon\trans}{4}- e^{-(\bm{\mathcal{M}}^{\rm no}_t)\trans} \frac{ \bm\Upsilon + \bm\Upsilon\trans}{4} e^{-\bm{\mathcal{M}}^{\rm no}_t} \Bigg] \cdot \bm{R}~{\tilde{\rho}}_{\rm no}\nonumber\\
    &+{\bm{R}}^\dagger \cdot\Bigg[ -\frac{i}{4}\bigg(\bm{\mathcal{T}}-e^{\bm{-(\mathcal{M}}^{\rm no}_t)\trans}\bm{\mathcal{T}}e^{-\bm{\mathcal{M}}^{\rm no}_t}\bigg)\Bigg] \cdot {\bm{R}}\,{\tilde{\rho}}_{\rm no} - \tr\left(\left[\bm \Upsilon+ [\bm \Upsilon-\bm \Phi] e^{-\bm{\mathcal{M}}^{\rm no}_t}\right](i\bm\Omega)\right)\,{\tilde{\rho}}_{\rm no}.
\end{align}
Combining Eq. \eqref{before_last_bos} with Eq. \eqref{major1_quadratures},
\begin{equation}\label{zeta-pairing-evolution_bos}
    \frac{d}{dt}\ln{\zeta}_t = \tr\left(\left[\bm \Upsilon+ [\bm \Upsilon-\bm \Phi] e^{-\bm{\mathcal{M}}^{\rm no}_t}\right](i\bm\Omega)\right),
\end{equation}
\begin{equation}\label{pairing_eM_eq_tmp2_bos}
    (i\bm\Omega) \frac{d e^{\bm{\mathcal{M}}^{\rm no}_t}}{dt} = -i \Big(\bm{\mathcal{T}}e^{\bm{\mathcal{M}}^{\rm no}_t} -e^{-(\bm{\mathcal{M}}^{\rm no}_t)\trans}\bm{\mathcal{T}}\Big) +2[\bm\Upsilon-\bm\Phi] + 2 e^{-(\bm{\mathcal{M}}^{\rm no}_t)\trans} [\bm\Upsilon-\bm\Phi]\trans e^{\bm{\mathcal{M}}^{\rm no}_t} - (\bm\Upsilon+\bm\Upsilon\trans) e^{\bm{\mathcal{M}}^{\rm no}_t} - e^{-(\bm{\mathcal{M}}^{\rm no}_t)\trans} (\bm\Upsilon + \bm\Upsilon\trans).
\end{equation}
Again, the matrix $\bm{\mathcal{M}}^{\rm no}_t$ is not so convenient to work with. For this reason, we recall the definition of the conditional covariance matrix $(\bm{\Theta}_{\rm no})_{ij} = \frac{1}{2}\tr\Big\{\{ R_i, R_j\}\rho(t)\Big\}$.  Given the Gaussian ansatz \eqref{gaussian-ansatz2_bos}, we use the formula  
\begin{equation}\label{pairing_theta_M_relation_bos}
    \bm\Theta_{\rm no} = -\coth(\bm{\mathcal{M}}_t^{\rm no}/2)\,(i\bm\Omega),
    \qquad 
    e^{\bm{\mathcal{M}}^{\rm no}_t}=\frac{\bm\Theta_{\rm no}(i\bm\Omega)-\mathds{1}}{\bm\Theta_{\rm no}(i\bm\Omega)+\mathds{1}}.
\end{equation}
which relates the matrix $\bm{\mathcal{M}}^{\rm no}_t$ to the correlation matrix $\bm{\Theta}_{\rm no}$. One can also readily show that 
\begin{equation}\label{pairing_theta_M_derivative_bos}
    \frac{d\bm\Theta_{\rm no}}{dt}(i\bm\Omega) = \frac{1}{2} \left(\mathds{1} + \bm\Theta_{\rm no}(i\bm\Omega)\right) \frac{d e^{\bm{\mathcal{M}}^{\rm no}_{t}}}{dt} \left(\mathds{1}+\bm\Theta_{\rm no}(i\bm\Omega)\right).
\end{equation}
Using Eq.~\eqref{pairing_theta_M_derivative_bos} to rewrite Eq. \eqref{pairing_eM_eq_tmp2_bos}, we get
\begin{equation}
    \frac{d\bm\Theta_{\rm no}}{dt} = - \big(\bm{\mathcal{W}} \bm\Theta_{\rm no} + \bm\Theta_{\rm no} \bm{\mathcal{W}}^\dagger\big) + \bm{\mathcal{F}} - (i\bm\Omega-\bm\Theta_{\rm no})\, \bm\Phi\trans \,(i\bm\Omega-\bm\Theta_{\rm no}) - (i\bm\Omega+\bm\Theta_{\rm no})\,\bm\Phi \,(i\bm\Omega+\bm\Theta_{\rm no}),
\end{equation}
where $\bm{\mathcal{W}}$ and $\bm{\mathcal{F}}$ are the matrices into Eqs.~(\ref{doppiaW_bos},\ref{F_bos}).
\\

\textbf{No-jump probability}: According to Eq. \eqref{zeta-pairing-evolution_bos}, $\zeta_t$ satisfies 
\begin{equation}\label{normalization_pairing1_bos}
    \frac{d}{dt} \ln \zeta_t = \tr\Big\{ (i\bm\Omega)\bm\Upsilon + e^{-{\bm{\mathcal{M}}_t^{\rm no}}} (i\bm\Omega)(\bm\Upsilon - \bm\Phi) \Big\},
\end{equation}
Thanks to the Jacobi formula, the partition function satisfies 
\begin{equation}\label{normalization_pairing2_bos}
    \frac{d}{dt} \ln Z_t^{\rm no} = \frac{1}{2} \tr \left\{ (1-e^{\bm{\mathcal{M}}^{\rm no}_t})^{-1} \frac{de^{\bm{\mathcal{M}}^{\rm no}_t}}{dt} \right\}.
\end{equation}
Combining Eqs.~(\ref{pairing_theta_M_relation_bos},~\ref{pairing_eM_eq_tmp2_bos},~\ref{normalization_pairing1_bos},~\ref{normalization_pairing2_bos},~\ref{identity_Mno_bos}) together with the invariance of the trace under transpose, we arrive at 
\begin{equation}
    \frac{d}{dt} \ln \zeta_t/Z_t^{\rm no} =\tr\Big\{ \bm\Phi (i\bm\Omega+\bm\Theta_{\rm no})\Big\}.
\end{equation}
The no-jump probability finally acquires the compact form
\begin{equation}
 P_{\rm no}(t) = \exp\left\{ - \int\limits_0^t dt'~\tr\Big[ \bm\Phi(i\bm\Omega+\bm\Theta_{\rm no}(t'))\Big] \right\}.
 \end{equation}

\end{widetext}

\bibliography{library2}

\begin{thebibliography}{52}%
\makeatletter
\providecommand \@ifxundefined [1]{%
 \@ifx{#1\undefined}
}%
\providecommand \@ifnum [1]{%
 \ifnum #1\expandafter \@firstoftwo
 \else \expandafter \@secondoftwo
 \fi
}%
\providecommand \@ifx [1]{%
 \ifx #1\expandafter \@firstoftwo
 \else \expandafter \@secondoftwo
 \fi
}%
\providecommand \natexlab [1]{#1}%
\providecommand \enquote  [1]{``#1''}%
\providecommand \bibnamefont  [1]{#1}%
\providecommand \bibfnamefont [1]{#1}%
\providecommand \citenamefont [1]{#1}%
\providecommand \href@noop [0]{\@secondoftwo}%
\providecommand \href [0]{\begingroup \@sanitize@url \@href}%
\providecommand \@href[1]{\@@startlink{#1}\@@href}%
\providecommand \@@href[1]{\endgroup#1\@@endlink}%
\providecommand \@sanitize@url [0]{\catcode `\\12\catcode `\$12\catcode
  `\&12\catcode `\#12\catcode `\^12\catcode `\_12\catcode `\%12\relax}%
\providecommand \@@startlink[1]{}%
\providecommand \@@endlink[0]{}%
\providecommand \url  [0]{\begingroup\@sanitize@url \@url }%
\providecommand \@url [1]{\endgroup\@href {#1}{\urlprefix }}%
\providecommand \urlprefix  [0]{URL }%
\providecommand \Eprint [0]{\href }%
\providecommand \doibase [0]{http://dx.doi.org/}%
\providecommand \selectlanguage [0]{\@gobble}%
\providecommand \bibinfo  [0]{\@secondoftwo}%
\providecommand \bibfield  [0]{\@secondoftwo}%
\providecommand \translation [1]{[#1]}%
\providecommand \BibitemOpen [0]{}%
\providecommand \bibitemStop [0]{}%
\providecommand \bibitemNoStop [0]{.\EOS\space}%
\providecommand \EOS [0]{\spacefactor3000\relax}%
\providecommand \BibitemShut  [1]{\csname bibitem#1\endcsname}%
\let\auto@bib@innerbib\@empty
\bibitem [{\citenamefont {Landi}\ \emph {et~al.}(2022)\citenamefont {Landi},
  \citenamefont {Poletti},\ and\ \citenamefont
  {Schaller}}]{landi2022nonequilibrium}%
  \BibitemOpen
  \bibfield  {author} {\bibinfo {author} {\bibfnamefont {G.~T.}\ \bibnamefont
  {Landi}}, \bibinfo {author} {\bibfnamefont {D.}~\bibnamefont {Poletti}}, \
  and\ \bibinfo {author} {\bibfnamefont {G.}~\bibnamefont {Schaller}},\
  }\href@noop {} {\bibfield  {journal} {\bibinfo  {journal} {Reviews of Modern
  Physics}\ }\textbf {\bibinfo {volume} {94}},\ \bibinfo {pages} {045006}
  (\bibinfo {year} {2022})}\BibitemShut {NoStop}%
\bibitem [{\citenamefont {Daley}(2014)}]{daley2014quantum}%
  \BibitemOpen
  \bibfield  {author} {\bibinfo {author} {\bibfnamefont {A.~J.}\ \bibnamefont
  {Daley}},\ }\href@noop {} {\bibfield  {journal} {\bibinfo  {journal}
  {Advances in Physics}\ }\textbf {\bibinfo {volume} {63}},\ \bibinfo {pages}
  {77} (\bibinfo {year} {2014})}\BibitemShut {NoStop}%
\bibitem [{\citenamefont {Wiseman}(1996)}]{wiseman1996quantum}%
  \BibitemOpen
  \bibfield  {author} {\bibinfo {author} {\bibfnamefont {H.~M.}\ \bibnamefont
  {Wiseman}},\ }\href@noop {} {\bibfield  {journal} {\bibinfo  {journal}
  {Quantum and Semiclassical Optics: Journal of the European Optical Society
  Part B}\ }\textbf {\bibinfo {volume} {8}},\ \bibinfo {pages} {205} (\bibinfo
  {year} {1996})}\BibitemShut {NoStop}%
\bibitem [{\citenamefont {Dalibard}\ \emph {et~al.}(1992)\citenamefont
  {Dalibard}, \citenamefont {Castin},\ and\ \citenamefont
  {M{\o}lmer}}]{dalibard1992wave}%
  \BibitemOpen
  \bibfield  {author} {\bibinfo {author} {\bibfnamefont {J.}~\bibnamefont
  {Dalibard}}, \bibinfo {author} {\bibfnamefont {Y.}~\bibnamefont {Castin}}, \
  and\ \bibinfo {author} {\bibfnamefont {K.}~\bibnamefont {M{\o}lmer}},\
  }\href@noop {} {\bibfield  {journal} {\bibinfo  {journal} {Physical review
  letters}\ }\textbf {\bibinfo {volume} {68}},\ \bibinfo {pages} {580}
  (\bibinfo {year} {1992})}\BibitemShut {NoStop}%
\bibitem [{\citenamefont {Gardiner}\ \emph {et~al.}(1992)\citenamefont
  {Gardiner}, \citenamefont {Parkins},\ and\ \citenamefont
  {Zoller}}]{gardiner1992wave}%
  \BibitemOpen
  \bibfield  {author} {\bibinfo {author} {\bibfnamefont {C.~W.}\ \bibnamefont
  {Gardiner}}, \bibinfo {author} {\bibfnamefont {A.~S.}\ \bibnamefont
  {Parkins}}, \ and\ \bibinfo {author} {\bibfnamefont {P.}~\bibnamefont
  {Zoller}},\ }\href@noop {} {\bibfield  {journal} {\bibinfo  {journal}
  {Physical Review A}\ }\textbf {\bibinfo {volume} {46}},\ \bibinfo {pages}
  {4363} (\bibinfo {year} {1992})}\BibitemShut {NoStop}%
\bibitem [{\citenamefont {Carollo}\ \emph {et~al.}(2019)\citenamefont
  {Carollo}, \citenamefont {Jack},\ and\ \citenamefont
  {Garrahan}}]{carollo2019unraveling}%
  \BibitemOpen
  \bibfield  {author} {\bibinfo {author} {\bibfnamefont {F.}~\bibnamefont
  {Carollo}}, \bibinfo {author} {\bibfnamefont {R.~L.}\ \bibnamefont {Jack}}, \
  and\ \bibinfo {author} {\bibfnamefont {J.~P.}\ \bibnamefont {Garrahan}},\
  }\href@noop {} {\bibfield  {journal} {\bibinfo  {journal} {Physical review
  letters}\ }\textbf {\bibinfo {volume} {122}},\ \bibinfo {pages} {130605}
  (\bibinfo {year} {2019})}\BibitemShut {NoStop}%
\bibitem [{\citenamefont {Levitov}\ and\ \citenamefont
  {Lesovik}(1993)}]{levitov1993charge}%
  \BibitemOpen
  \bibfield  {author} {\bibinfo {author} {\bibfnamefont {L.~S.}\ \bibnamefont
  {Levitov}}\ and\ \bibinfo {author} {\bibfnamefont {G.~B.}\ \bibnamefont
  {Lesovik}},\ }\href@noop {} {\bibfield  {journal} {\bibinfo  {journal} {JETP
  LETTERS C/C OF PIS'MA V ZHURNAL EKSPERIMENTAL'NOI TEORETICHESKOI FIZIKI}\
  }\textbf {\bibinfo {volume} {58}},\ \bibinfo {pages} {230} (\bibinfo {year}
  {1993})}\BibitemShut {NoStop}%
\bibitem [{\citenamefont {Esposito}\ \emph {et~al.}(2007)\citenamefont
  {Esposito}, \citenamefont {Harbola},\ and\ \citenamefont
  {Mukamel}}]{esposito2007fluctuation}%
  \BibitemOpen
  \bibfield  {author} {\bibinfo {author} {\bibfnamefont {M.}~\bibnamefont
  {Esposito}}, \bibinfo {author} {\bibfnamefont {U.}~\bibnamefont {Harbola}}, \
  and\ \bibinfo {author} {\bibfnamefont {S.}~\bibnamefont {Mukamel}},\
  }\href@noop {} {\bibfield  {journal} {\bibinfo  {journal} {Physical Review
  B}\ }\textbf {\bibinfo {volume} {75}},\ \bibinfo {pages} {155316} (\bibinfo
  {year} {2007})}\BibitemShut {NoStop}%
\bibitem [{\citenamefont {Esposito}\ \emph {et~al.}(2009)\citenamefont
  {Esposito}, \citenamefont {Harbola},\ and\ \citenamefont
  {Mukamel}}]{esposito2009nonequilibrium}%
  \BibitemOpen
  \bibfield  {author} {\bibinfo {author} {\bibfnamefont {M.}~\bibnamefont
  {Esposito}}, \bibinfo {author} {\bibfnamefont {U.}~\bibnamefont {Harbola}}, \
  and\ \bibinfo {author} {\bibfnamefont {S.}~\bibnamefont {Mukamel}},\
  }\href@noop {} {\bibfield  {journal} {\bibinfo  {journal} {Reviews of modern
  physics}\ }\textbf {\bibinfo {volume} {81}},\ \bibinfo {pages} {1665}
  (\bibinfo {year} {2009})}\BibitemShut {NoStop}%
\bibitem [{\citenamefont {Landi}\ \emph {et~al.}(2023)\citenamefont {Landi},
  \citenamefont {Kewming}, \citenamefont {Mitchison},\ and\ \citenamefont
  {Potts}}]{landi2023current}%
  \BibitemOpen
  \bibfield  {author} {\bibinfo {author} {\bibfnamefont {G.~T.}\ \bibnamefont
  {Landi}}, \bibinfo {author} {\bibfnamefont {M.~J.}\ \bibnamefont {Kewming}},
  \bibinfo {author} {\bibfnamefont {M.~T.}\ \bibnamefont {Mitchison}}, \ and\
  \bibinfo {author} {\bibfnamefont {P.~P.}\ \bibnamefont {Potts}},\ }\href@noop
  {} {\bibfield  {journal} {\bibinfo  {journal} {arXiv preprint
  arXiv:2303.04270}\ } (\bibinfo {year} {2023})}\BibitemShut {NoStop}%
\bibitem [{\citenamefont {Brandes}(2008)}]{brandes2008waiting}%
  \BibitemOpen
  \bibfield  {author} {\bibinfo {author} {\bibfnamefont {T.}~\bibnamefont
  {Brandes}},\ }\href@noop {} {\bibfield  {journal} {\bibinfo  {journal}
  {Annalen der Physik}\ }\textbf {\bibinfo {volume} {520}},\ \bibinfo {pages}
  {477} (\bibinfo {year} {2008})}\BibitemShut {NoStop}%
\bibitem [{\citenamefont {Landi}(2021)}]{landi2021waiting}%
  \BibitemOpen
  \bibfield  {author} {\bibinfo {author} {\bibfnamefont {G.~T.}\ \bibnamefont
  {Landi}},\ }\href@noop {} {\bibfield  {journal} {\bibinfo  {journal}
  {Physical Review B}\ }\textbf {\bibinfo {volume} {104}},\ \bibinfo {pages}
  {195408} (\bibinfo {year} {2021})}\BibitemShut {NoStop}%
\bibitem [{\citenamefont {Plenio}\ and\ \citenamefont
  {Knight}(1998)}]{plenio1998quantum}%
  \BibitemOpen
  \bibfield  {author} {\bibinfo {author} {\bibfnamefont {M.~B.}\ \bibnamefont
  {Plenio}}\ and\ \bibinfo {author} {\bibfnamefont {P.~L.}\ \bibnamefont
  {Knight}},\ }\href@noop {} {\bibfield  {journal} {\bibinfo  {journal}
  {Reviews of Modern Physics}\ }\textbf {\bibinfo {volume} {70}},\ \bibinfo
  {pages} {101} (\bibinfo {year} {1998})}\BibitemShut {NoStop}%
\bibitem [{\citenamefont {Purkayastha}(2022)}]{Purkayastha2022}%
  \BibitemOpen
  \bibfield  {author} {\bibinfo {author} {\bibfnamefont {A.}~\bibnamefont
  {Purkayastha}},\ }\href {\doibase 10.1103/physreva.105.062204} {\bibfield
  {journal} {\bibinfo  {journal} {Physical Review A}\ }\textbf {\bibinfo
  {volume} {105}} (\bibinfo {year} {2022}),\
  10.1103/physreva.105.062204}\BibitemShut {NoStop}%
\bibitem [{\citenamefont {Turkeshi}\ and\ \citenamefont
  {Schir{\'o}}(2021)}]{turkeshi2021diffusion}%
  \BibitemOpen
  \bibfield  {author} {\bibinfo {author} {\bibfnamefont {X.}~\bibnamefont
  {Turkeshi}}\ and\ \bibinfo {author} {\bibfnamefont {M.}~\bibnamefont
  {Schir{\'o}}},\ }\href@noop {} {\bibfield  {journal} {\bibinfo  {journal}
  {Physical Review B}\ }\textbf {\bibinfo {volume} {104}},\ \bibinfo {pages}
  {144301} (\bibinfo {year} {2021})}\BibitemShut {NoStop}%
\bibitem [{\citenamefont {Coppola}\ \emph {et~al.}(2023)\citenamefont
  {Coppola}, \citenamefont {Landi},\ and\ \citenamefont
  {Karevski}}]{coppola2023wigner}%
  \BibitemOpen
  \bibfield  {author} {\bibinfo {author} {\bibfnamefont {M.}~\bibnamefont
  {Coppola}}, \bibinfo {author} {\bibfnamefont {G.~T.}\ \bibnamefont {Landi}},
  \ and\ \bibinfo {author} {\bibfnamefont {D.}~\bibnamefont {Karevski}},\
  }\href@noop {} {\bibfield  {journal} {\bibinfo  {journal} {Physical Review
  A}\ }\textbf {\bibinfo {volume} {107}},\ \bibinfo {pages} {052213} (\bibinfo
  {year} {2023})}\BibitemShut {NoStop}%
\bibitem [{\citenamefont {Genoni}\ \emph {et~al.}(2016)\citenamefont {Genoni},
  \citenamefont {Lami},\ and\ \citenamefont {Serafini}}]{Genoni2016}%
  \BibitemOpen
  \bibfield  {author} {\bibinfo {author} {\bibfnamefont {M.~G.}\ \bibnamefont
  {Genoni}}, \bibinfo {author} {\bibfnamefont {L.}~\bibnamefont {Lami}}, \ and\
  \bibinfo {author} {\bibfnamefont {A.}~\bibnamefont {Serafini}},\ }\href
  {\doibase 10.1080/00107514.2015.1125624} {\bibfield  {journal} {\bibinfo
  {journal} {Contemporary Physics}\ }\textbf {\bibinfo {volume} {57}},\
  \bibinfo {pages} {331} (\bibinfo {year} {2016})}\BibitemShut {NoStop}%
\bibitem [{\citenamefont {Belenchia}\ \emph {et~al.}(2022)\citenamefont
  {Belenchia}, \citenamefont {Paternostro},\ and\ \citenamefont
  {Landi}}]{Belenchia2022}%
  \BibitemOpen
  \bibfield  {author} {\bibinfo {author} {\bibfnamefont {A.}~\bibnamefont
  {Belenchia}}, \bibinfo {author} {\bibfnamefont {M.}~\bibnamefont
  {Paternostro}}, \ and\ \bibinfo {author} {\bibfnamefont {G.~T.}\ \bibnamefont
  {Landi}},\ }\href {\doibase 10.1103/physreva.105.022213} {\bibfield
  {journal} {\bibinfo  {journal} {Physical Review A}\ }\textbf {\bibinfo
  {volume} {105}} (\bibinfo {year} {2022}),\
  10.1103/physreva.105.022213}\BibitemShut {NoStop}%
\bibitem [{\citenamefont {Castro-Alvaredo}\ \emph {et~al.}(2016)\citenamefont
  {Castro-Alvaredo}, \citenamefont {Doyon},\ and\ \citenamefont
  {Yoshimura}}]{castro2016emergent}%
  \BibitemOpen
  \bibfield  {author} {\bibinfo {author} {\bibfnamefont {O.~A.}\ \bibnamefont
  {Castro-Alvaredo}}, \bibinfo {author} {\bibfnamefont {B.}~\bibnamefont
  {Doyon}}, \ and\ \bibinfo {author} {\bibfnamefont {T.}~\bibnamefont
  {Yoshimura}},\ }\href@noop {} {\bibfield  {journal} {\bibinfo  {journal}
  {Physical Review X}\ }\textbf {\bibinfo {volume} {6}},\ \bibinfo {pages}
  {041065} (\bibinfo {year} {2016})}\BibitemShut {NoStop}%
\bibitem [{\citenamefont {Ruggiero}\ \emph {et~al.}(2020)\citenamefont
  {Ruggiero}, \citenamefont {Calabrese}, \citenamefont {Doyon},\ and\
  \citenamefont {Dubail}}]{ruggiero2020quantum}%
  \BibitemOpen
  \bibfield  {author} {\bibinfo {author} {\bibfnamefont {P.}~\bibnamefont
  {Ruggiero}}, \bibinfo {author} {\bibfnamefont {P.}~\bibnamefont {Calabrese}},
  \bibinfo {author} {\bibfnamefont {B.}~\bibnamefont {Doyon}}, \ and\ \bibinfo
  {author} {\bibfnamefont {J.}~\bibnamefont {Dubail}},\ }\href@noop {}
  {\bibfield  {journal} {\bibinfo  {journal} {Physical review letters}\
  }\textbf {\bibinfo {volume} {124}},\ \bibinfo {pages} {140603} (\bibinfo
  {year} {2020})}\BibitemShut {NoStop}%
\bibitem [{\citenamefont {Doyon}(2020)}]{doyon2020lecture}%
  \BibitemOpen
  \bibfield  {author} {\bibinfo {author} {\bibfnamefont {B.}~\bibnamefont
  {Doyon}},\ }\href@noop {} {\bibfield  {journal} {\bibinfo  {journal} {SciPost
  Physics Lecture Notes}\ ,\ \bibinfo {pages} {018}} (\bibinfo {year}
  {2020})}\BibitemShut {NoStop}%
\bibitem [{\citenamefont {Bastianello}\ \emph {et~al.}(2019)\citenamefont
  {Bastianello}, \citenamefont {Alba},\ and\ \citenamefont
  {Caux}}]{bastianello2019generalized}%
  \BibitemOpen
  \bibfield  {author} {\bibinfo {author} {\bibfnamefont {A.}~\bibnamefont
  {Bastianello}}, \bibinfo {author} {\bibfnamefont {V.}~\bibnamefont {Alba}}, \
  and\ \bibinfo {author} {\bibfnamefont {J.-S.}\ \bibnamefont {Caux}},\
  }\href@noop {} {\bibfield  {journal} {\bibinfo  {journal} {Physical Review
  Letters}\ }\textbf {\bibinfo {volume} {123}},\ \bibinfo {pages} {130602}
  (\bibinfo {year} {2019})}\BibitemShut {NoStop}%
\bibitem [{\citenamefont {Capizzi}\ \emph {et~al.}(2022)\citenamefont
  {Capizzi}, \citenamefont {Scopa}, \citenamefont {Rottoli},\ and\
  \citenamefont {Calabrese}}]{capizzi2022domain}%
  \BibitemOpen
  \bibfield  {author} {\bibinfo {author} {\bibfnamefont {L.}~\bibnamefont
  {Capizzi}}, \bibinfo {author} {\bibfnamefont {S.}~\bibnamefont {Scopa}},
  \bibinfo {author} {\bibfnamefont {F.}~\bibnamefont {Rottoli}}, \ and\
  \bibinfo {author} {\bibfnamefont {P.}~\bibnamefont {Calabrese}},\ }\href@noop
  {} {\bibfield  {journal} {\bibinfo  {journal} {Europhysics Letters}\ }
  (\bibinfo {year} {2022})}\BibitemShut {NoStop}%
\bibitem [{\citenamefont {Scopa}\ \emph {et~al.}(2022)\citenamefont {Scopa},
  \citenamefont {Calabrese},\ and\ \citenamefont {Dubail}}]{scopa2022exact}%
  \BibitemOpen
  \bibfield  {author} {\bibinfo {author} {\bibfnamefont {S.}~\bibnamefont
  {Scopa}}, \bibinfo {author} {\bibfnamefont {P.}~\bibnamefont {Calabrese}}, \
  and\ \bibinfo {author} {\bibfnamefont {J.}~\bibnamefont {Dubail}},\
  }\href@noop {} {\bibfield  {journal} {\bibinfo  {journal} {SciPost Physics}\
  }\textbf {\bibinfo {volume} {12}},\ \bibinfo {pages} {207} (\bibinfo {year}
  {2022})}\BibitemShut {NoStop}%
\bibitem [{\citenamefont {Fagotti}(2017)}]{fagotti2017higher}%
  \BibitemOpen
  \bibfield  {author} {\bibinfo {author} {\bibfnamefont {M.}~\bibnamefont
  {Fagotti}},\ }\href@noop {} {\bibfield  {journal} {\bibinfo  {journal}
  {Physical Review B}\ }\textbf {\bibinfo {volume} {96}},\ \bibinfo {pages}
  {220302} (\bibinfo {year} {2017})}\BibitemShut {NoStop}%
\bibitem [{\citenamefont {Scopa}\ \emph {et~al.}(2021)\citenamefont {Scopa},
  \citenamefont {Krajenbrink}, \citenamefont {Calabrese},\ and\ \citenamefont
  {Dubail}}]{scopa2021exact}%
  \BibitemOpen
  \bibfield  {author} {\bibinfo {author} {\bibfnamefont {S.}~\bibnamefont
  {Scopa}}, \bibinfo {author} {\bibfnamefont {A.}~\bibnamefont {Krajenbrink}},
  \bibinfo {author} {\bibfnamefont {P.}~\bibnamefont {Calabrese}}, \ and\
  \bibinfo {author} {\bibfnamefont {J.}~\bibnamefont {Dubail}},\ }\href@noop {}
  {\bibfield  {journal} {\bibinfo  {journal} {Journal of Physics A:
  Mathematical and Theoretical}\ }\textbf {\bibinfo {volume} {54}},\ \bibinfo
  {pages} {404002} (\bibinfo {year} {2021})}\BibitemShut {NoStop}%
\bibitem [{\citenamefont {Dubail}\ \emph {et~al.}(2017)\citenamefont {Dubail},
  \citenamefont {St{\'e}phan}, \citenamefont {Viti},\ and\ \citenamefont
  {Calabrese}}]{dubail2017conformal}%
  \BibitemOpen
  \bibfield  {author} {\bibinfo {author} {\bibfnamefont {J.}~\bibnamefont
  {Dubail}}, \bibinfo {author} {\bibfnamefont {J.-M.}\ \bibnamefont
  {St{\'e}phan}}, \bibinfo {author} {\bibfnamefont {J.}~\bibnamefont {Viti}}, \
  and\ \bibinfo {author} {\bibfnamefont {P.}~\bibnamefont {Calabrese}},\
  }\href@noop {} {\bibfield  {journal} {\bibinfo  {journal} {SciPost Physics}\
  }\textbf {\bibinfo {volume} {2}},\ \bibinfo {pages} {002} (\bibinfo {year}
  {2017})}\BibitemShut {NoStop}%
\bibitem [{\citenamefont {Collura}\ \emph {et~al.}(2018)\citenamefont
  {Collura}, \citenamefont {De~Luca},\ and\ \citenamefont
  {Viti}}]{collura2018analytic}%
  \BibitemOpen
  \bibfield  {author} {\bibinfo {author} {\bibfnamefont {M.}~\bibnamefont
  {Collura}}, \bibinfo {author} {\bibfnamefont {A.}~\bibnamefont {De~Luca}}, \
  and\ \bibinfo {author} {\bibfnamefont {J.}~\bibnamefont {Viti}},\ }\href@noop
  {} {\bibfield  {journal} {\bibinfo  {journal} {Physical Review B}\ }\textbf
  {\bibinfo {volume} {97}},\ \bibinfo {pages} {081111} (\bibinfo {year}
  {2018})}\BibitemShut {NoStop}%
\bibitem [{\citenamefont {Collura}\ \emph {et~al.}(2020)\citenamefont
  {Collura}, \citenamefont {De~Luca}, \citenamefont {Calabrese},\ and\
  \citenamefont {Dubail}}]{collura2020domain}%
  \BibitemOpen
  \bibfield  {author} {\bibinfo {author} {\bibfnamefont {M.}~\bibnamefont
  {Collura}}, \bibinfo {author} {\bibfnamefont {A.}~\bibnamefont {De~Luca}},
  \bibinfo {author} {\bibfnamefont {P.}~\bibnamefont {Calabrese}}, \ and\
  \bibinfo {author} {\bibfnamefont {J.}~\bibnamefont {Dubail}},\ }\href@noop {}
  {\bibfield  {journal} {\bibinfo  {journal} {Physical Review B}\ }\textbf
  {\bibinfo {volume} {102}},\ \bibinfo {pages} {180409} (\bibinfo {year}
  {2020})}\BibitemShut {NoStop}%
\bibitem [{\citenamefont {Alba}\ \emph {et~al.}(2021)\citenamefont {Alba},
  \citenamefont {Bertini}, \citenamefont {Fagotti}, \citenamefont {Piroli},\
  and\ \citenamefont {Ruggiero}}]{alba2021generalized}%
  \BibitemOpen
  \bibfield  {author} {\bibinfo {author} {\bibfnamefont {V.}~\bibnamefont
  {Alba}}, \bibinfo {author} {\bibfnamefont {B.}~\bibnamefont {Bertini}},
  \bibinfo {author} {\bibfnamefont {M.}~\bibnamefont {Fagotti}}, \bibinfo
  {author} {\bibfnamefont {L.}~\bibnamefont {Piroli}}, \ and\ \bibinfo {author}
  {\bibfnamefont {P.}~\bibnamefont {Ruggiero}},\ }\href@noop {} {\bibfield
  {journal} {\bibinfo  {journal} {Journal of Statistical Mechanics: Theory and
  Experiment}\ }\textbf {\bibinfo {volume} {2021}},\ \bibinfo {pages} {114004}
  (\bibinfo {year} {2021})}\BibitemShut {NoStop}%
\bibitem [{\citenamefont {Bouchoule}\ and\ \citenamefont
  {Dubail}(2022)}]{bouchoule2022generalized}%
  \BibitemOpen
  \bibfield  {author} {\bibinfo {author} {\bibfnamefont {I.}~\bibnamefont
  {Bouchoule}}\ and\ \bibinfo {author} {\bibfnamefont {J.}~\bibnamefont
  {Dubail}},\ }\href@noop {} {\bibfield  {journal} {\bibinfo  {journal}
  {Journal of Statistical Mechanics: Theory and Experiment}\ }\textbf {\bibinfo
  {volume} {2022}},\ \bibinfo {pages} {014003} (\bibinfo {year}
  {2022})}\BibitemShut {NoStop}%
\bibitem [{\citenamefont {Bulchandani}\ \emph {et~al.}(2017)\citenamefont
  {Bulchandani}, \citenamefont {Vasseur}, \citenamefont {Karrasch},\ and\
  \citenamefont {Moore}}]{bulchandani2017solvable}%
  \BibitemOpen
  \bibfield  {author} {\bibinfo {author} {\bibfnamefont {V.~B.}\ \bibnamefont
  {Bulchandani}}, \bibinfo {author} {\bibfnamefont {R.}~\bibnamefont
  {Vasseur}}, \bibinfo {author} {\bibfnamefont {C.}~\bibnamefont {Karrasch}}, \
  and\ \bibinfo {author} {\bibfnamefont {J.~E.}\ \bibnamefont {Moore}},\
  }\href@noop {} {\bibfield  {journal} {\bibinfo  {journal} {Physical review
  letters}\ }\textbf {\bibinfo {volume} {119}},\ \bibinfo {pages} {220604}
  (\bibinfo {year} {2017})}\BibitemShut {NoStop}%
\bibitem [{\citenamefont {Bulchandani}\ \emph {et~al.}(2018)\citenamefont
  {Bulchandani}, \citenamefont {Vasseur}, \citenamefont {Karrasch},\ and\
  \citenamefont {Moore}}]{bulchandani2018bethe}%
  \BibitemOpen
  \bibfield  {author} {\bibinfo {author} {\bibfnamefont {V.~B.}\ \bibnamefont
  {Bulchandani}}, \bibinfo {author} {\bibfnamefont {R.}~\bibnamefont
  {Vasseur}}, \bibinfo {author} {\bibfnamefont {C.}~\bibnamefont {Karrasch}}, \
  and\ \bibinfo {author} {\bibfnamefont {J.~E.}\ \bibnamefont {Moore}},\
  }\href@noop {} {\bibfield  {journal} {\bibinfo  {journal} {Physical Review
  B}\ }\textbf {\bibinfo {volume} {97}},\ \bibinfo {pages} {045407} (\bibinfo
  {year} {2018})}\BibitemShut {NoStop}%
\bibitem [{\citenamefont {Doyon}\ \emph {et~al.}(2018)\citenamefont {Doyon},
  \citenamefont {Yoshimura},\ and\ \citenamefont {Caux}}]{doyon2018soliton}%
  \BibitemOpen
  \bibfield  {author} {\bibinfo {author} {\bibfnamefont {B.}~\bibnamefont
  {Doyon}}, \bibinfo {author} {\bibfnamefont {T.}~\bibnamefont {Yoshimura}}, \
  and\ \bibinfo {author} {\bibfnamefont {J.-S.}\ \bibnamefont {Caux}},\
  }\href@noop {} {\bibfield  {journal} {\bibinfo  {journal} {Physical review
  letters}\ }\textbf {\bibinfo {volume} {120}},\ \bibinfo {pages} {045301}
  (\bibinfo {year} {2018})}\BibitemShut {NoStop}%
\bibitem [{\citenamefont {Schemmer}\ \emph {et~al.}(2019)\citenamefont
  {Schemmer}, \citenamefont {Bouchoule}, \citenamefont {Doyon},\ and\
  \citenamefont {Dubail}}]{schemmer2019generalized}%
  \BibitemOpen
  \bibfield  {author} {\bibinfo {author} {\bibfnamefont {M.}~\bibnamefont
  {Schemmer}}, \bibinfo {author} {\bibfnamefont {I.}~\bibnamefont {Bouchoule}},
  \bibinfo {author} {\bibfnamefont {B.}~\bibnamefont {Doyon}}, \ and\ \bibinfo
  {author} {\bibfnamefont {J.}~\bibnamefont {Dubail}},\ }\href@noop {}
  {\bibfield  {journal} {\bibinfo  {journal} {Physical review letters}\
  }\textbf {\bibinfo {volume} {122}},\ \bibinfo {pages} {090601} (\bibinfo
  {year} {2019})}\BibitemShut {NoStop}%
\bibitem [{\citenamefont {Malvania}\ \emph {et~al.}(2021)\citenamefont
  {Malvania}, \citenamefont {Zhang}, \citenamefont {Le}, \citenamefont
  {Dubail}, \citenamefont {Rigol},\ and\ \citenamefont
  {Weiss}}]{malvania2021generalized}%
  \BibitemOpen
  \bibfield  {author} {\bibinfo {author} {\bibfnamefont {N.}~\bibnamefont
  {Malvania}}, \bibinfo {author} {\bibfnamefont {Y.}~\bibnamefont {Zhang}},
  \bibinfo {author} {\bibfnamefont {Y.}~\bibnamefont {Le}}, \bibinfo {author}
  {\bibfnamefont {J.}~\bibnamefont {Dubail}}, \bibinfo {author} {\bibfnamefont
  {M.}~\bibnamefont {Rigol}}, \ and\ \bibinfo {author} {\bibfnamefont {D.~S.}\
  \bibnamefont {Weiss}},\ }\href@noop {} {\bibfield  {journal} {\bibinfo
  {journal} {Science}\ }\textbf {\bibinfo {volume} {373}},\ \bibinfo {pages}
  {1129} (\bibinfo {year} {2021})}\BibitemShut {NoStop}%
\bibitem [{\citenamefont {Collura}\ \emph {et~al.}(2012)\citenamefont
  {Collura}, \citenamefont {Aufderheide}, \citenamefont {Roux},\ and\
  \citenamefont {Karevski}}]{collura2012entangling}%
  \BibitemOpen
  \bibfield  {author} {\bibinfo {author} {\bibfnamefont {M.}~\bibnamefont
  {Collura}}, \bibinfo {author} {\bibfnamefont {H.}~\bibnamefont
  {Aufderheide}}, \bibinfo {author} {\bibfnamefont {G.}~\bibnamefont {Roux}}, \
  and\ \bibinfo {author} {\bibfnamefont {D.}~\bibnamefont {Karevski}},\
  }\href@noop {} {\bibfield  {journal} {\bibinfo  {journal} {Physical Review
  A}\ }\textbf {\bibinfo {volume} {86}},\ \bibinfo {pages} {013615} (\bibinfo
  {year} {2012})}\BibitemShut {NoStop}%
\bibitem [{\citenamefont {Wendenbaum}\ \emph {et~al.}(2013)\citenamefont
  {Wendenbaum}, \citenamefont {Collura},\ and\ \citenamefont
  {Karevski}}]{wendenbaum2013hydrodynamic}%
  \BibitemOpen
  \bibfield  {author} {\bibinfo {author} {\bibfnamefont {P.}~\bibnamefont
  {Wendenbaum}}, \bibinfo {author} {\bibfnamefont {M.}~\bibnamefont {Collura}},
  \ and\ \bibinfo {author} {\bibfnamefont {D.}~\bibnamefont {Karevski}},\
  }\href@noop {} {\bibfield  {journal} {\bibinfo  {journal} {Physical Review
  A}\ }\textbf {\bibinfo {volume} {87}},\ \bibinfo {pages} {023624} (\bibinfo
  {year} {2013})}\BibitemShut {NoStop}%
\bibitem [{\citenamefont {Cao}\ \emph {et~al.}(2019)\citenamefont {Cao},
  \citenamefont {Tilloy},\ and\ \citenamefont {De~Luca}}]{cao2019entanglement}%
  \BibitemOpen
  \bibfield  {author} {\bibinfo {author} {\bibfnamefont {X.}~\bibnamefont
  {Cao}}, \bibinfo {author} {\bibfnamefont {A.}~\bibnamefont {Tilloy}}, \ and\
  \bibinfo {author} {\bibfnamefont {A.}~\bibnamefont {De~Luca}},\ }\href@noop
  {} {\bibfield  {journal} {\bibinfo  {journal} {SciPost Physics}\ }\textbf
  {\bibinfo {volume} {7}},\ \bibinfo {pages} {024} (\bibinfo {year}
  {2019})}\BibitemShut {NoStop}%
\bibitem [{\citenamefont {Jin}\ \emph {et~al.}(2021)\citenamefont {Jin},
  \citenamefont {Gauti{\'e}}, \citenamefont {Krajenbrink}, \citenamefont
  {Ruggiero},\ and\ \citenamefont {Yoshimura}}]{jin2021interplay}%
  \BibitemOpen
  \bibfield  {author} {\bibinfo {author} {\bibfnamefont {T.}~\bibnamefont
  {Jin}}, \bibinfo {author} {\bibfnamefont {T.}~\bibnamefont {Gauti{\'e}}},
  \bibinfo {author} {\bibfnamefont {A.}~\bibnamefont {Krajenbrink}}, \bibinfo
  {author} {\bibfnamefont {P.}~\bibnamefont {Ruggiero}}, \ and\ \bibinfo
  {author} {\bibfnamefont {T.}~\bibnamefont {Yoshimura}},\ }\href@noop {}
  {\bibfield  {journal} {\bibinfo  {journal} {Journal of Physics A:
  Mathematical and Theoretical}\ }\textbf {\bibinfo {volume} {54}},\ \bibinfo
  {pages} {404001} (\bibinfo {year} {2021})}\BibitemShut {NoStop}%
\bibitem [{\citenamefont {Bouchoule}\ \emph {et~al.}(2020)\citenamefont
  {Bouchoule}, \citenamefont {Doyon},\ and\ \citenamefont
  {Dubail}}]{bouchoule2020effect}%
  \BibitemOpen
  \bibfield  {author} {\bibinfo {author} {\bibfnamefont {I.}~\bibnamefont
  {Bouchoule}}, \bibinfo {author} {\bibfnamefont {B.}~\bibnamefont {Doyon}}, \
  and\ \bibinfo {author} {\bibfnamefont {J.}~\bibnamefont {Dubail}},\
  }\href@noop {} {\bibfield  {journal} {\bibinfo  {journal} {SciPost Physics}\
  }\textbf {\bibinfo {volume} {9}},\ \bibinfo {pages} {044} (\bibinfo {year}
  {2020})}\BibitemShut {NoStop}%
\bibitem [{\citenamefont {Dast}\ \emph {et~al.}(2014)\citenamefont {Dast},
  \citenamefont {Haag}, \citenamefont {Cartarius},\ and\ \citenamefont
  {Wunner}}]{dast2014quantum}%
  \BibitemOpen
  \bibfield  {author} {\bibinfo {author} {\bibfnamefont {D.}~\bibnamefont
  {Dast}}, \bibinfo {author} {\bibfnamefont {D.}~\bibnamefont {Haag}}, \bibinfo
  {author} {\bibfnamefont {H.}~\bibnamefont {Cartarius}}, \ and\ \bibinfo
  {author} {\bibfnamefont {G.}~\bibnamefont {Wunner}},\ }\href@noop {}
  {\bibfield  {journal} {\bibinfo  {journal} {Physical Review A}\ }\textbf
  {\bibinfo {volume} {90}},\ \bibinfo {pages} {052120} (\bibinfo {year}
  {2014})}\BibitemShut {NoStop}%
\bibitem [{\citenamefont {Alba}\ and\ \citenamefont
  {Carollo}(2022{\natexlab{a}})}]{alba2022noninteracting}%
  \BibitemOpen
  \bibfield  {author} {\bibinfo {author} {\bibfnamefont {V.}~\bibnamefont
  {Alba}}\ and\ \bibinfo {author} {\bibfnamefont {F.}~\bibnamefont {Carollo}},\
  }\href@noop {} {\bibfield  {journal} {\bibinfo  {journal} {Physical Review
  B}\ }\textbf {\bibinfo {volume} {105}},\ \bibinfo {pages} {054303} (\bibinfo
  {year} {2022}{\natexlab{a}})}\BibitemShut {NoStop}%
\bibitem [{\citenamefont {Alba}\ and\ \citenamefont
  {Carollo}(2022{\natexlab{b}})}]{alba2022hydrodynamics}%
  \BibitemOpen
  \bibfield  {author} {\bibinfo {author} {\bibfnamefont {V.}~\bibnamefont
  {Alba}}\ and\ \bibinfo {author} {\bibfnamefont {F.}~\bibnamefont {Carollo}},\
  }\href@noop {} {\bibfield  {journal} {\bibinfo  {journal} {Journal of Physics
  A: Mathematical and Theoretical}\ }\textbf {\bibinfo {volume} {55}},\
  \bibinfo {pages} {074002} (\bibinfo {year} {2022}{\natexlab{b}})}\BibitemShut
  {NoStop}%
\bibitem [{\citenamefont {Carollo}\ and\ \citenamefont
  {Alba}(2022)}]{carollo2022dissipative}%
  \BibitemOpen
  \bibfield  {author} {\bibinfo {author} {\bibfnamefont {F.}~\bibnamefont
  {Carollo}}\ and\ \bibinfo {author} {\bibfnamefont {V.}~\bibnamefont {Alba}},\
  }\href@noop {} {\bibfield  {journal} {\bibinfo  {journal} {Physical Review
  B}\ }\textbf {\bibinfo {volume} {105}},\ \bibinfo {pages} {144305} (\bibinfo
  {year} {2022})}\BibitemShut {NoStop}%
\bibitem [{\citenamefont {Aspelmeyer}\ \emph {et~al.}(2014)\citenamefont
  {Aspelmeyer}, \citenamefont {Kippenberg},\ and\ \citenamefont
  {Marquardt}}]{aspelmeyer2014cavity}%
  \BibitemOpen
  \bibfield  {author} {\bibinfo {author} {\bibfnamefont {M.}~\bibnamefont
  {Aspelmeyer}}, \bibinfo {author} {\bibfnamefont {T.~J.}\ \bibnamefont
  {Kippenberg}}, \ and\ \bibinfo {author} {\bibfnamefont {F.}~\bibnamefont
  {Marquardt}},\ }\href@noop {} {\bibfield  {journal} {\bibinfo  {journal}
  {Reviews of Modern Physics}\ }\textbf {\bibinfo {volume} {86}},\ \bibinfo
  {pages} {1391} (\bibinfo {year} {2014})}\BibitemShut {NoStop}%
\bibitem [{\citenamefont {Bowen}\ and\ \citenamefont
  {Milburn}(2015)}]{bowen2015quantum}%
  \BibitemOpen
  \bibfield  {author} {\bibinfo {author} {\bibfnamefont {W.~P.}\ \bibnamefont
  {Bowen}}\ and\ \bibinfo {author} {\bibfnamefont {G.~J.}\ \bibnamefont
  {Milburn}},\ }\href@noop {} {\emph {\bibinfo {title} {Quantum
  optomechanics}}}\ (\bibinfo  {publisher} {CRC press},\ \bibinfo {year}
  {2015})\BibitemShut {NoStop}%
\bibitem [{\citenamefont {Peschel}(2003)}]{peschel2003calculation}%
  \BibitemOpen
  \bibfield  {author} {\bibinfo {author} {\bibfnamefont {I.}~\bibnamefont
  {Peschel}},\ }\href@noop {} {\bibfield  {journal} {\bibinfo  {journal}
  {Journal of Physics A: Mathematical and General}\ }\textbf {\bibinfo {volume}
  {36}},\ \bibinfo {pages} {L205} (\bibinfo {year} {2003})}\BibitemShut
  {NoStop}%
\bibitem [{\citenamefont {Peschel}\ and\ \citenamefont
  {Eisler}(2009)}]{peschel2009reduced}%
  \BibitemOpen
  \bibfield  {author} {\bibinfo {author} {\bibfnamefont {I.}~\bibnamefont
  {Peschel}}\ and\ \bibinfo {author} {\bibfnamefont {V.}~\bibnamefont
  {Eisler}},\ }\href@noop {} {\bibfield  {journal} {\bibinfo  {journal}
  {Journal of physics a: mathematical and theoretical}\ }\textbf {\bibinfo
  {volume} {42}},\ \bibinfo {pages} {504003} (\bibinfo {year}
  {2009})}\BibitemShut {NoStop}%
\bibitem [{\citenamefont {Peschel}(2012)}]{peschel2012entanglement}%
  \BibitemOpen
  \bibfield  {author} {\bibinfo {author} {\bibfnamefont {I.}~\bibnamefont
  {Peschel}},\ }\href@noop {} {\bibfield  {journal} {\bibinfo  {journal}
  {Brazilian Journal of Physics}\ }\textbf {\bibinfo {volume} {42}},\ \bibinfo
  {pages} {267} (\bibinfo {year} {2012})}\BibitemShut {NoStop}%
\bibitem [{\citenamefont {Bravyi}(2004)}]{bravyi2004lagrangian}%
  \BibitemOpen
  \bibfield  {author} {\bibinfo {author} {\bibfnamefont {S.}~\bibnamefont
  {Bravyi}},\ }\href@noop {} {\bibfield  {journal} {\bibinfo  {journal} {arXiv
  preprint quant-ph/0404180}\ } (\bibinfo {year} {2004})}\BibitemShut {NoStop}%
\bibitem [{\citenamefont {Eisler}\ and\ \citenamefont
  {Zimbor{\'a}s}(2015)}]{eisler2015partial}%
  \BibitemOpen
  \bibfield  {author} {\bibinfo {author} {\bibfnamefont {V.}~\bibnamefont
  {Eisler}}\ and\ \bibinfo {author} {\bibfnamefont {Z.}~\bibnamefont
  {Zimbor{\'a}s}},\ }\href@noop {} {\bibfield  {journal} {\bibinfo  {journal}
  {New Journal of Physics}\ }\textbf {\bibinfo {volume} {17}},\ \bibinfo
  {pages} {053048} (\bibinfo {year} {2015})}\BibitemShut {NoStop}%
\end{thebibliography}%

\end{document}